\documentclass[12pt]{article}

\usepackage[english]{babel}

\usepackage{amssymb}
\usepackage{graphicx}
\usepackage{amsmath}
\usepackage{hyperref}
\usepackage{caption}
\usepackage{subcaption}
\usepackage{cite}
\usepackage{float}
\usepackage{pdflscape}
\usepackage{booktabs}
\usepackage[title]{appendix}
\usepackage{placeins}

\title{Noncommutative effects in Bianchi I cosmology with reduced relativistic gas}

\author{\small
	T. M. Abreu\thanks{\texttt{thiago\_moralles@ufjf.br}}, 
	G. Oliveira-Neto\thanks{\texttt{gilneto@fisica.ufjf.br}}, 
	A. C. R. Mendes\thanks{\texttt{albertcrm@gmail.com}}, 
	and S. C. Reis\thanks{\texttt{simplim15@hotmail.com}} \\
	\small\textit{Universidade Federal de Juiz de Fora, Brazil}
}

\date{\today}

\begin{document}

\maketitle

\begin{abstract}
In this work, we conduct a noncommutative analysis of a Bianchi I model coupled to a reduced relativistic gas. The classical field equations are modified by introducing noncommutativity through a generalized symplectic formalism. We observe that the noncommutative version of this model exhibits several significant differences compared to its commutative counterpart, as it alters the dynamics of the universe's expansion and isotropization.
	
\end{abstract}

\section{Introduction}
The study of noncommutative (NC) geometry  has profound implications in various fields of physics, particularly in cosmology \cite{MarcolliPierpaoli2010} and quantum gravity \cite{DouglasNekrasov2001}. Noncommutative geometry introduces modifications to the underlying space-time structure, which can lead to new insights and potentially resolve outstanding problems in theoretical physics. In this work, we explore the effects of noncommutativity on a Bianchi I (BI) cosmological model coupled to a reduced relativistic gas (RRG), with respect to accelerated expansion, the transition between cosmological eras, and isotropization.

The Bianchi I model is a type of anisotropic and homogeneous universe model that generalizes the flat homogeneous and isotropic Friedmann-Robertson-Walker (FRW) model, being useful for studying a possible transition of the universe from a previous anisotropic stage to the current isotropic stage. The model is part of the Bianchi classification, which categorizes nine types of homogeneous and anisotropic universe models that are solutions of Einstein’s field equations \cite{Wald1984, Ryan1975, Ellis1969}.

The reduced relativistic gas model \cite{BerredoPeixoto2005, Fabris2009, Fabris2012} consists of an approach to the matter fluid in the form of a gas of relativistic particles, which is inserted between the radiation and matter solutions. The model is based on two main components of energy density that describe the contributions of relativistic and non-relativistic particles. Thus, the RRG represents a solution that interpolates between the radiation-dominated universe (relativistic particles) and the matter-dominated universe (non-relativistic particles), that is, the transition from the \emph{radiation era} to the \emph{matter era}.

By introducing noncommutativity, we aim to investigate how the classical dynamics of this model are altered and what new physical phenomena may arise. We intend to verify whether noncommutativity can be a good candidate to describe the accelerated expansion of the universe, playing the role of dark energy. Our approach is based on a series of previously published papers \cite{correa2009, abreu2012, oliveira2017a, monerat2017, oliveira2017b, oliveira2019, abreu2019, oliveira2021, oliveira2024} and we seek to go beyond previous studies and investigate how the noncommutative Bianchi I model can describe the transition from an anisotropic universe to the current isotropic universe, without the need for a cosmological constant term.

For the study of noncommutativities, we use a generalized symplectic formalism \cite{Andrade2024}, founded on the original ideas of Faddeev and Jackiw \cite{FJ} and further developed in the form of an algorithm by Barcelos-Neto \& Wotzasek \cite{Barcelos,Barcelos2}, designed for the study of constrained systems. This formalism allows us to modify the classical field equations by introducing noncommutative variables\cite{Abreu2006}, providing a coherent framework to study the resulting dynamics, now incorporating noncommutative effects into classical field theories. The Faddeev–Jackiw symplectic approach is widely used in various fields, such as classical and quantum mechanics, quantum field theory, gauge field theories, condensed matter and string theory \cite{parra2024, paulin2024, manavella2023, dai2021, abreu2018, escalante2018, dengiz2018, dengiz2016, ramos2016, toms2015, horvathy2006, liao2006, long2003}. A recent list of publications involving the symplectic formalism and noncommutativity, with a focus on gravity and cosmology, can be found in \cite{gomez2024, liang2024, zen2024, socorro2024, rodriguez2024, kuhfittig2022, touati2021, berkane2021, kanazawa2019, ma2018, rodriguez2018, rodrigues2018, escalante2017, moia2017, cartas2017, escalante2016}. 

Anyway, it is important to emphasize that the symplectic method is rooted in the ideas of deformation quantization. In this theoretical approach, the traditional process of canonical quantization, which generates an algebra of quantum observables, is replaced by another one, in which quantum observables are generated from their classical equivalents, obeying a star product that deforms the algebra of classical observables \cite{bayen1978,gutt1979}. This allows us to consistently introduce noncommutativity and study its effects on several models.

We developed our work as follows. We started with the classical field equations governing the dynamics of the BI universe coupled to a reduced relativistic gas. These equations were then modified by incorporating the noncommutative variables. We examined the resulting modified field equations to identify the differences and new features arising from noncommutativity. This analysis included solving the evolution equations and studying the behavior of the scale functions in the noncommutative model, comparing the results with the commutative case. The results were discussed, and we highlighted how the noncommutative modifications influence the evolution of the Bianchi I universe. In particular, we show that after some time the isotropization of the model occurs and the expansion rate of the universe may increase, due to the NC parameters.

We have organized this paper as follows. In Section \ref{formalism}, we present the first notions involving the inclusion of noncommutativity and a concise description of the Faddeev-Jackiw theory. In Section \ref{foundations}, we introduce the Bianchi I (BI) cosmological model and the reduced relativistic gas (RRG). In Section \ref{integrating}, we integrate the RRG into the BI model. In Section \ref{FJ_applied}, we apply the Faddeev-Jackiw formalism to our model. In Section \ref{NCTY_included}, we incorporate noncommutativity into the complete model. In Section \ref{equationsmodel}, we derive the evolution equations of the model. In Section \ref{evolution}, we study the time evolution of the model. In Section \ref{estimation}, we estimate some parameters of the noncommutative model based on current cosmological observations. Finally, in Section \ref{conclusions}, we outline our concluding remarks. Appendix \ref{appendix} gathers the List of Tables.

\sloppy
\section{Generalized symplectic formalism for constrained systems and noncommutativities}
\label{formalism}

In the theoretical approach of deformation quantization, the process of canonical quantization is such that quantum observables are generated from their classical equivalents through  a star product ($\star_\hbar$), also known as the Moyal-Weyl product\cite{Moyal1949, Vey1975, Flato1975, Flato1976, Bayen1978a}, that deforms the algebra ${\cal A}_0$ of classical observables into another algebra ${\cal A}_{\hbar}$, and it is given by,
\begin{equation}
\lbrace f, g \rbrace_{\hbar} = f \star_\hbar g - g \star_\hbar f,
\end{equation}
where
\begin{equation}
(f \star_\hbar g)(\xi) = \exp\left\{\frac{i}{2}\hbar \omega_{ab} \partial^a_{(\xi_1)} \partial^b_{(\xi_2)}\right\}f(\xi_1)g(\xi_2)\bigg|_{\xi_1=\xi_2=\xi},
\end{equation}
with $\xi$ being the phase space of the functions and $\omega_{ab}$ called the \emph{classical symplectic structure}, that is,
\begin{equation}
\omega_{ab} = \left( \begin{array}{cc}
0 & \delta_{ij} \\
-\delta_{ij} & 0
\end{array} \right),
\end{equation}
and obeying the relation $\omega^{ab} \omega_{bc} = \delta^{a}_{c}$. A detailed exposition of deformation quantization can be found in \cite{Gutt2005}. Here, only the basic tools for an adequate understanding of the content to be explored in the following sections will be presented. In this sense, it is important to begin by highlighting the operationality of the star product, which is written as,
\begin{equation}
	f \star_\hbar g = fg + \hbar B_1(f,g) + \hbar^2 B_2(f,g) + \ldots , 
\end{equation}
where \( B_i (f, g) \) are bidifferential operators, i.e., differential operators that act on pairs of functions \((f, g)\) by applying partial derivatives simultaneously to these pairs. This product is associative if the parameter \( \hbar \) is constant.

It is possible to include more deformations by modifying the usual symplectic structure $\omega_{ab}$ to a more general $\alpha_{ab}$, generating an algebra \({\cal A}_{\hbar , \alpha}\) endowed with the following star product \cite{DjemaiSmail2003}, 
\begin{equation}
(f \star_{\hbar , \alpha} g)(\xi) = \exp\left\{\frac{i}{2}\hbar \alpha_{ab} \partial^a_{(\xi_1)} \partial^b_{(\xi_2)}\right\}f(\xi_1)g(\xi_2)\bigg|_{\xi_1=\xi_2=\xi},
\end{equation}   
such that the new symplectic structure is given by, 
\begin{equation}
\alpha_{ab} = \left( \begin{array}{cc}
\Theta_{ij} & \delta_{ij} + \Sigma_{ij}\\
-\delta_{ij} - \Sigma_{ij} & \Pi_{ij}
\end{array} \right),
\label{structure}
\end{equation}
with the following Poisson structure,
\begin{eqnarray}
\left\{ q_{i} , q_{j} \right\}_{\hbar,\alpha} = i\hbar \Theta_{ij}~~,~~\left\{ q_{i} , p_{j} \right\}_{\hbar,\alpha} = i\hbar (\delta_{ij} + \Sigma_{ij})~~,~~\left\{ p_{i} , p_{j} \right\}_{\hbar,\alpha} = i\hbar \Pi_{ij} , 
\end{eqnarray}
with the inclusion of noncommutativities through the antisymmetric matrices $\Theta_{ij}$ and $\Pi_{ij}$ and the symmetric matrix $\Sigma_{ij}$. The coefficients of these matrices are the noncommutative parameters of the theory. 

In the strictly classical case, there is a Poisson algebra \({\cal A}_{\alpha}\) obeying the following star product, without $i \hbar$, given by \cite{Djemai2004}, 
\begin{equation}
(f \star_{\alpha} g)(\xi) = \exp\left\{\frac{1}{2} \alpha_{ab} \partial^a_{(\xi_1)} \partial^b_{(\xi_2)}\right\}f(\xi_1)g(\xi_2)\bigg|_{\xi_1=\xi_2=\xi},
\end{equation}  
which generates the following Poisson structure,
\begin{eqnarray}
\left\{ q_{i} , q_{j} \right\}_{\alpha} =  \Theta_{ij}~~,~~\left\{ q_{i} , p_{j} \right\}_{\alpha} = \delta_{ij} + \Sigma_{ij}~~,~~\left\{ p_{i} , p_{j} \right\}_{\alpha} = \Pi_{ij} . 
\end{eqnarray}

The Faddeev-Jackiw (FJ) formalism makes use of that concept of symplectic structure and provides an efficient approach for the Hamiltonian formulation and quantization of dynamical systems. It focuses directly on the canonical structure without categorizing constraints. The formulation is based on a first-order Lagrangian in time derivatives. This choice is far from restrictive, since second-order Lagrangians can be converted to first-order by introducing conjugate variables, which preserves the generality of the formalism. Thus, let the Lagrangian be given by,
\begin{equation}
L = p_i \dot{q}^i - H(p, q), \quad i = 1, \ldots, n ,
\end{equation}
where \( p_i \) and \( q^i \) are canonical conjugate variables and \( H(p, q) \) is the Hamiltonian of the system. By introducing the 2n-component phase space coordinate \( \xi^a = (q^i, p_i) \), the Lagrangian can be rewritten as,
\begin{equation}
L \, dt = A_i \, d\xi^i - V(\xi) \, dt ,
\label{L_FJ}
\end{equation}
where \( A_i := A_{\xi^{i}}\) is the component of a connection 1-form in the phase space formulation and \( V(\xi) = H(p, q) \) is the symplectic potential.

From the connection 1-form \( A_i \), the symplectic structure \( f_{ij} \) is defined as,
\begin{equation}
f_{ij} = \frac{\partial A_j}{\partial \xi^i} - \frac{\partial A_i}{\partial \xi^j} , 
\label{eq_def_f}
\end{equation}
which leads to equations of motion derived from the Euler-Lagrange equations being given by,
\begin{equation}
f_{ij} \dot{\xi}^j = \frac{\partial V}{\partial \xi^i}.
\end{equation}

When \( f_{ij} \) is nonsingular, there exists an inverse \( f_{ij}^{-1} \) such that,
\begin{equation}
\dot{\xi}^i = f_{ij}^{-1} \frac{\partial V}{\partial \xi^j} ,
\end{equation}
and the structure of generalized Poisson brackets is, in turn, defined as,
\begin{equation}
\{\xi^i, \xi^j\} = f_{ij}^{-1}.
\label{def_f_inverse}
\end{equation}

On the other hand, when \( f_{ij} \) is singular, that is, it does not have an inverse, constraints arise. These constraints can be handled variationally until an unconstrained Lagrangian is obtained, in a process in which non-essential variables are identified and eliminated through the use of the equations of motion associated with the constraints, and the symplectic structure \( f_{ij} \) is rediagonalized to remove these variables and maintain the canonical structure. Thus, for a first-order Lagrangian with $M$ constraints,
\begin{equation}
L = p_i \dot{q}^i - H(p, q) - \lambda^l \phi_l(p, q) ,
\end{equation}
where \( \lambda^l \) are Lagrange multipliers and \( \phi_l(p, q) \) represent the constraints, with \(l = 1, ... , M\), and the equations of motion associated with the constraints are used to eliminate the non-essential variables. This process can be iterated until an unconstrained Lagrangian is obtained, similar to the initial simplified Lagrangian.

The Faddeev-Jackiw formalism provides an efficient and straightforward way to handle the quantization of dynamical systems. It simplifies the process by focusing on the canonical structure and avoiding the traditional categorization of constraints, being particularly useful for systems where the direct application of Dirac's formalism is complicated or impractical \cite{dirac1950,dirac1958}. In the context of noncommutativity, the Faddeev-Jackiw formalism allows the possibility of deforming the original algebra of the classical phase space through the construction of the symplectic matrix \( f_{ij}^{-1} \) without the need to perform the cumbersome calculations of the deformed Poisson brackets \(\{\xi^i, \xi^j\}\).

We can incorporate noncommutativity in phase space through an extended symplectic structure\cite{Abreu2006}. Let $ \tilde{\xi}^a = (\tilde{q}^i, \tilde{p}_i) $ be the coordinates of the phase space with an extended symplectic structure given by,
\begin{equation}
\tilde{f}_{ab} = \left( \begin{array}{cc}
\theta_{ij} & \delta_{ij} + \eta_{ij} \\
- \delta_{ij} - \eta_{ij} & \pi_{ij}
\end{array} \right). \label{f_blocks}
\end{equation}

For generality, we adopt a Faddeev–Jackiw formulation for fields and continuous systems. In the case where $\tilde{f}_{ab}$ is non-singular (invertible), the inverse symplectic matrix is,
\begin{equation}
\int \tilde{f}_{ab}(x, y) \, \tilde{f}^{bc}(y, z) \, dy = \delta^c_a \delta(x - z). \label{f_inverse}
\end{equation}
    
We thus seek a first-order Lagrangian $\tilde{\mathcal{L}}$ whose symplectic structure is exactly equal to $\tilde{f}_{ab}$. Assuming the FJ ansatz, we have,
\begin{equation}
\tilde{\mathcal{L}} = A_{\tilde{\xi}^{a}}  \, \dot{\tilde{\xi}}^a - V(\tilde{\xi}), \label{NC_lagrangian}
\end{equation}    
which generalizes Eq. \eqref{L_FJ} to the NC case. Thus, by assuming the Lagrangian in Eq. \eqref{NC_lagrangian}, we can functionally vary the corresponding action with $\tilde{\xi}^a \rightarrow \tilde{\xi}^a + \delta\tilde{\xi}^a$, integrating by parts and identifying $\tilde{f}_{ab}$ as the kernel of the symplectic structure, we obtain,
\begin{equation}
\tilde{f}_{ab}(x,y) = \frac{\delta A_{\tilde{\xi}^{b}}(x)}{\delta \tilde{\xi}^a(y)} - \frac{\delta A_{\tilde{\xi}^{a}}(x)}{\delta \tilde{\xi}^b(y)}. \label{f_symplectic}
\end{equation}   

Thus, the assumption of Eq. \eqref{NC_lagrangian} together with the definition in Eq. \eqref{f_symplectic} ensures that $\tilde{f}_{ab}$ represents the symplectic structure of the system. Nevertheless, the quantities $A_{\tilde{\xi}^{a}}$ are yet to be determined. To this end, we impose that the structure defined in Eq. \eqref{f_symplectic} must coincide with the block-structured noncommutative formulation given in Eq. \eqref{f_blocks}. Accordingly, by employing Eqs. \eqref{f_blocks}, \eqref{f_inverse} and \eqref{f_symplectic}, we obtain a system of equations, which allows for the determination of the quantities $A_{\tilde{\xi}^{a}}$, that is,
\begin{align}
\theta^{ij} B_{jk}(x,y) + (\delta^{ij} + \eta^{ij}) A_{jk}(x,y) &= \delta^i_k \delta(x - y), \\
A_{jk}(x,y) \theta^{ji} + (\delta^{ij} + \eta^{ij}) C_{jk}(x,y) &= 0, \\
- (\delta^{ij} + \eta^{ij}) B_{jk}(x,y) + \pi^{ij} A_{jk}(x,y) &= 0, \\
A_{kj}(x,y) (\delta^{ji} + \eta^{ji}) + \pi^{ij} C_{jk}(x,y) &= \delta^i_k \delta(x - y),
\label{system_symplectic}
\end{align}
where
\begin{align}
A_{jk}(x,y) &= \frac{\delta A_{\tilde{p}_j}(x)}{\delta \tilde{q}_k(y)} - \frac{\delta A_{\tilde{q}_k}(x)}{\delta \tilde{p}_j(y)}, \\
B_{jk}(x,y) &= \frac{\delta A_{\tilde{q}_j}(x)}{\delta \tilde{q}_k(y)} - \frac{\delta A_{\tilde{q}_k}(x)}{\delta \tilde{q}_j(y)}, \\
C_{jk}(x,y) &= \frac{\delta A_{\tilde{p}_j}(x)}{\delta \tilde{p}_k(y)} - \frac{\delta A_{\tilde{p}_k}(x)}{\delta \tilde{p}_j(y)}.
\end{align}

As a consequence, we can construct a first-order Lagrangian, given by, 
\begin{equation}
\mathcal{L}_{NC} = A_{\tilde{\xi}^{a}}  \, \dot{\tilde{\xi}}^a - V(\tilde{\xi}), \label{NC_L} 
\end{equation}
where the quantities $A_{\tilde{\xi}^{a}}$  are now determined from the solution of the system of equations in Eq. \eqref{system_symplectic}. In subsequent sections, we will apply this procedure to derive a system of equations that enables the determination of the noncommutative solution corresponding to the model under consideration.

\sloppy
\section{Foundations of Bianchi I (BI) and reduced relativistic gas (RRG)}
\label{foundations}

The Bianchi I model is part of the class of \emph{Bianchi models}, which describe homogeneous universes with different types of anisotropy. The model specifically assumes zero spatial curvature (\(k=0\)) and assigns different anisotropies to each spatial direction through different scale factors \(a(t)\), \(b(t)\), and \( c (t)\), rendering it an anisotropic counterpart to the flat universe described by the Friedmann-Robertson-Walker (FRW) model. This characteristic makes it particularly valuable for studying the isotropization of the universe, especially given that cosmographic measurements indicate our universe is predominantly flat \cite{george,anton,vagnozzi1,vagnozzi2,vagnozzi3}. Furthermore, this model may be relevant for describing possible anisotropic eras in the early universe and its transition to an isotropic universe, as well as perhaps explaining the small anisotropies observed in the cosmic microwave background (CMB). The metric of the Bianchi I model is expressed as,
\begin{equation}
ds^2 = -N^2 dt^2 + a^2(t) dx^2 + b^2(t) dy^2 + c^2(t) dz^2,
\end{equation}
where \(a(t)\), \(b(t)\), and \(c(t)\) are the scale factors associated with different directions, that vary independently with time, $N(t)$ is the lapse function from the ADM formalism \cite{ADM} and ($x$, $y$, $z$) are the Cartesian coordinates with their usual domains. We also adopt natural units, that is, $8\pi G = c = 1$, where $G$ is the gravitational constant and $c$ is the speed of light in vacuum.

For simplification purposes, the Bianchi I model is commonly studied using the Misner parameterization. In this parameterization its metric is given by \cite{misner1969},
\begin{equation}
ds^2 = - N^2 (t) dt^2 + a^2 (t) \left [e^{2 (\beta_+ + \sqrt 3 \beta_-)} dx^2 + e^{2( \beta_+ - \sqrt 3 \beta_-)} dy^2 + e^{ - 4 \beta_+ } dz^2 \right ],
\label{newmetric}
\end{equation}
where now $a(t)$ is the isotropic scale factor, while $\beta_{\pm} (t)$ are the anisotropic scale functions.

Now, let us introduce the reduced relativistic gas (RRG) model \cite{BerredoPeixoto2005}, which is a simplified description of a gas of relativistic particles. The model is based on two main components of energy density that describe the contributions of relativistic and non-relativistic particles, which causes the energy density \(\rho\) to be given by \cite{BerredoPeixoto2005},
\begin{equation}
\rho = \left [\rho_{1}^{2} \left ( \frac{a_{0}}{a} \right )^6 + \rho_{2}^{2} \left ( \frac{ a_{0}}{a} \right )^8 \right ]^{1/2},
\label{eqRRG}
\end{equation}
where \(\rho_{1}\) is an integration constant associated with the energy density of the non-relativistic component, \(\rho_{2}\) is an integration constant associated with the energy density of the relativistic component, \(a_{0}\)  is the initial scale factor, and \(a\) is the scale factor of the universe, which varies with time.

The energy density that decays in the form $a^{-3}$ (corresponding to the first term involving $\rho_{1}$) is characteristic of non-relativistic particles (or cold matter), where the energy density decreases with volume. On the other hand, the energy density that decays in the form $a^{-4}$ (corresponding to the second term involving $\rho_{2}$) is characteristic of relativistic particles, where the energy density decreases more rapidly due to the cosmological redshift effect on photon energy.


Combining the Bianchi I (BI) model for the geometric sector with the reduced relativistic gas (RRG) model for the matter sector can be a significant step toward understanding the evolution of the primordial universe. The BI model allows for the study of the universe's dynamics under anisotropic conditions, providing a detailed view of how initial anisotropic characteristics could have evolved into the isotropic universe observed today. Simultaneously, the RRG model robustly and simplistically can describe the particle dominance transition between the radiation and matter eras. Integrating these models enables a comprehensive analysis of the interactions between anisotropic spacetime geometry and the behavior of relativistic matter, enabling the elucidation of cosmological evolution and perhaps offering testable predictions to cosmographic data.

\section{Integrating RRG into the BI Model}
\label{integrating}
In order to construct the model that couples RRG to the Bianchi I metric, we will consider that the total action $S$ is the sum of the Einstein-Hilbert action for the geometric sector and the action of the matter sector described by a perfect fluid of RRG, so that,
\begin{equation}
S = \int \, d^4 x \,  \sqrt{- g}\left(\frac{R}{2} - \rho \right),
\end{equation}
where $R$ is the Ricci scalar for the BI case and $\rho$ the energy density given by Eq. \eqref{eqRRG}, which holds due to pressure isotropy in the BI model\cite{Simpliciano}.

Computing the total action and performing integrations by parts, we obtain the following expression,
\begin{equation}
S = \int d^4x \, a^3 \left \{ \left ( -\frac{3}{N}\frac{\dot{a}^2}{a^2} + \frac{3}{N} {\dot{\beta}^{2}_{+}} + \frac{3}{N} {\dot{\beta}^{2}_{-}} \right ) - N \left [\rho_{1}^{2} \left ( \frac{a_{0}}{a} \right )^6 + \rho_{2}^{2} \left ( \frac{a_{0}}{a} \right )^8 \right ]^{1/2} \right \}.
\end{equation}

From the above expression, we can obtain the following canonically conjugate momenta to $a$, $\beta_{+}$ and $\beta_{-}$, given by,
\begin{equation}
p_a =  - 6 \frac{\dot{a} a}{N}, \quad
p_{+} =  6 \frac{a^3 \dot{\beta}_{+}}{N}, \quad
p_{-} =  6 \frac{a^3 \dot{\beta}_{-}}{N},
\end{equation}
which can be used to obtain the total Hamiltonian density $\mathcal H$, through the Legendre transformation, which results in the expression,
\begin{equation}
\mathcal H = N \left \{ - \frac{p_{a}^2}{12 a} + \frac{p_{+}^2}{12 a^3} + \frac{p_{-}^2}{12 a^3} + a^3 \left [ \rho_{1}^{2} \left ( \frac{a_{0}}{a} \right )^6 + \rho_{2}^{2} \left ( \frac{a_{0}}{a} \right )^8 \right ]^{1/2} \right \}.
\label{H}
\end{equation}
\section{Faddeev-Jackiw formalism applied to the BI RRG model}
\label{FJ_applied}
For the current model, that couples RRG to the BI metric, we employ the Faddeev-Jackiw (FJ) formalism, which aims to treat constrained systems by eliminating superfluous degrees of freedom. Beyond its effectiveness  in this context, the FJ approach fixes in a non-arbitrary manner the transformation from noncommutative to commutative variables, in contrast with approaches like \cite{monerat2017, oliveira2017b, oliveira2021, oliveira2024}, where such transformations are introduced \textit{ad hoc}. The initial step in FJ procedure involves obtaining the primary Lagrangian density from the provided Hamiltonian density, termed as the zero-Lagrangian, which for the studied model is
\begin{equation}
\mathcal L^{(0)} = p_{a} \dot{a} + p_{+} \dot{\beta}_{+} + p_{-} \dot{\beta}_{-} - V^{(0)} (a, p_{a},{ \beta}_{+}, p_{+}, {\beta}_{-}, p_{-}, N) , 
\label{L-zero}
\end{equation}
where $V^{(0)}$ is the symplectic potential, given by,
\begin{equation}
V^{(0)} = N \Omega = N \left \{ - \frac{p_{a}^2}{12 a} + \frac{p_{+}^2}{12 a^3} + \frac{p_{-}^2}{12 a^3} + a^3 
\left[ \rho_{1}^{2} \left ( \frac{a_{0}}{a} \right )^6 + \rho_{2}^{2} \left ( \frac{a_{0}}{a} \right )^8  \right ]^{1/2} \right \}.
\label{V_0}
\end{equation}

The symplectic variables ${\xi }_{i}^{(0)}$ = $(a, p_{a},{ \beta}_{+}, p_{+}, {\beta}_{-}, p_{-}, N) $ are identified, and then the 1-form elements in Eq. \eqref{L-zero} of the zero iteration are given by,
\begin{equation}
A^{(0)}_{a} = p_{a}, \,\,   A^{(0)}_{p_{a}} = 0, \,\, A^{(0)}_{\beta_{\pm}} = p_{\pm} , \,\, A^{(0)}_{p_{\pm}} = 0 , \,\, A^{(0)}_{N} = 0 . 
\end{equation}
The elements of the symplectic matrix are obtained from the definition given in Eq. \eqref{eq_def_f}, which results in the following symplectic matrix for the zero iteration, which is singular,
\begin{equation}
f^{(0)} = \begin{bmatrix}
0 & -1 & 0 & 0 & 0 & 0 & 0 \\ 
1 & 0 & 0 & 0 & 0 & 0 & 0 \\ 
0 & 0 & 0 & -1 & 0 & 0 & 0 \\ 
0 & 0 & 1 & 0 & 0 & 0 & 0 \\ 
0 & 0 & 0 & 0 & 0 & -1 & 0 \\ 
0 & 0 & 0 & 0 & 1 & 0 & 0 \\ 
0 & 0 & 0 & 0 & 0 & 0 & 0
\end{bmatrix},
\end{equation}
whose zero mode is given by,
\begin{equation}
{{\nu}}^{(0)} = \begin{bmatrix}
0 & 0 & 0 & 0 & 0 & 0 & 1
\end{bmatrix} .
\label{zeromode}
\end{equation}
The multiplication of the zero mode, Eq. \eqref{zeromode}, with the gradient of the symplectic potential, Eq. \eqref{V_0}, results in,
\begin{equation}
\Omega = 0. 
\end{equation}
This relation corresponds to a constraint, which must then be introduced into the first-order Lagrangian density through a Lagrange multiplier $\tau$, resulting in the first-iteration Lagrangian density,
\begin{equation}
\mathcal{L}^{(1)} = p_{a} \dot{a} + p_{+} \dot{\beta}_{+} + p_{-} \dot{\beta}_{-} + \Omega \dot{\tau} - V^{(1)} (a, p_{a}, \beta_{+}, p_{+}, \beta_{-}, p_{-}, N),
\label{L-um}
\end{equation}
where now $ V^{(1)} = V^{(0)} = N \Omega$.  The set of first iteration symplectic variables is now given by $
\xi_{i}^{(1)}$ = $(a, p_{a}, \beta_{+}, p_{+}, \beta_{-}, p_{-}, N, \tau)$ and the 1-form elements for Eq. \eqref{L-um} are,
\begin{equation}
A^{(1)}_{a} = p_{a}, \,\, A^{(1)}_{p_{a}} = 0, \,\, A^{(1)}_{\beta_{\pm}} = p_{\pm}, \,\, A^{(1)}_{p_{\pm}} = 0, \,\, A^{(1)}_{N} = 0, \,\, A^{(1)}_{\tau} = \Omega.
\end{equation}

Proceeding again with the definition in Eq. \eqref{eq_def_f}, it follows that the first-iteration symplectic matrix is given by,
\begin{equation}
\renewcommand{\arraystretch}{1.5} 
f^{(1)} = \begin{bmatrix}
0 & -1 & 0 & 0 & 0 & 0 & 0 & \frac{\partial \Omega}{\partial a} \\ 
1 & 0 & 0 & 0 & 0 & 0 & 0 & \frac{\partial \Omega}{\partial p_{a}} \\ 
0 & 0 & 0 & -1 & 0 & 0 & 0 & \frac{\partial \Omega}{\partial \beta_{+}} \\ 
0 & 0 & 1 & 0 & 0 & 0 & 0 & \frac{\partial \Omega}{\partial p_{+}} \\ 
0 & 0 & 0 & 0 & 0 & -1 & 0 & \frac{\partial \Omega}{\partial \beta_{-}} \\ 
0 & 0 & 0 & 0 & 1 & 0 & 0 & \frac{\partial \Omega}{\partial p_{-}} \\ 
0 & 0 & 0 & 0 & 0 & 0 & 0 & 0 \\ 
-\frac{\partial \Omega}{\partial a} & -\frac{\partial \Omega}{\partial p_{a}} & -\frac{\partial \Omega}{\partial \beta_{+}} & -\frac{\partial \Omega}{\partial p_{+}} & -\frac{\partial \Omega}{\partial \beta_{-}} & -\frac{\partial \Omega}{\partial p_{-}} & 0 & 0
\end{bmatrix},
\end{equation}
whose zero mode is,
\begin{equation}
\renewcommand{\arraystretch}{1.5} 
{{\nu}}^{(1)} = \begin{bmatrix}
- \frac{\partial \Omega}{\partial p_{a}} & \frac{\partial \Omega}{\partial a} & - \frac{\partial \Omega}{\partial p_{+}} & \frac{\partial \Omega}{\partial \beta_{+}} & - \frac{\partial \Omega}{\partial p_{-}} & \frac{\partial \Omega}{\partial \beta_{-}} & 1 & 1
\end{bmatrix} .
\label{zeromode1}
\end{equation}
The multiplication of the zero-mode of the first iteration, Eq. \eqref{zeromode1}, with the gradient of the symplectic potential $V^{(1)}$, gives the same constraint $\Omega = 0$ obtained for the zero iteration. This indicates that the system has a gauge symmetry that must be fixed and introduced into the zero Lagrangian, Eq. \eqref{L-zero}. To fix this symmetry, we introduce a gauge fixing term $\Sigma$ within the zero Lagrangian, Eq. \eqref{L-zero}, through a Lagrange multiplier $\eta$, in order to obtain a new zero Lagrangian density $\tilde{\mathcal {L}}^{(0)}$ \cite{correa2009}, now given by
\begin{equation}
\tilde{\mathcal {L}}^{(0)} = p_{a} \dot{a} + p_{+} \dot{\beta}_{+} + p_{-} \dot{\beta}_{-} + \Sigma \dot{\eta} - \tilde{V}^{(0)} (a, p_{a},{ \beta}_{+}, p_{+}, {\beta}_{-}, p_{-}, N) ,	
\label{L-zero-nova}
\end{equation}
with $\Sigma = N - 1$, which implies that $N = 1$. The symplectic potential is now $\tilde{V}^{(0)} = N \Omega$ and the set of symplectic variables of the system is $
{\tilde{\xi }}_{i}^{(0)}$ = $(a, p_{a}, \beta_{+}, p_{+}, \beta_{-}, p_{-}, N, \eta)$. Thus, the 1-form elements in Eq. \eqref{L-zero-nova} are,
\begin{equation}
\tilde{A}^{(0)}_{a} = p_{a}, \,\, \tilde{A}^{(0)}_{p_{a}} = 0, \,\, \tilde{A}^{(0)}_{\beta_{\pm}} = p_{\pm}, \,\, \tilde{A}^{(0)}_{p_{\pm}} = 0, \,\, \tilde{A}^{(0)}_{N} = 0, \,\, \tilde{A}^{(0)}_{\eta} = \Sigma := N - 1,
\end{equation}
and the new matrix of the zero iteration is then non-singular and given by,
\begin{equation}
{\tilde{f}}^{(0)} = \begin{bmatrix}
0 & -1 & 0 & 0 & 0 & 0 & 0 & 0 \\ 
1 & 0 & 0 & 0 & 0 & 0 & 0 & 0 \\ 
0 & 0 & 0 & -1 & 0 & 0 & 0 & 0 \\ 
0 & 0 & 1 & 0 & 0 & 0 & 0 & 0 \\ 
0 & 0 & 0 & 0 & 0 & -1 & 0 & 0 \\ 
0 & 0 & 0 & 0 & 1 & 0 & 0 & 0 \\ 
0 & 0 & 0 & 0 & 0 & 0 & 0 & 1 \\
0 & 0 & 0 & 0 & 0 & 0 & -1 & 0
\end{bmatrix},
\end{equation}
Therefore, its inverse can be computed as,
\begin{equation}
{f}^{-1}_{(0)} = \begin{bmatrix}
0 & 1 & 0 & 0 & 0 & 0 & 0 & 0 \\ 
-1 & 0 & 0 & 0 & 0 & 0 & 0 & 0 \\ 
0 & 0 & 0 & 1 & 0 & 0 & 0 & 0 \\ 
0 & 0 & -1 & 0 & 0 & 0 & 0 & 0 \\ 
0 & 0 & 0 & 0 & 0 & 1 & 0 & 0 \\ 
0 & 0 & 0 & 0 & -1 & 0 & 0 & 0 \\ 
0 & 0 & 0 & 0 & 0 & 0 & 0 & -1 \\
0 & 0 & 0 & 0 & 0 & 0 & 1 & 0
\end{bmatrix},
\label{matrizfcomutativa}
\end{equation}
which is the symplectic matrix, whose elements are the Poisson brackets of the commutative theory, with the non-vanishing brackets given by,
\begin{equation} \left \{ a, p_a \right\} = \left\{ \beta_{+}, p_{+} \right\} = \left\{ \beta_{-}, p_{-} \right\} = - \left\{ N, \eta \right\} = 1. 
\end{equation}

\sloppy
\section{NC BI Model with RRG}
\label{NCTY_included}
To construct our NC Bianchi I model with RRG, we will impose noncommutative parameters on the symplectic structure of the commutative model, which involves deforming the Poisson brackets. Specifically, we will introduce the following noncommutativities:
\begin{equation}
\left \{ a , \beta_{\pm} \right \} = \varepsilon_{\pm} , \,\ \left \{ \beta_{+} , \beta_{-} \right \} = \gamma , \,\ \left \{ p_{a} , p_{\pm} \right \} = \sigma_{\pm} ,  \,\ \left \{ p_{+} , p_{-} \right \} = \chi ,
\end{equation}
where $\varepsilon_{+}$, $\varepsilon_{-}$, $\gamma$, $\sigma_{+}$, $\sigma_{-}$ and $\chi$ are constants. 
This transforms the commutative symplectic matrix Eq. \eqref{matrizfcomutativa} into,
\begin{equation}
{f}^{-1} = \begin{bmatrix}
0 & 1 & \varepsilon_{+} & 0 & \varepsilon_{-} & 0 & 0 & 0 \\ 
-1 & 0 & 0 & \sigma_{+} & 0 & \sigma_{-} & 0 & 0 \\ 
-\varepsilon_{+} & 0 & 0 & 1 & \gamma & 0 & 0 & 0 \\ 
0 & - \sigma_{+} & -1 & 0 & 0 & \chi & 0 & 0 \\ 
-\varepsilon_{-} & 0 & - \gamma & 0 & 0 & 1 & 0 & 0 \\ 
0 & \sigma_{-} & 0 & - \chi & -1 & 0 & 0 & 0 \\ 
0 & 0 & 0 & 0 & 0 & 0 & 0 & -1 \\
0 & 0 & 0 & 0 & 0 & 0 & 1 & 0
\end{bmatrix},
\label{matrizfNC}
\end{equation}
the elements of which are now the Poisson brackets of the noncommutative theory. The calculation of the inverse of matrix Eq. \eqref{matrizfNC} yields,
\begin{equation}
f = \frac{1}{\Gamma} \begin{bmatrix} 
0 &  1 - \gamma \chi & - \sigma_{+} & \gamma \sigma_{-} & - \sigma_{-} & - \gamma \sigma_{+} & 0 & 0 \\ 
\gamma \chi - 1 & 0 & - \chi \varepsilon_{-}  & - \varepsilon_{+} & \chi \varepsilon_{+} & - \varepsilon_{-} & 0 & 0 \\ 
\sigma_{+} & \chi \varepsilon_{-} & 0 & 1 - \varepsilon_{-} \sigma_{-}  & -\chi & \varepsilon_{-} \sigma_{+} & 0 & 0 \\ 
- \gamma \sigma_{-} & \varepsilon_{+} & \varepsilon_{-}\sigma_{-} -1 & 0 & -\varepsilon_{+} \sigma_{-} & -\gamma & 0 & 0 \\ 
\sigma_{-} & - \chi \varepsilon_{+} & \chi & \varepsilon_{+} \sigma_{-} & 0 & 1 - \varepsilon_{+} \sigma_{+} & 0 & 0 \\ 
\gamma \sigma_{+}  & \varepsilon_{-} & - \varepsilon_{-} \sigma_{+} & \gamma & \varepsilon_{+} \sigma_{+} -1 & 0 & 0 & 0 \\ 
0 & 0 & 0 & 0 & 0 & 0 & 0 & \Gamma \\
0 & 0 & 0 & 0 & 0 & 0 & -\Gamma  & 0
\end{bmatrix},
\end{equation}
where $ \Gamma =  \varepsilon_{+} \sigma_{+} +  \varepsilon_{-} \sigma_{-} + \gamma \chi - 1 $ and $ \Gamma \neq 0$. The elements of the matrix $f$ are the terms $f_{ij}$, which are subject to the condition Eq. \eqref{eq_def_f}.  By applying it to the terms of the matrix $f$, it results in an extensive system of equations in the 1-forms $A_{i}$ (omitted here). However, we want the model to remain first-order in velocities, so the canonical momenta 1-forms $A_{p_{a}}$, $A_{p_{+}}$, $A_{p_{-}}$, and $A_{N}$ must be zero; otherwise, terms like $A_{p_{a}} \dot{p}_{a}$, $A_{p_{+}} \dot{p}_{+}$, $A_{p_{-}} \dot{p}_{-}$, and $A_{N} \dot{N}$ would arise, which are second-order in velocities. Furthermore, due to this, the noncommutativity parameters $\varepsilon_{+}$, $\varepsilon_{-}$, and $\gamma$, although present in the structure of the deformed symplectic matrix, will not appear in the resulting dynamical system. Therefore, the previous system of equations is further simplified, and its solution is given by,
\begin{equation}
\begin{aligned}
&A_{a} = - \frac{p_{a}}{\Gamma}, \quad
A_{\beta_{+}} =  \frac{1}{\Gamma} \left (- \sigma_{+} a - p_{+} + \frac{1}{2} \chi \beta_{-} \right), \\
&A_{\beta_{-}} =  \frac{1}{\Gamma} \left ( - \sigma_{-} a - p_{-} - \frac{1}{2} \chi \beta_{+} \right), \quad
A_{\eta} = \Sigma := N - 1,
\end{aligned}
\label{EqsA}
\end{equation}
where in the gauge $N = 1$, we will have $A_{\eta} = 0$. 

Thus, the first-order Lagrangian density in velocities, is given by,
\begin{equation}
\begin{split}
\mathcal{L}_{NC} = &-\frac{p_{a}}{\Gamma}  \dot{a} 
+ \frac{1}{\Gamma} \left (- \sigma_{+} a - p_{+} + \frac{1}{2} \chi \beta_{-} \right) \dot{\beta_{+}} \\
&+ \frac{1}{\Gamma} \left ( - \sigma_{-} a - p_{-} - \frac{1}{2} \chi \beta_{+} \right) \dot{\beta_{-}} \\
&+ \Sigma \dot{\eta} - V_{NC} (a, p_{a}, p_{+}, p_{-}, N).
\end{split}
\label{L_NC}
\end{equation}
where $V_{NC} = N \Omega$. 

We can express this Lagrangian in terms of a new set of commutative variables that satisfy the usual canonical Poisson brackets up to first order in the noncommutative parameters, by introducing the following coordinate transformation,
\begin{equation}
\begin{aligned}
&\tilde{a} = a, \quad \tilde{\beta}_{+} = \beta_{+}, \quad \tilde{\beta}_{-} = \beta_{-}, \quad \tilde{p}_{a} = - \frac{p_{a}}{\Gamma}, \\ 
& \tilde{p}_{+} = \frac{1}{\Gamma} \left (- \sigma_{+} a - p_{+} + \frac{1}{2} \chi \beta_{-} \right), \quad
\tilde{p}_{-} = \frac{1}{\Gamma} \left ( - \sigma_{-} a - p_{-} - \frac{1}{2} \chi \beta_{+} \right).
\label{Eqs_tilde}
\end{aligned}
\end{equation}

This new set of variables $\{\tilde{a}, \tilde{\beta}_{+}, \tilde{\beta}_{-}, \tilde{p}_{a}, \tilde{p}_{+}, \tilde{p}_{-} \}$ incorporates all noncommutative effects into the parameters $\chi$, $\sigma_{+}$, and $\sigma_{-}$, while preserving the usual Poisson bracket algebra. These transformations arise naturally from the symplectic structure associated with the Lagrangian in Eq.~\eqref{L_NC}, as derived from the symplectic formalism adopted here. Consequently, in terms of the new variables, the Lagrangian \eqref{L_NC} becomes
\begin{equation}
\mathcal{L}_{NC} = {\tilde{p}_{a}} \dot{a} + {\tilde{p}_{+}} \dot{\beta_{+}} + {\tilde{p}_{-}} \dot{\beta_{-}} + \Sigma \dot{\eta} - N \Omega .
\end{equation}

By restricting to first-order terms in the noncommutative parameters, we can invert the transformation in Eq.~\eqref{Eqs_tilde}, obtaining
\begin{equation}
p_{a} = \tilde{p}_{a}, \quad
p_{+} = \tilde{p}_{+} - \sigma_{+} a + \frac{1}{2} \chi \beta_{-}, \quad
p_{-} = \tilde{p}_{-} - \sigma_{-} a - \frac{1}{2} \chi \beta_{+}.
\end{equation}

Therefore, from Eq. \eqref{H} for $N = 1$, we obtain
\begin{eqnarray}
\mathcal{\tilde{H}} &=& - \frac{\tilde{p}_{a}^2}{12 a} 
+ \frac{(\tilde{p}_{+} - \sigma_{+} a + \frac{1}{2} \chi \beta_{-} )^2}{12 a^3} 
+ \frac{(\tilde{p}_{-} - \sigma_{-} a - \frac{1}{2} \chi \beta_{+})^2}{12 a^3} \nonumber \\ 
&& + a^3 
\left[ \rho_{1}^{2} \left( \frac{a_{0}}{a} \right)^{6} 
+ \rho_{2}^{2} \left( \frac{a_{0}}{a} \right)^{8} \right]^{1/2} ,
\label{H_NC}
\end{eqnarray}
which is the Hamiltonian of the system modified by the noncommutative contribution. Note that when $\sigma_{+}$ = $\sigma_{+}$ = $\chi$ = 0 we recover the original super-Hamiltonian $\mathcal{H}$.  

\sloppy
\section{Evolution Equations for the NC BI Model with RRG}
\label{equationsmodel}

To study the evolution of the NC BI RRG model over time, we start with the Hamilton equations, which are given in function of the Hamiltonian Eq. \eqref{H_NC}. Thus, we have
\begin{equation}
\dot{a} = \left \{ a, \mathcal{\tilde{H}} \right \} = \frac{\partial \mathcal{\tilde{H}}}{\partial \tilde{p}_{a}} = - \frac{1}{6a} \tilde{p}_{a} , 
\label{Heq1}
\end{equation}
\begin{eqnarray}
\dot{\tilde{p}}_{a} &=& \left \{ \tilde{p}_{a} , \mathcal{\tilde{H}} \right \} = - \frac{\partial \mathcal{\tilde{H}}}{\partial a}  \nonumber \\
&=&  - \frac{\tilde{p}_{a}^2}{12 a^2} + \frac{1}{4 a^4} \left (\tilde{p}_{+} - \sigma_{+}a + \frac{1}{2} \chi \beta_{-} \right )^2 + \frac{1}{6 a^3} \left (\tilde{p}_{+} - \sigma_{+}a + \frac{1}{2} \chi \beta_{-} \right )\sigma_{+} \nonumber \\
&+& \frac{1}{4 a^4} \left (\tilde{p}_{-} - \sigma_{-}a - \frac{1}{2} \chi \beta_{+} \right )^2 + \frac{1}{6 a^3} \left (\tilde{p}_{-} - \sigma_{-}a - \frac{1}{2} \chi \beta_{+} \right )\sigma_{-} \nonumber \\
&-&  3a^2 \left [ \rho_{1}^2 \left ( \frac{a_{0}}{a} \right )^{6} + \rho_{2}^2 \left ( \frac{a_{0}}{a} \right )^{8} \right ]^{1/2} + a^2 \frac{ \left [ 3 \rho_{1}^2 \left ( \frac{a_{0}}{a} \right )^{6} + 4 \rho_{2}^2 \left ( \frac{a_{0}}{a} \right )^{8} \right ]}{ \left [ \rho_{1}^2 \left ( \frac{a_{0}}{a} \right )^{6} + \rho_{2}^2 \left ( \frac{a_{0}}{a} \right )^{8} \right ]^{1/2}} . 
\label{Heq2}
\end{eqnarray}
\begin{equation}
{\dot{\beta}}_{+} = \left \{ \beta_{+} , \mathcal{\tilde{H}}  \right \} = \frac{\partial \mathcal{\tilde{H}}}{\partial \tilde{p}_{+}} = \frac{1}{6 a^3} \left ( \tilde{p}_{+} - \sigma_{+}a + \frac{1}{2} \chi \beta_{-} \right ) ,
\label{Heq3}
\end{equation}
\begin{equation}
\dot{\tilde{p}}_{+} = \left \{ \tilde{p}_{+} , \mathcal{\tilde{H}}  \right \} = -  \frac{\partial \mathcal{\tilde{H}}}{\partial {\beta}_{+}} = \frac{1}{12 a^3} \left ( \tilde{p}_{-} - \sigma_{-}a - \frac{1}{2} \chi \beta_{+} \right ) \chi , 
\label{Heq4}
\end{equation}
\begin{equation}
{\dot{\beta}}_{-} = \left \{ \beta_{-} , \mathcal{\tilde{H}}  \right \} = \frac{\partial \mathcal{\tilde{H}} }{\partial \tilde{p}_{-}} = \frac{1}{6 a^3} \left ( \tilde{p}_{-} - \sigma_{-}a - \frac{1}{2} \chi \beta_{+} \right ) ,
\label{Heq5}
\end{equation}
\begin{equation}
\dot{\tilde{p}}_{-} = \left \{ \tilde{p}_{-} , \mathcal{\tilde{H}}  \right \} = -  \frac{\partial \mathcal{\tilde{H}}}{\partial {\beta}_{-}} = - \frac{1}{12 a^3} \left ( \tilde{p}_{+} - \sigma_{+}a + \frac{1}{2} \chi \beta_{-} \right ) \chi . 
\label{Heq6}
\end{equation}
It can be observed from Eq. \eqref{Heq1} that,
\begin{equation}
\tilde{p}_{a}  = - 6 a \dot{a} .
\label{pEq1}
\end{equation}
On the other hand, from Eq. \eqref{Heq3} and Eq. \eqref{Heq6}, we see that, 
\begin{equation}
\tilde{p}_{-} = C_{1} - \frac{1}{2} \chi \beta_{+}, 
\label{pEq2}
\end{equation}
where $C_{1}$ is an integration constant.

Similarly, from Eq. \eqref{Heq4} and Eq. \eqref{Heq5}, we have that, 
\begin{equation}
\tilde{p}_{+} = C_{2} + \frac{1}{2} \chi \beta_{-}, 
\label{pEq3}
\end{equation}
where $C_{2}$ is another integration constant.

We need three evolution equations for the scale functions $ a $, $ \beta_{+} $ and $\beta_{-} $. The first of these arises from considering the super-Hamiltonian constraint, which is the generalized Friedmann-like equation. Thus, imposing $\mathcal{\tilde{H}} = 0$, with $\tilde{H}$ given by Eq. \eqref{H_NC}, and considering terms up to the first order in the noncommutative parameters, we obtain,
\begin{eqnarray}
\left( \frac{\dot{a}}{a} \right)^{2} &=& \frac{1}{36 a^6}  \left ( C_{2}{}^{2} + 2 C_{2} \chi \beta_{-} + C_{1}{}^{2} - 2 C_{1} \chi \beta_{+} \right ) \nonumber \\
&-& \frac{1}{18 a^5} \left ( C_{1} \sigma_{-} + C_{2} \sigma_{+} \right )  
+ \frac{1}{3} \left[ \rho_{1}^2 \left( \frac{a_{0}}{a} \right)^{6} + \rho_{2}^2 \left( \frac{a_{0}}{a} \right)^{8} \right]^{1/2} \, .
\label{condition_a}
\end{eqnarray}
%

To find the second evolution equation, we can combine Eq. \eqref{Heq6} with Eq. \eqref{pEq2} and Eq. \eqref{pEq3}, considering up to the first order in the noncommutative parameters, obtaining,
\begin{equation}
\dot{\beta}_{+} = \frac{1}{6 a^3} \left ( C_{2} - \sigma_{+} {a} + \chi {\beta}_{-} \right ).
\label{condition_beta_+}
\end{equation}

Finally, to find the third evolution equation, we can combine Eq. \eqref{Heq4} with Eq. \eqref{pEq2} and Eq. \eqref{pEq3}, considering up to the first order in the noncommutative parameters, resulting in,
\begin{equation}
\dot{\beta}_{-} = \frac{1}{6 a^3} \left ( C_{1} - \sigma_{-} {a} - \chi {\beta}_{+} \right ).
\label{condition_beta_-}
\end{equation}

Thus, for the NC BI model with RRG, the following system of equations is obtained: three evolution equations for the scale functions, Eqs.\eqref{condition_a}, \eqref{condition_beta_+}, and \eqref{condition_beta_-}. In the next section, a numerical study of this solution will be performed.

\section{The evolution of the universe in the NC model}
\label{evolution}

This section presents the study of the numerical solutions of the system of Eqs. \eqref{condition_a}, \eqref{condition_beta_+}, and \eqref{condition_beta_-}, which describe the dynamics of our model. To this end, we performed an analysis of the evolution of the scale functions $a(t)$, $\beta_{+}(t)$, and $\beta_{-}(t)$ with respect to different parameters: $\chi$, $\sigma_{+}$, $\sigma_{-}$, $C_{1}$, $C_{2}$, $\rho_{1}$, $\rho_{2}$, $a_{0} := a(0)$, $\beta_{+0} := \beta_{+} (0)$ and $\beta_{-0} := \beta_{-} (0)$. This investigation of solutions was carried out by individually varying each parameter or initial condition of the model while keeping all other parameters or initial conditions fixed. In this way, we were able to compare the influence of different values of the given parameter or initial condition on the dynamics of $a$, $\beta_{+}$ and $\beta_{-}$. We conducted this study for a wide range of values of the different parameters and initial conditions. In doing so, we adopted the convention that whenever we studied a model parameter that was not an initial condition of the scale functions, we fixed the initial conditions as $a_{0}$ = $\beta_{+0}$ = $\beta_{-0}$ = 1. Nonetheless, in cases where a particular initial condition was studied, we varied it while keeping the other two fixed at the value of 1. 

As an example of the individualized analysis of each parameter or initial condition, we selected a set of five values and plotted the corresponding solutions as five curves in the $a \times t$, $\beta_{+} \times t$, and $\beta_{-} \times t$ graphs. These plots are presented below in the appropriate subsections. Additionally, we compiled tables displaying the numerical values of $a$, $\beta_{+}$, and $\beta_{-}$ for the five selected values of the studied parameter or initial condition at various time points, which are presented in Appendix \ref{appendix}. In all cases, the scale factor $a$ consistently exhibits an expansive behavior, whereas the functions $\beta_{+}$ and $\beta_{-}$ approach constant values after an initial phase of growth and/or decay—trends that are clearly distinguishable in the plots. The tables further illustrate the ever-increasing nature of $a$ and the stabilization of $\beta_{+}$ and $\beta_{-}$ at late times. Below is a summary of the results for each parameter or initial condition varied and the indication of the corresponding graphs and table.

\subsection{Varying the NC parameters}

\begin{figure}[H]
		\centering
		\begin{subfigure}{0.45\textwidth}
			\centering
			\includegraphics[width=\textwidth]{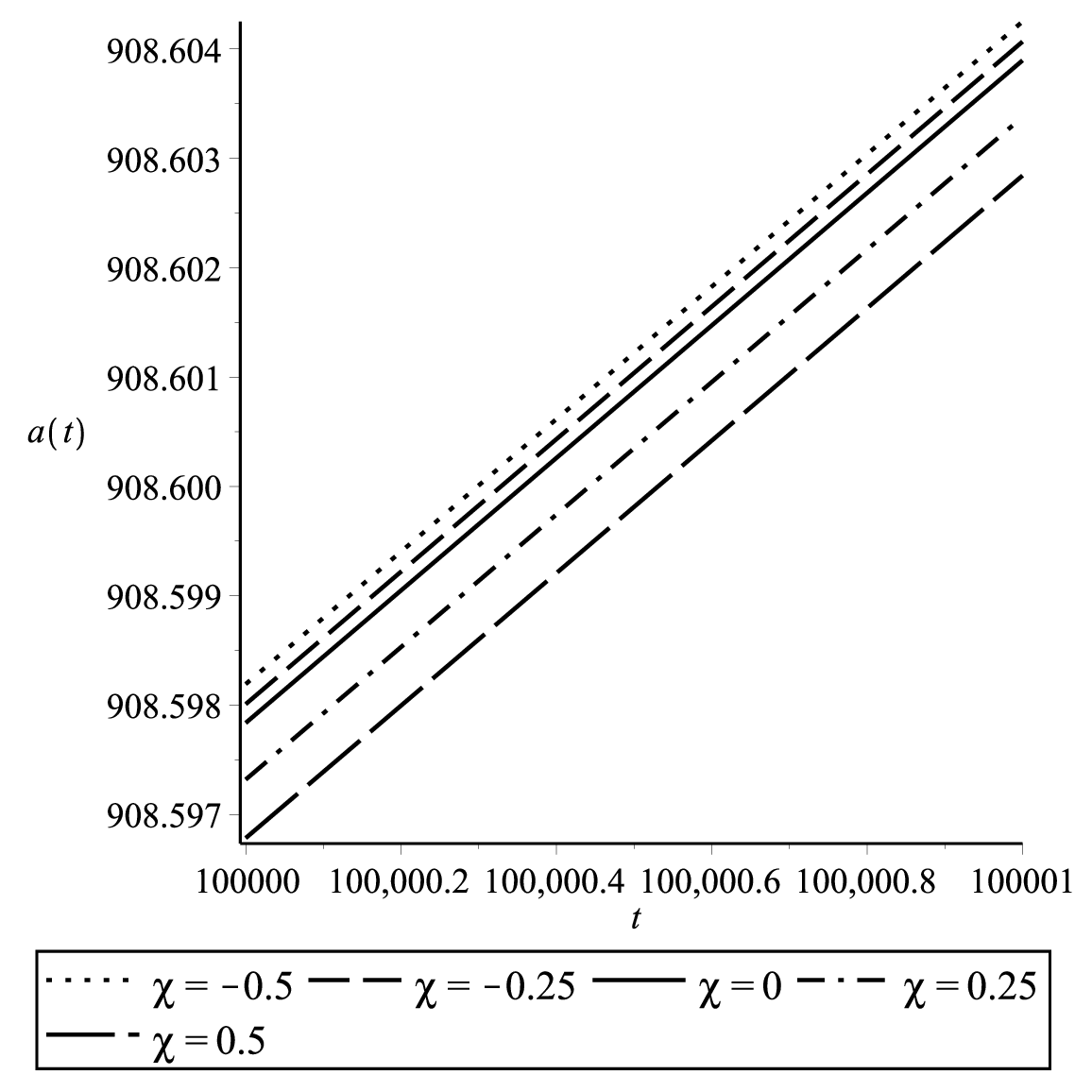}
			
			\label{fig:a_chi_short}
		\end{subfigure}
		\hfill
		\begin{subfigure}{0.45\textwidth}
			\centering
			\includegraphics[width=\textwidth]{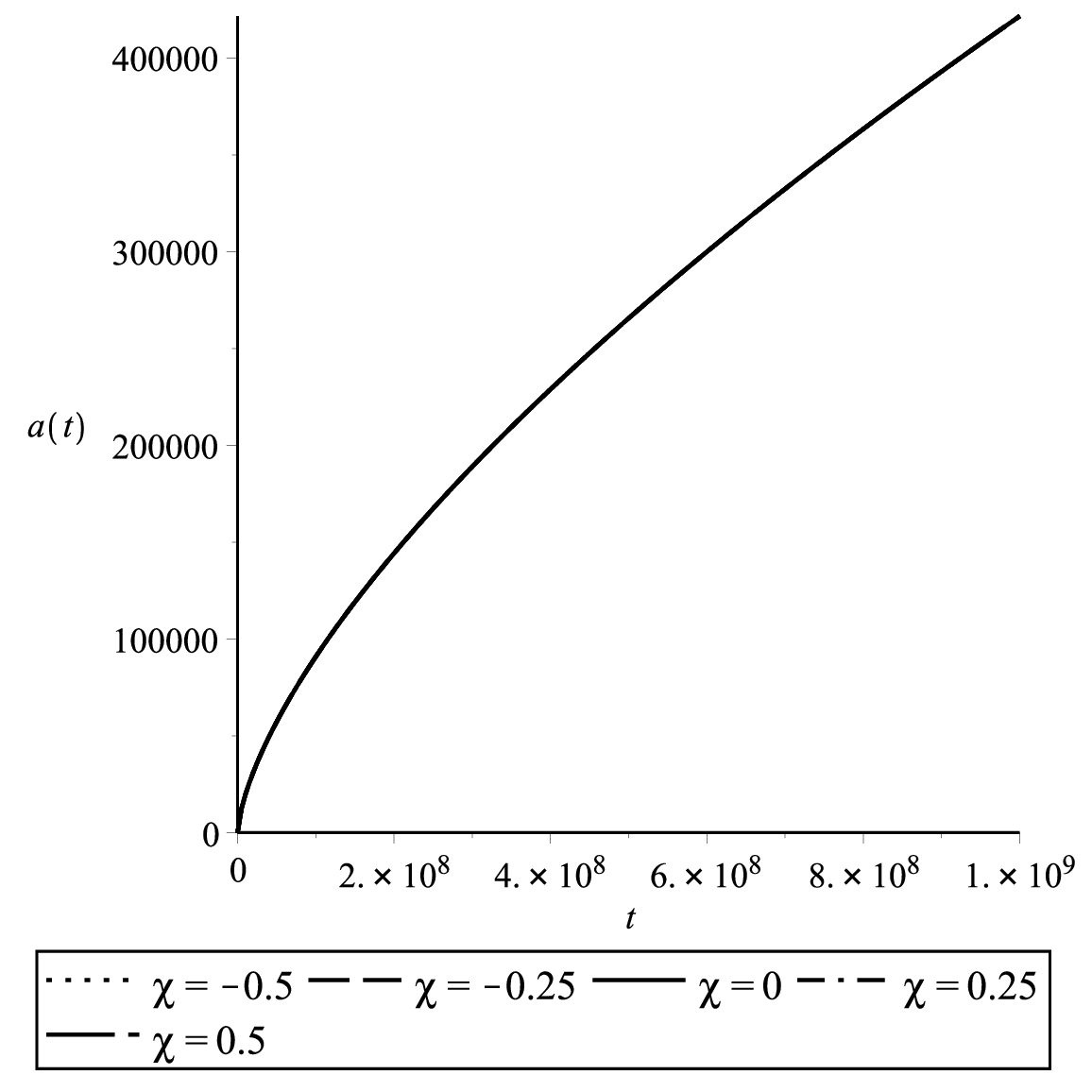}
			
			\label{fig:a_chi_long}
		\end{subfigure}
		\hfill
		\begin{subfigure}{0.45\textwidth}
			\centering
			\includegraphics[width=\textwidth]{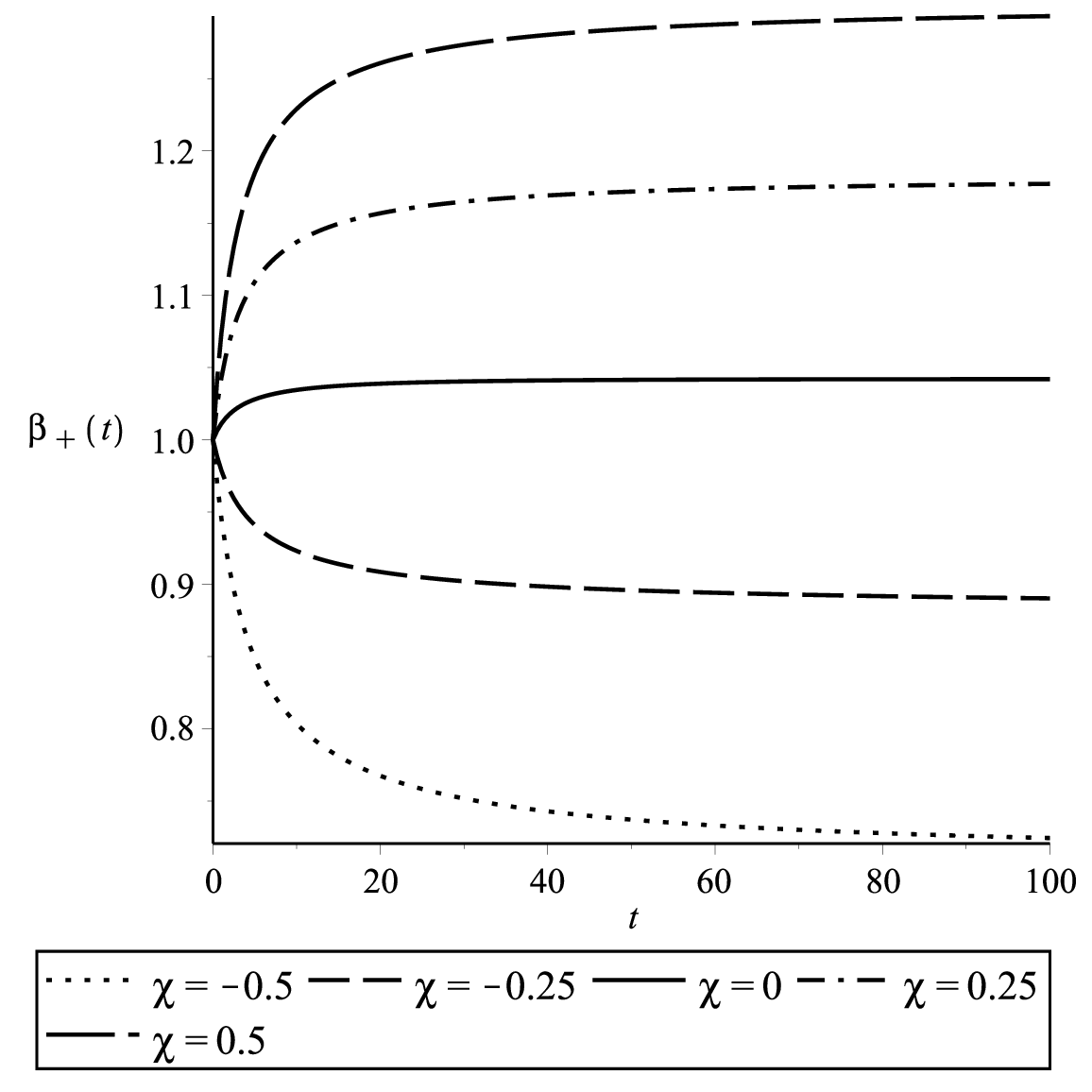}
			
			\label{fig:beta_plus_chi_short}
		\end{subfigure}
		\hfill
		\begin{subfigure}{0.45\textwidth}
			\centering
			\includegraphics[width=\textwidth]{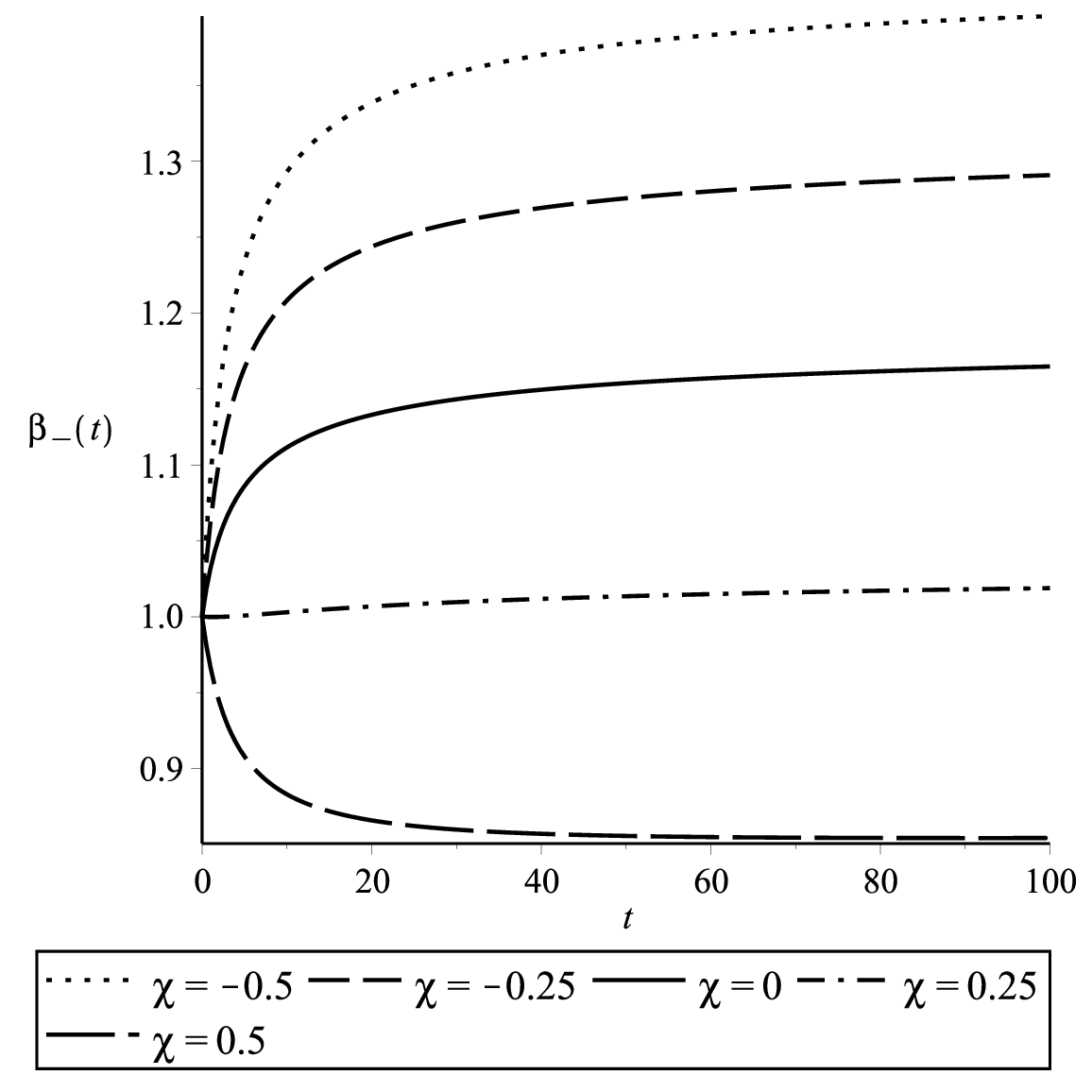}
			
			\label{fig:beta_minus_chi_short}
		\end{subfigure}
		
		\caption{\small Behavior of $a$, $\beta_{+}$ and $\beta_{-}$ with the time $t$ for different values of $\chi$. We take $\rho_1$ = $\rho_2$ = 0.1, $C_1$ = 0.2, $C_2$ = 0.1, $\sigma_{+}$ = 0.01, $\sigma_{-}$ = -0.05.}
		\label{fig:combined_chi}
	\end{figure}

\begin{figure}[H]
		\centering
		\begin{subfigure}{0.45\textwidth}
			\centering
			\includegraphics[width=\textwidth]{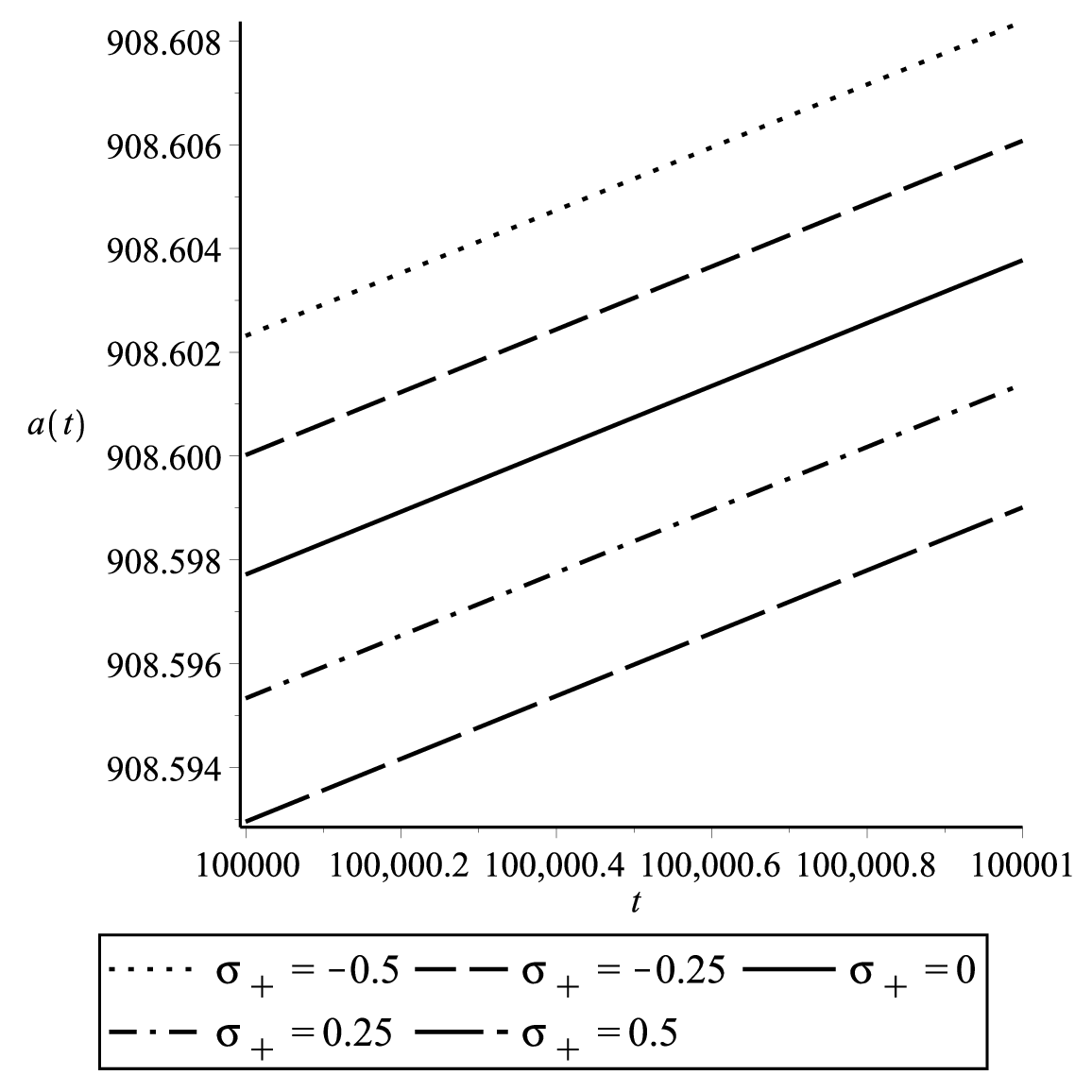}
			\label{fig:a_sigma_short}
		\end{subfigure}
		\hfill
		\begin{subfigure}{0.45\textwidth}
			\centering
			\includegraphics[width=\textwidth]{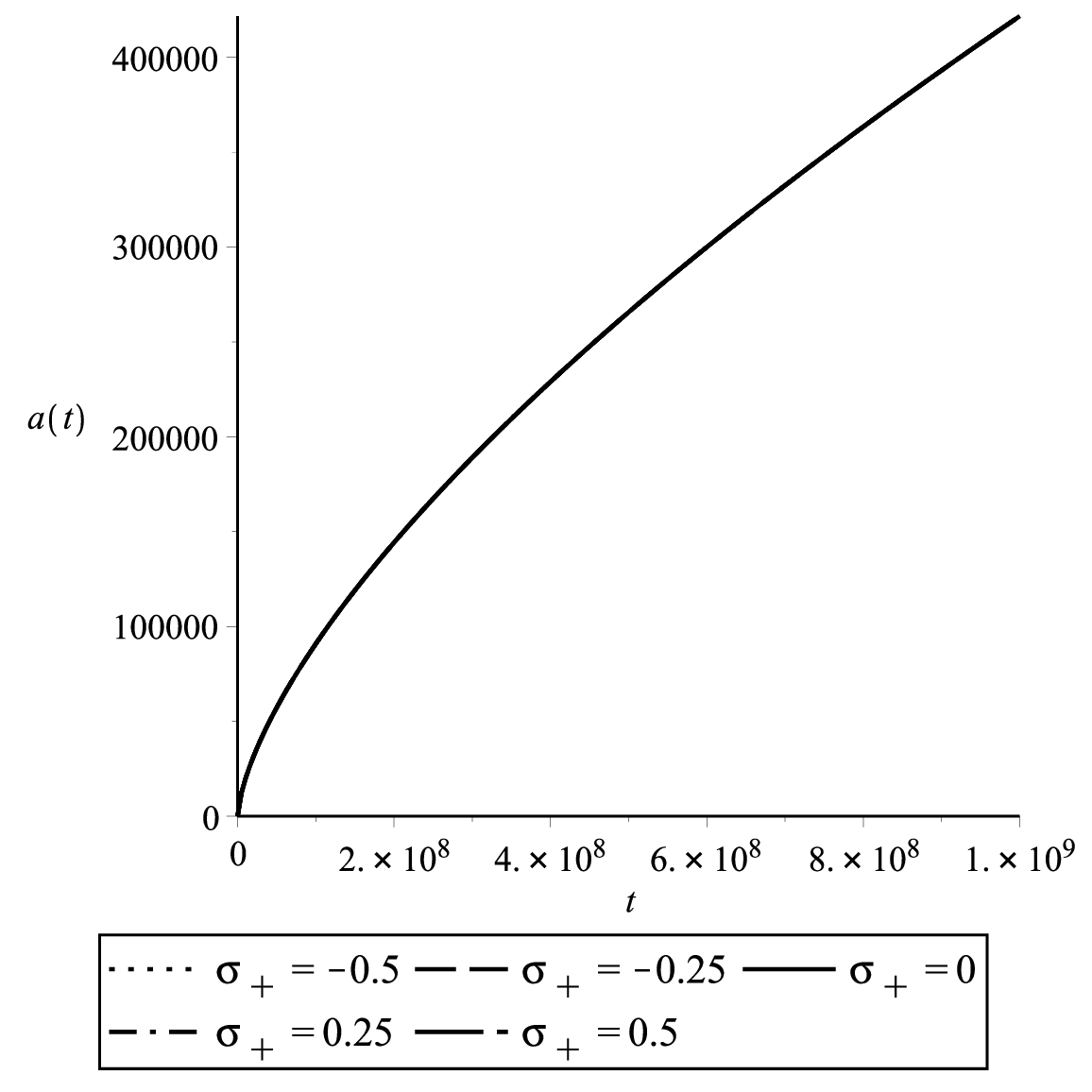}
			\label{fig:a_sigma_long}
		\end{subfigure}
		\hfill
		\begin{subfigure}{0.45\textwidth}
			\centering
			\includegraphics[width=\textwidth]{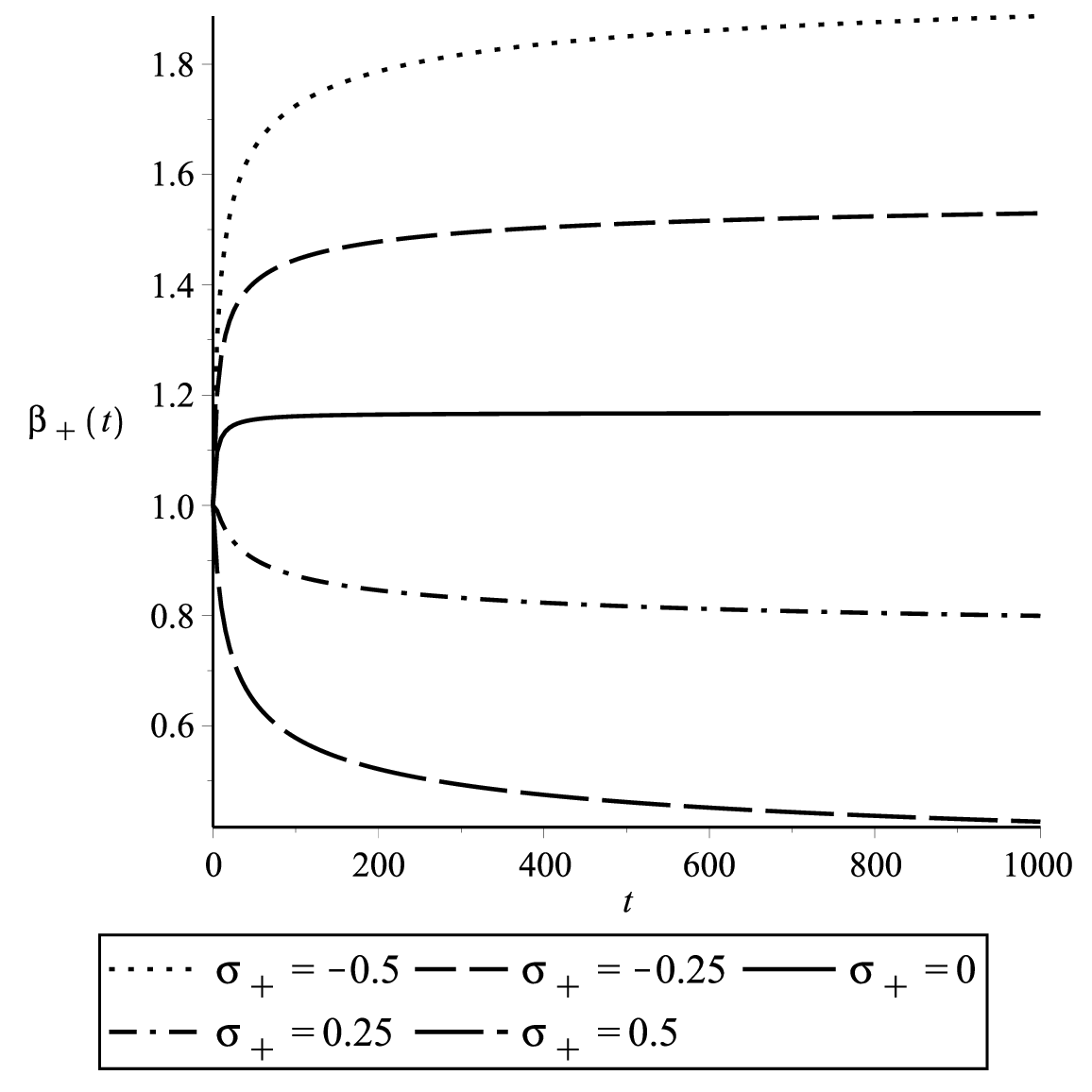}
			\label{fig:beta_plus_sigma_short}
		\end{subfigure}
		\hfill
		\begin{subfigure}{0.45\textwidth}
			\centering
			\includegraphics[width=\textwidth]{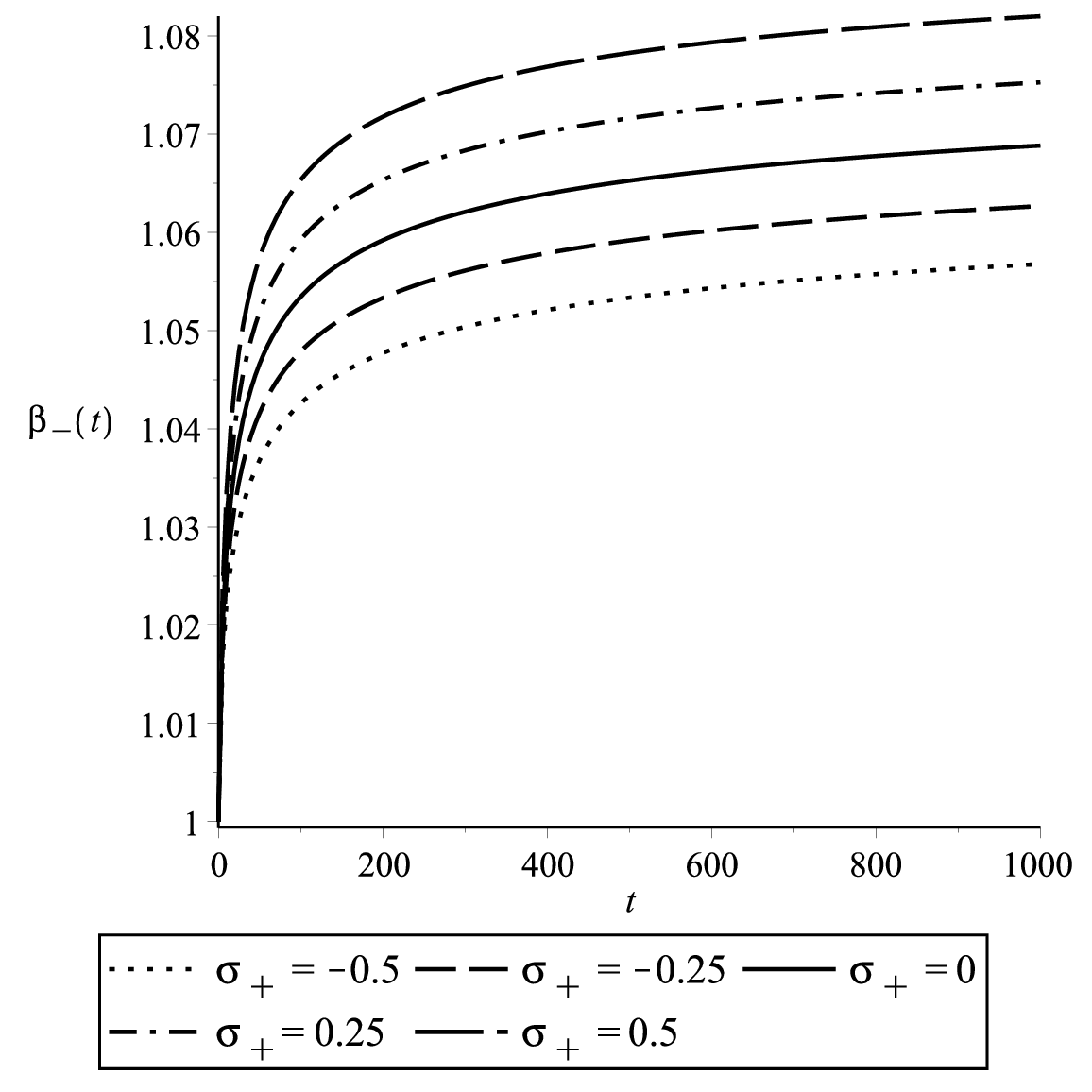}
			\label{fig:beta_minus_sigma_short}
		\end{subfigure}
		
		\caption{\small Behavior of $a$, $\beta_{+}$, and $\beta_{-}$ with the time $t$ for different values of $\sigma_{+}$. We take $\rho_1$ = $\rho_2$ = 0.1, $C_1$ = 0.1, $C_2$ = 0.2, $\chi$ = 0.1, and $\sigma_{-}$ = -0.05.}
		\label{fig:combined_sigma}
	\end{figure}

\begin{figure}[H]
		\centering
		\begin{subfigure}{0.45\textwidth}
			\centering
			\includegraphics[width=\textwidth]{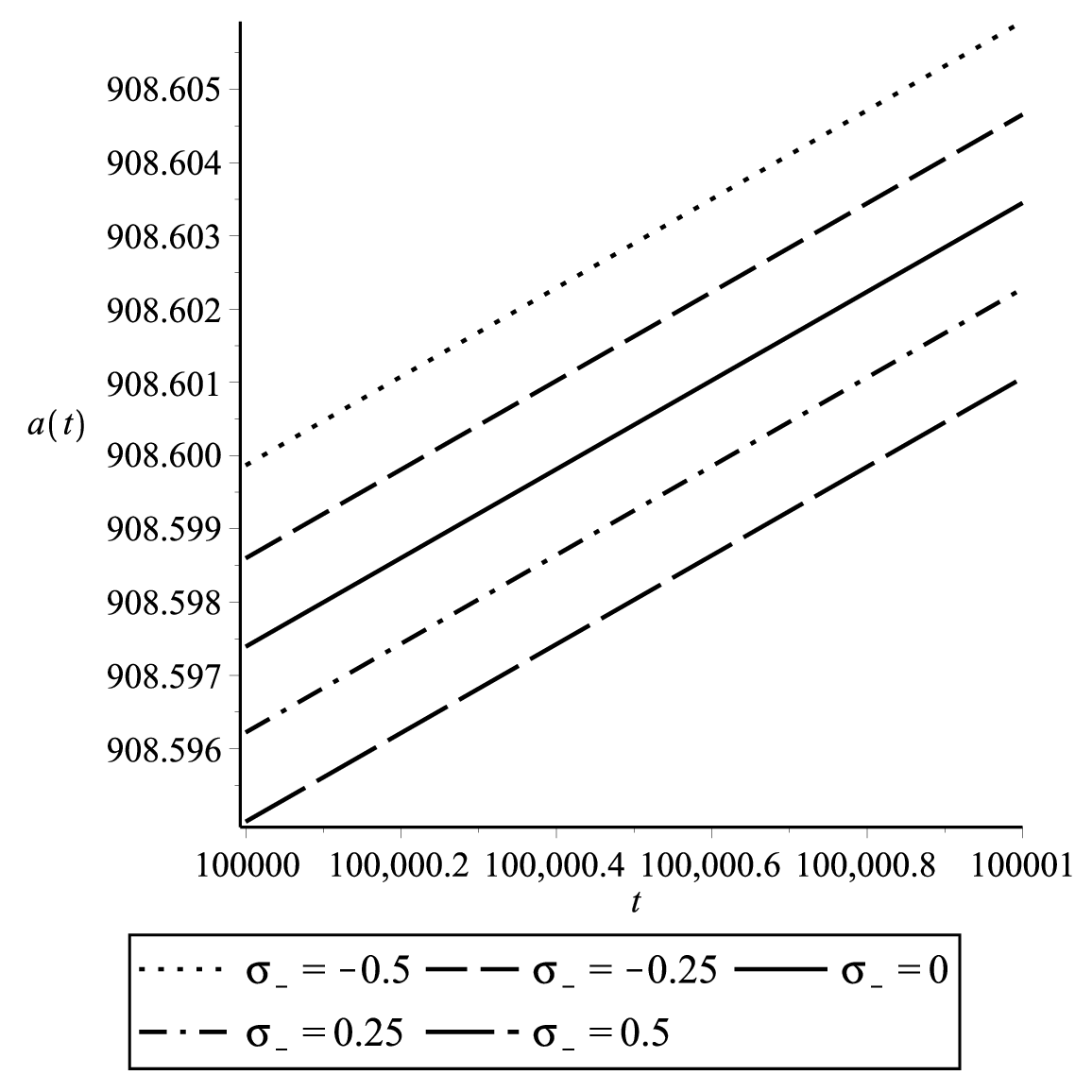}
			\label{fig:a_sigma_minus_short}
		\end{subfigure}
		\hfill
		\begin{subfigure}{0.45\textwidth}
			\centering
			\includegraphics[width=\textwidth]{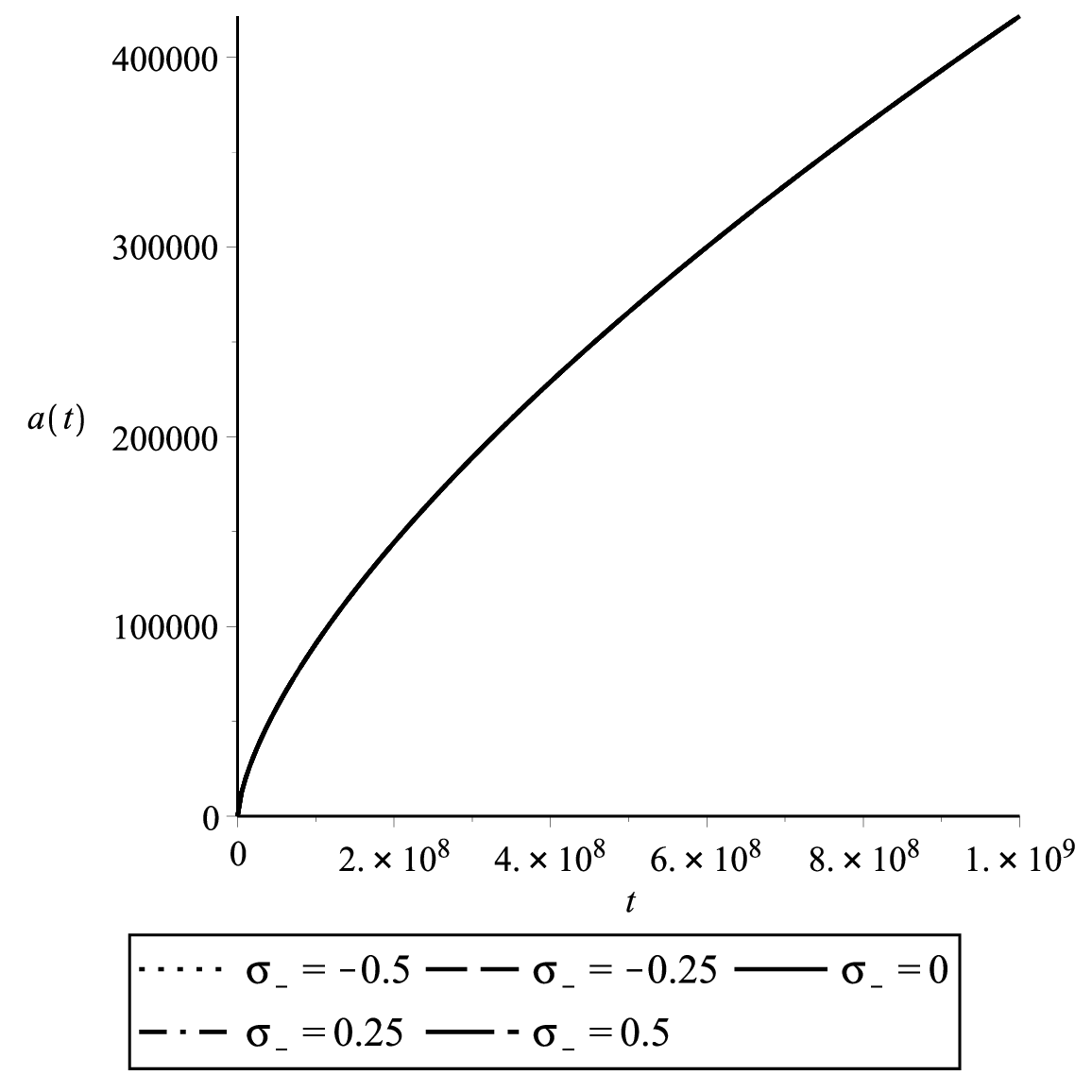}
			\label{fig:a_sigma_minus_long}
		\end{subfigure}
		\hfill
		\begin{subfigure}{0.45\textwidth}
			\centering
			\includegraphics[width=\textwidth]{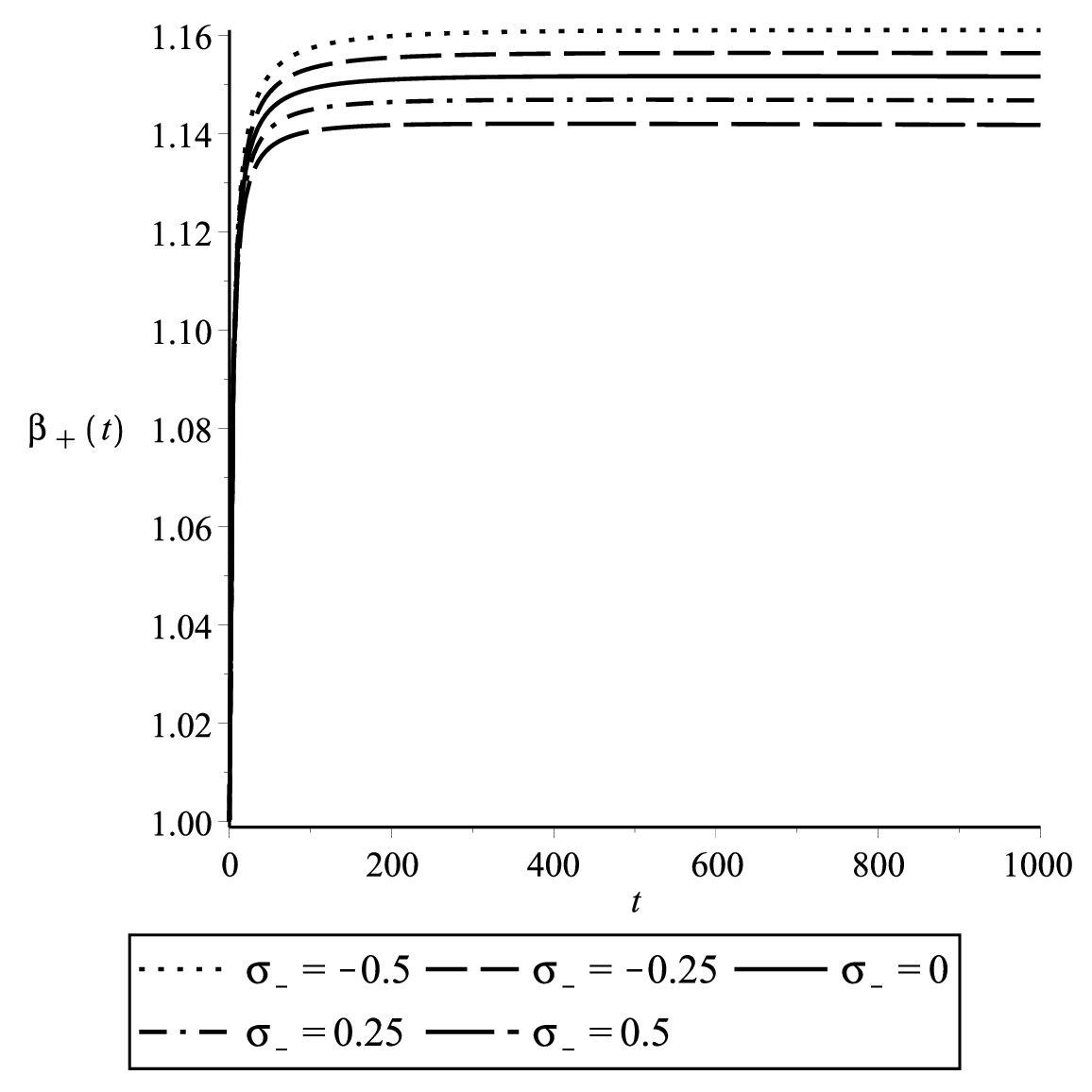}
			\label{fig:beta_plus_sigma_minus_short}
		\end{subfigure}
		\hfill
		\begin{subfigure}{0.45\textwidth}
			\centering
			\includegraphics[width=\textwidth]{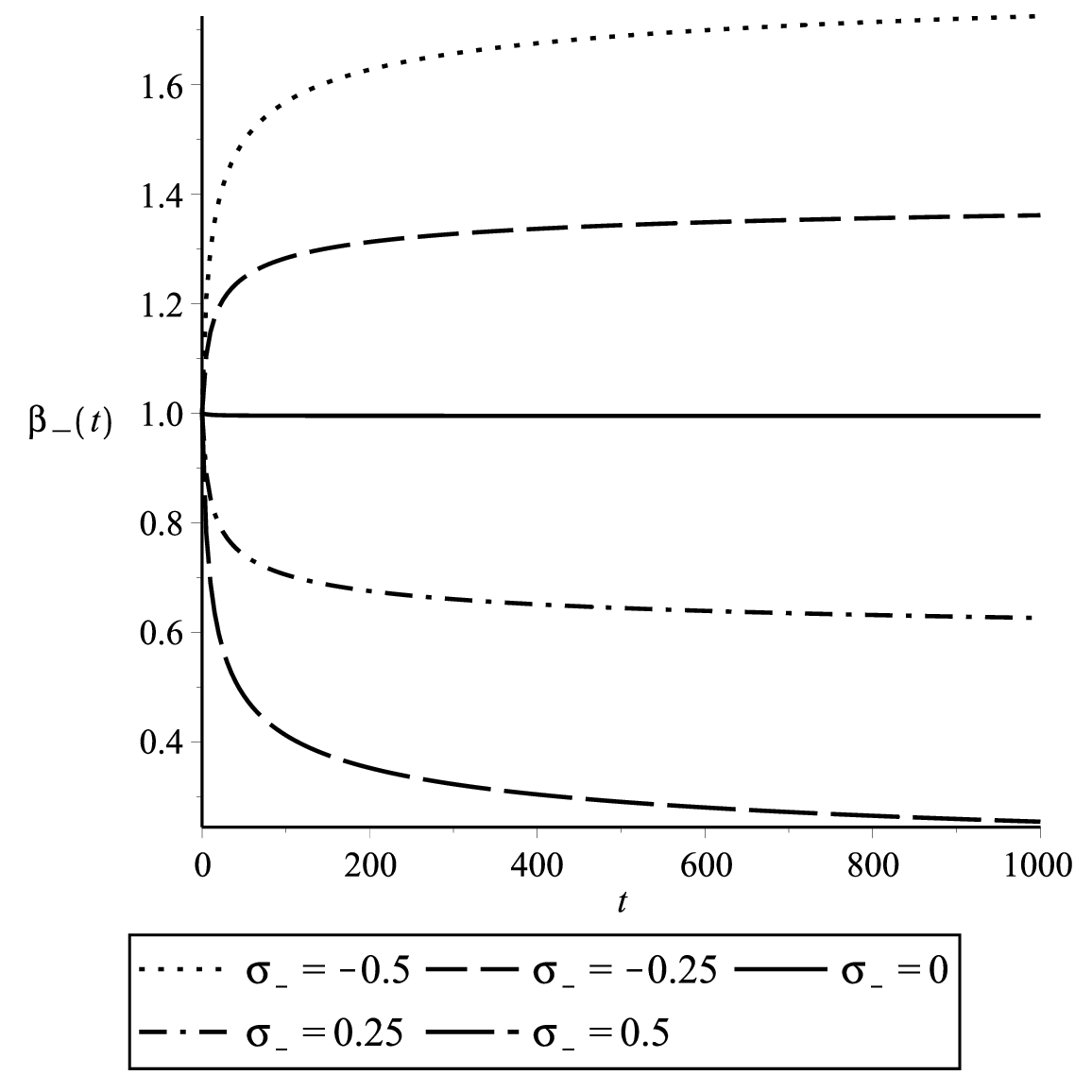}
			\label{fig:beta_minus_sigma_minus_short}
		\end{subfigure}
		
		\caption{\small Behavior of $a$, $\beta_{+}$, and $\beta_{-}$ with the time $t$ for different values of $\sigma_{-}$. We take $\rho_1$ = $\rho_2$ = 0.1, $C_1$ = 0.1, $C_2$ = 0.2, $\chi$ = 0.1, and $\sigma_{+}$ = 0.01.}
		\label{fig:combined_sigma_minus}
	\end{figure}

For the noncommutative parameters, we obtained the following results: For $\chi$, a more rapid expansion of the scale factor $a$ is observed for negative values of $\chi$, as compared to the case $\chi = 0$, whereas positive values of $\chi$ lead to a slower expansion. Additionally, $\beta_{+}$ reaches higher constant values as $\chi$ increases, while $\beta_{-}$ reaches higher constant values as $\chi$ decreases (see Figure~\ref{fig:combined_chi} and Table~\ref{tab:chi_values_full}). As for $\sigma_{+}$, its influence follows a similar pattern: the scale factor expands faster for negative values and more slowly for positive ones, relative to the case $\sigma_{+} = 0$. Furthermore, $\beta_{+}$ asymptotically increases as $\sigma_{+}$ becomes more negative, while $\beta_{-}$ grows with increasing values of $\sigma_{+}$ (See Figure \ref{fig:combined_sigma} and Table \ref{tab:sigma_plus_values_full}). Regarding $\sigma_{-}$, the scale factor $a$ expands more rapidly for negative values of $\sigma_{-}$, when compared to the case $\sigma_{-} = 0$, whereas positive values of $\sigma_{-}$ lead to a slower expansion. Additionally, both $\beta_{+}$ and $\beta_{-}$ reach higher constant values as $\sigma_{-}$ decreases (see Figure~\ref{fig:combined_sigma_minus} and Table~\ref{tab:sigma_minus_values_full}).

In this context, noncommutative parameters can, in principle, be employed to describe the current accelerated expansion of the universe without invoking the existence of dark energy. This is because such parameters can be chosen to induce an increase in the expansion rate of the isotropic scale factor $a(t)$. Furthermore, noncommutative parameters influence the time required for the complete isotropization of the early universe, governing the transition from the anisotropic to the isotropic regime, at which point the anisotropic parameters $ \beta_{+} (t) $ and $ \beta_{-} (t) $ reach constant values.

\subsection{Varying the anisotropy parameters}

\begin{figure}[H]
	\centering
	\begin{subfigure}{0.45\textwidth}
		\centering
		\includegraphics[width=\textwidth]{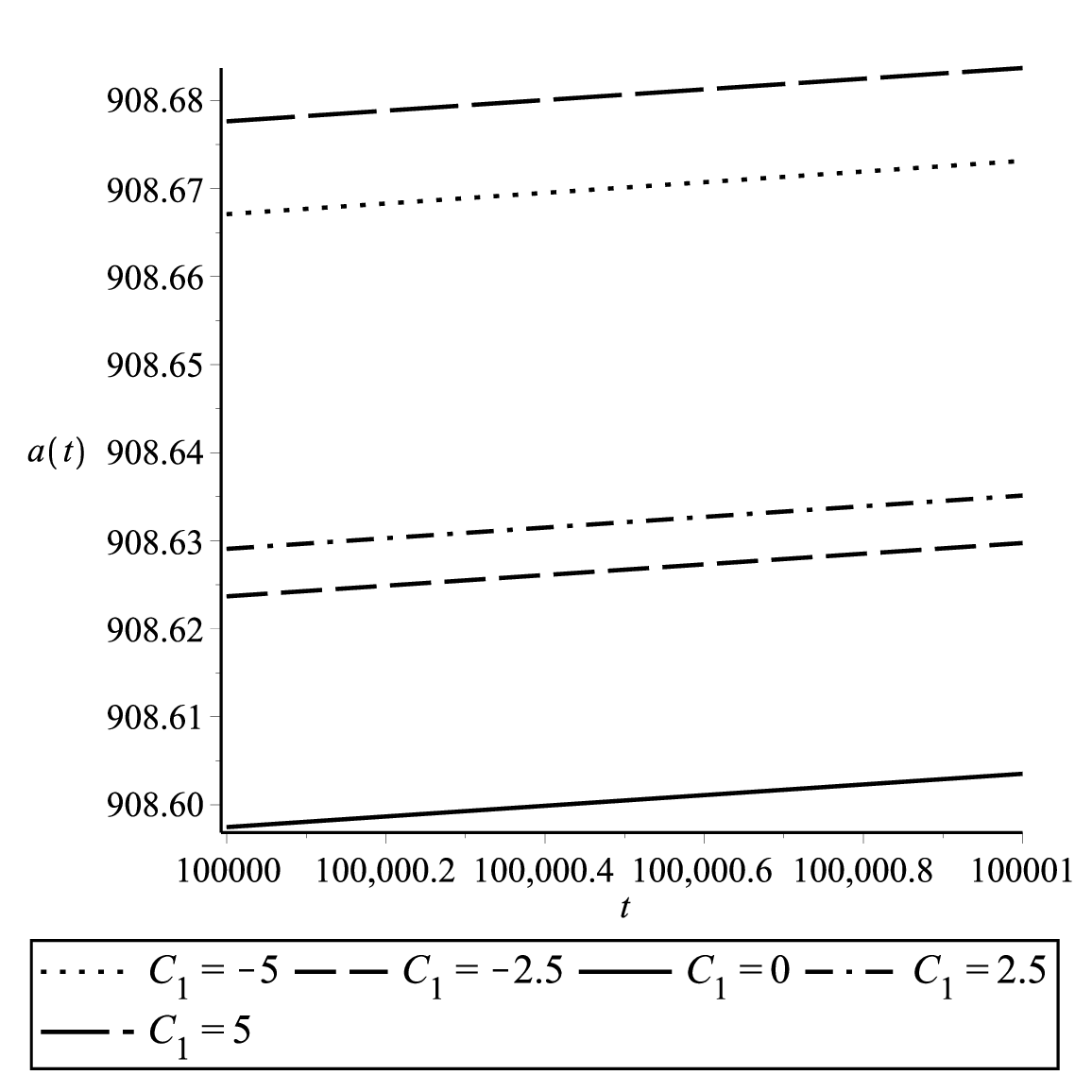}
		\label{fig:a_C1_short}
	\end{subfigure}
	\hfill
	\begin{subfigure}{0.45\textwidth}
		\centering
		\includegraphics[width=\textwidth]{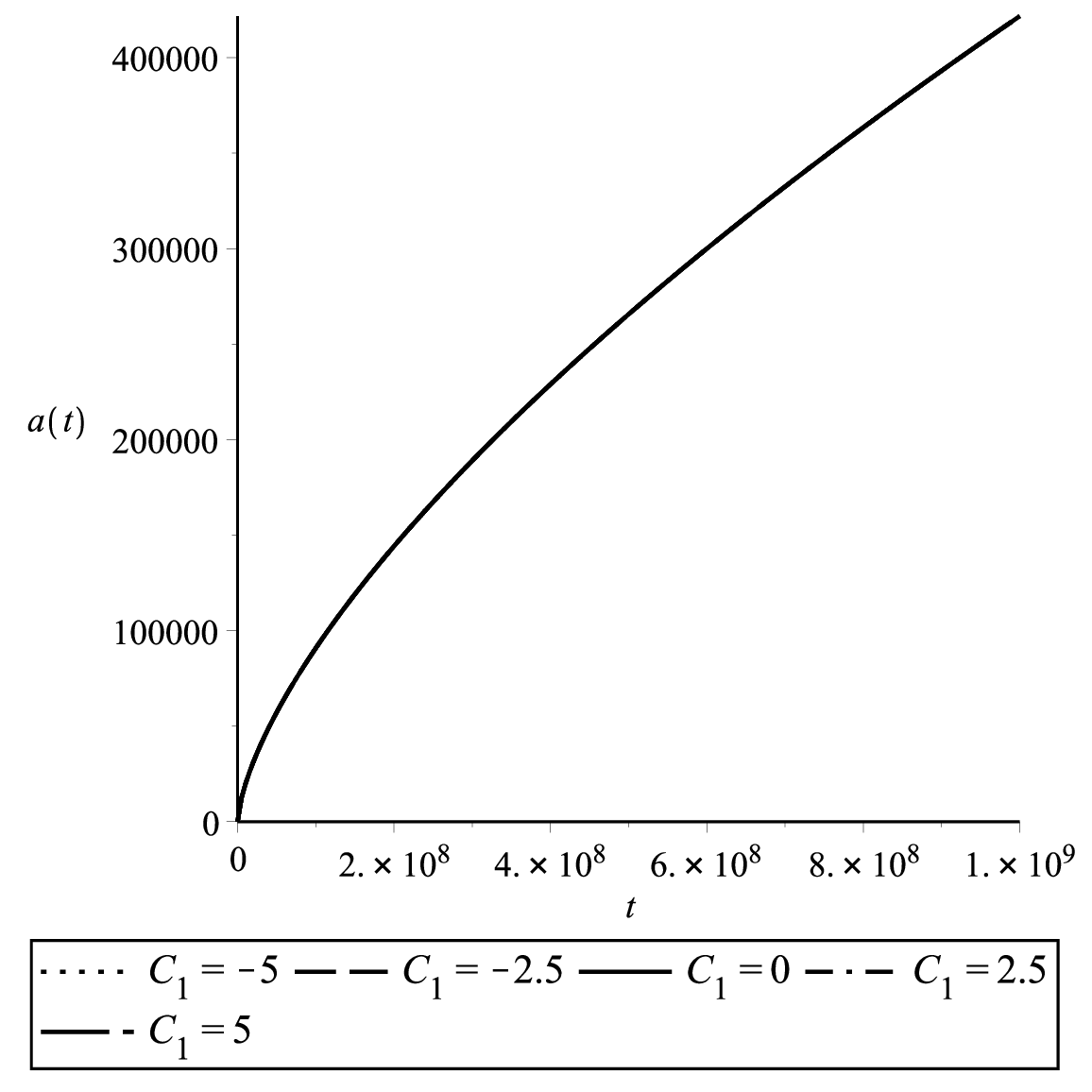}
		\label{fig:a_C1_long}
	\end{subfigure}
	\hfill
	\begin{subfigure}{0.45\textwidth}
		\centering
		\includegraphics[width=\textwidth]{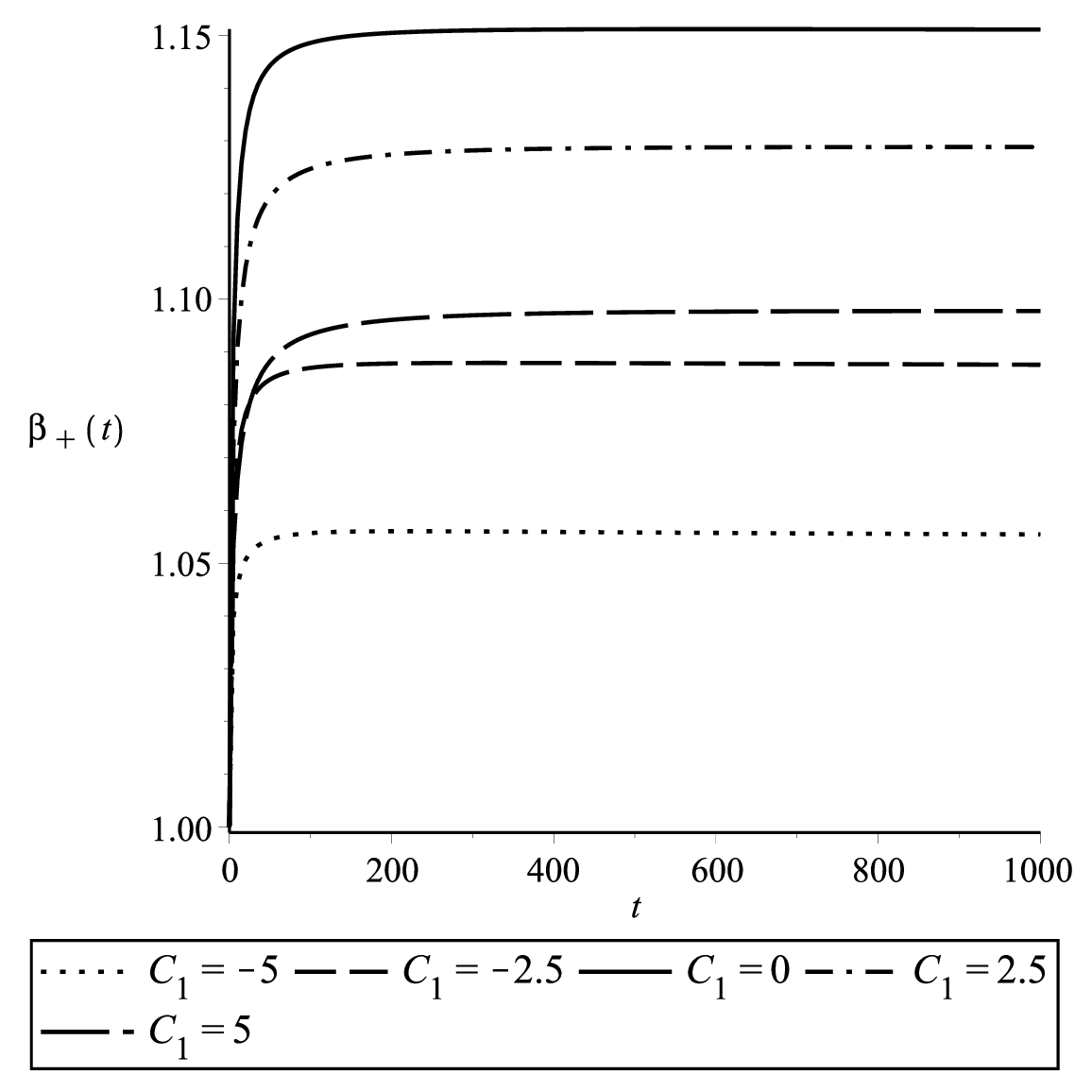}
		\label{fig:beta_plus_C1_short}
	\end{subfigure}
	\hfill
	\begin{subfigure}{0.45\textwidth}
		\centering
		\includegraphics[width=\textwidth]{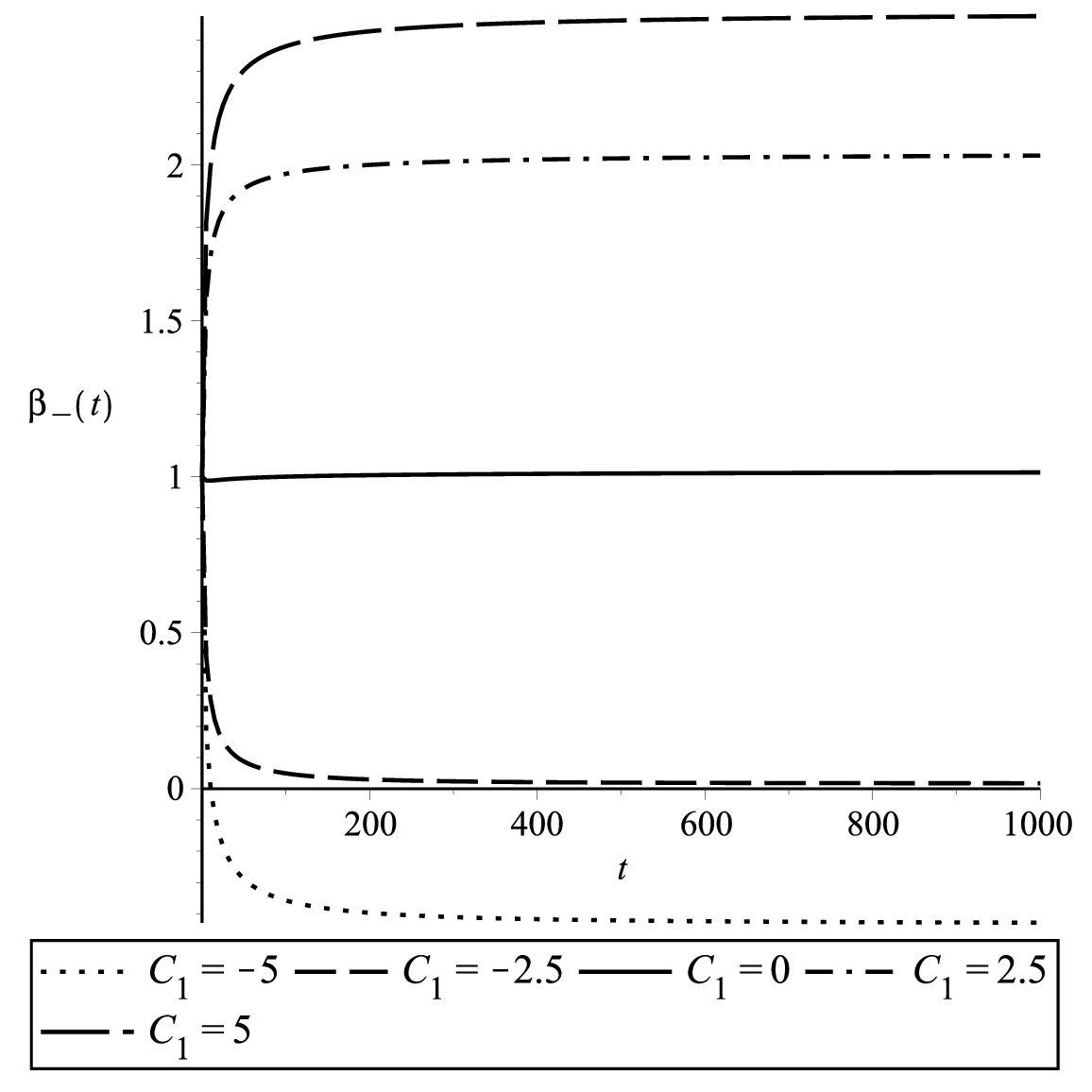}
		\label{fig:beta_minus_C1_short}
	\end{subfigure}
	
	\caption{\small Behavior of $a$, $\beta_{+}$, and $\beta_{-}$ with the time $t$ for different values of $C_{1}$. We take $\rho_1$ = $\rho_2$ = 0.1, $C_2$ = 0.2, $\chi$ = 0.1, $\sigma_{+}$ = 0.01, and $\sigma_{-}$ = -0.05.}
	\label{fig:combined_C1}
\end{figure}

For the parameters $C_{1}$ and $C_{2}$, associated with anisotropy, we obtained the following results. For $C_{1}$, we find that the scale factor $a$ expands more rapidly for larger absolute values, with positive values leading to greater expansion than their negative counterparts. Regarding the scale function $\beta_{+}$, we have: for positive values of $C_{1}$, the larger the value of $C_{1}$, the smaller the constant value reached by $\beta_{+}$; for negative values, the more negative the value of $C_{1}$, the smaller the final value attained by $\beta_{+}$; and the maximum constant value of $\beta_{+}$ is reached when $C_{1} = 0$. As for the scale function $\beta_{-}$, for positive values of $C_{1}$, the larger the value, the higher the constant value approached by $\beta_{-}$, which is always greater than the one attained for $C_{1} = 0$; for negative values of $C_{1}$, the smaller the value, the lower the constant value reached by $\beta_{-}$, and these are always below the value obtained for $C_{1} = 0$ (see Figure~\ref{fig:combined_C1} and Table~\ref{tab:C1_values_full}). 

\begin{figure}[H]
	\centering
	\begin{subfigure}{0.45\textwidth}
		\centering
		\includegraphics[width=\textwidth]{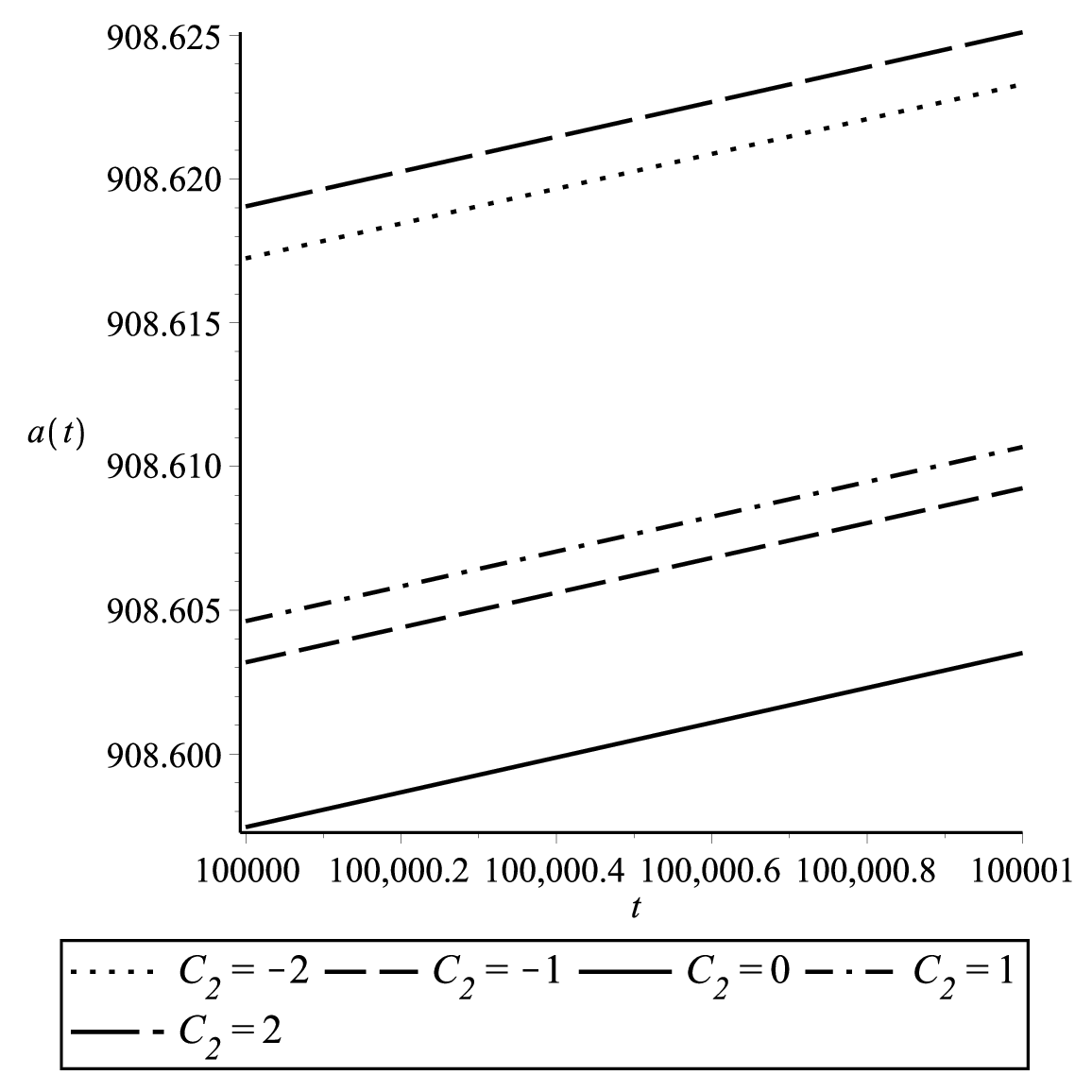}
		\label{fig:a_C2_short}
	\end{subfigure}
	\hfill
	\begin{subfigure}{0.45\textwidth}
		\centering
		\includegraphics[width=\textwidth]{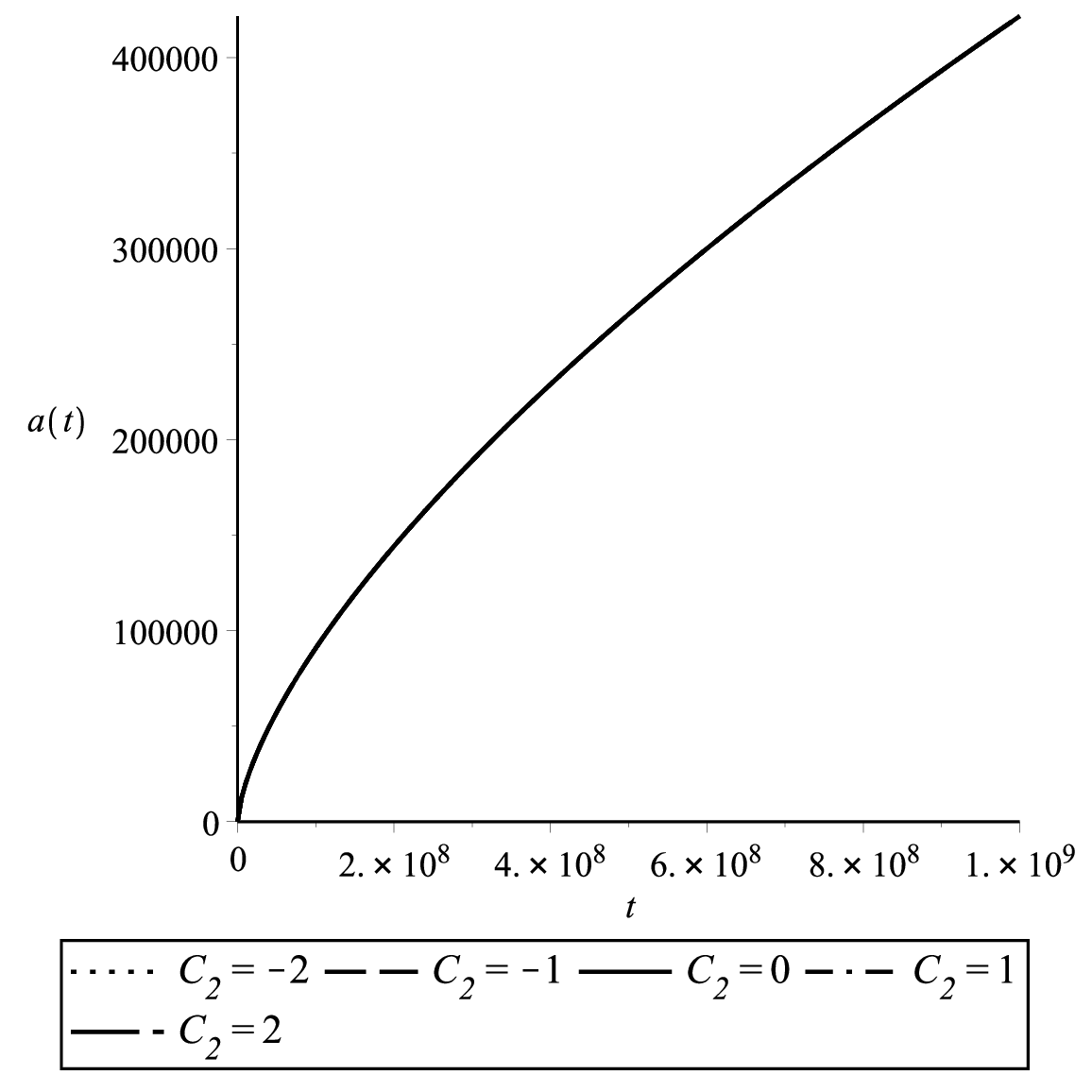}
		\label{fig:a_C2_long}
	\end{subfigure}
	\hfill
	\begin{subfigure}{0.45\textwidth}
		\centering
		\includegraphics[width=\textwidth]{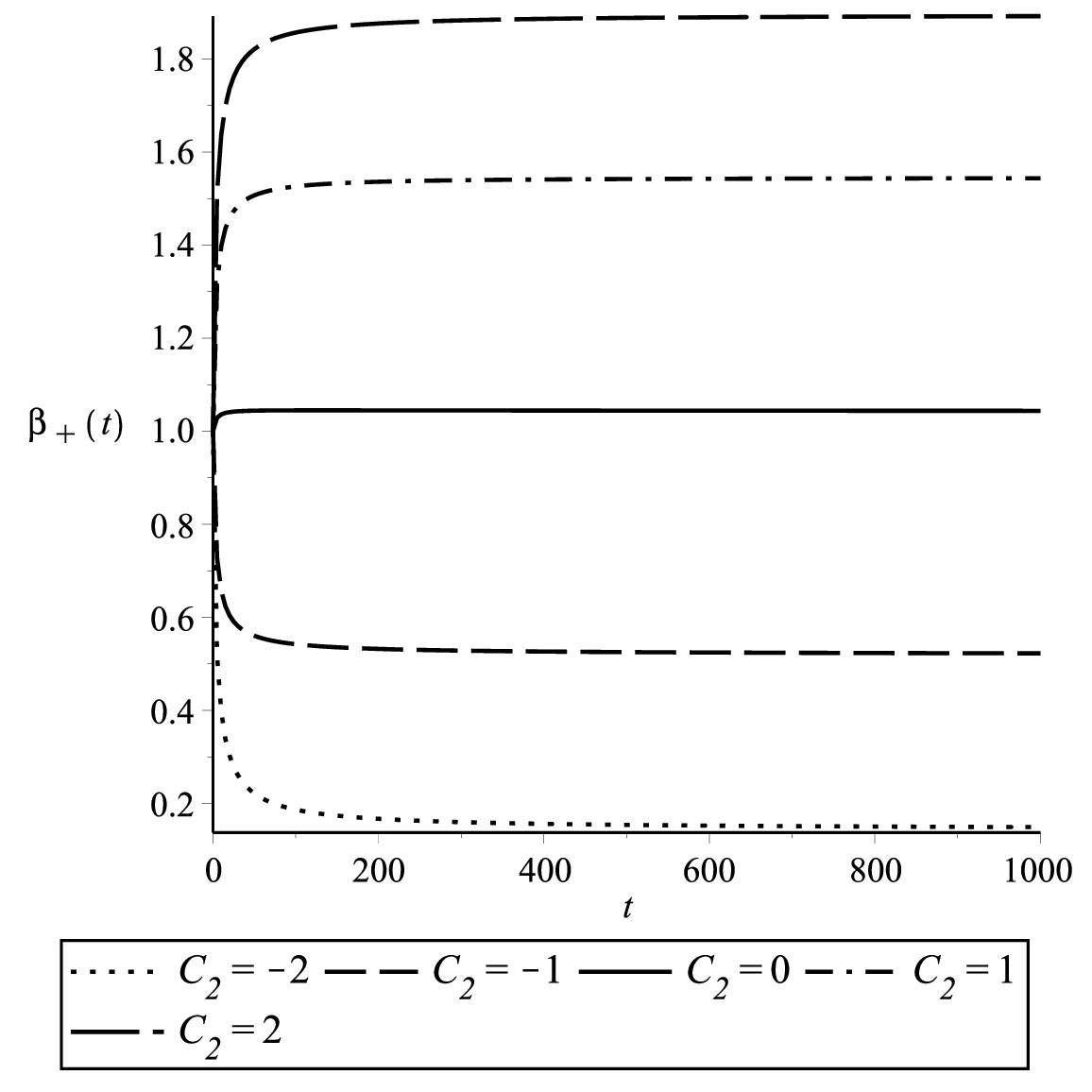}
		\label{fig:beta_plus_C2_short}
	\end{subfigure}
	\hfill
	\begin{subfigure}{0.45\textwidth}
		\centering
		\includegraphics[width=\textwidth]{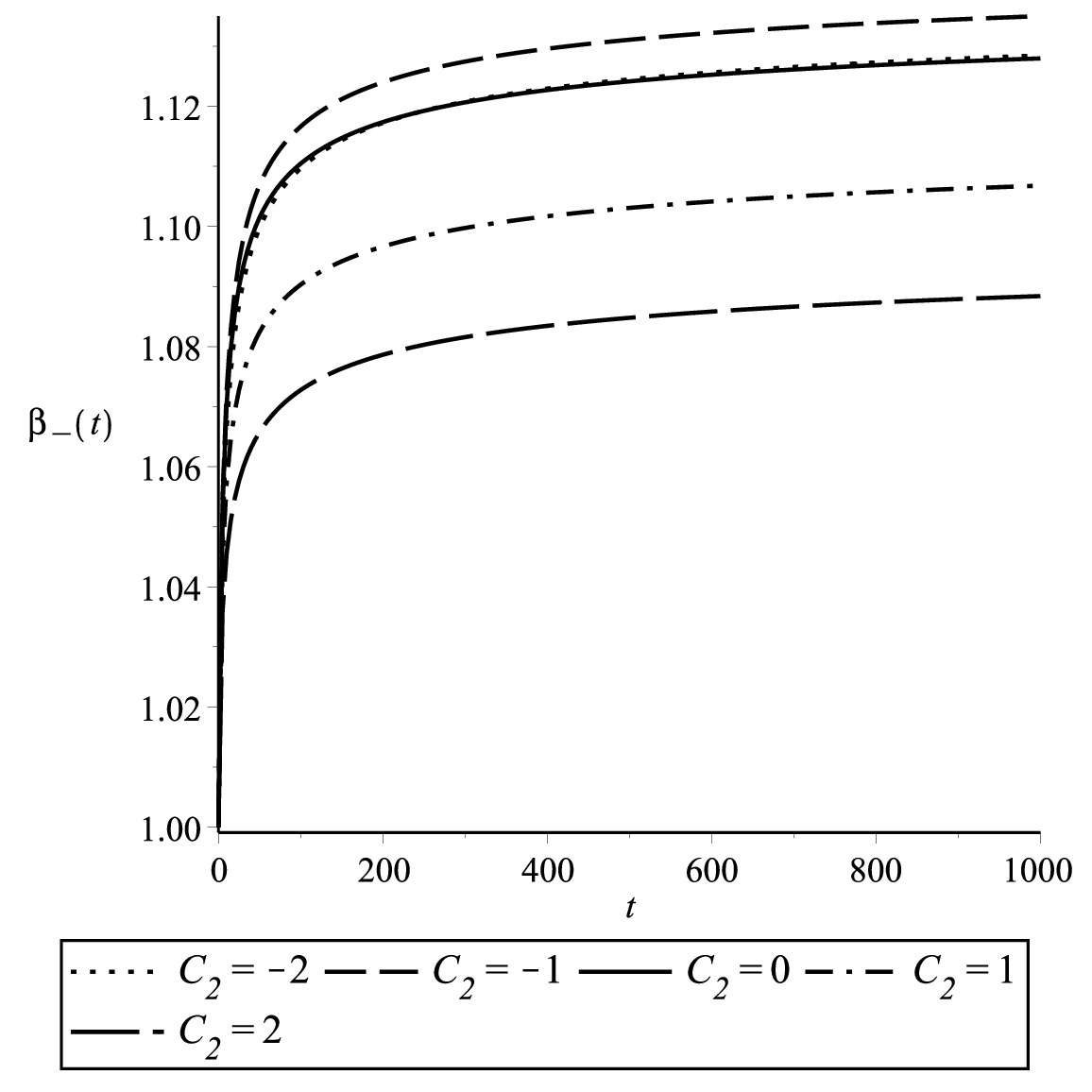}
		\label{fig:beta_minus_C2_short}
	\end{subfigure}
	
	\caption{\small Behavior of $a$, $\beta_{+}$, and $\beta_{-}$ with the time $t$ for different values of $C_{2}$. We take $\rho_1$ = $\rho_2$ = 0.1, $C_1$ = 0.2, $\chi$ = 0.1, $\sigma_{+}$ = 0.01, and $\sigma_{-}$ = -0.05.}
	\label{fig:combined_C2}
\end{figure}

For $C_{2}$, the scale factor $a$ increases for larger absolute values, with positive values leading to greater expansion than their negative counterparts. The behavior of the scale function $\beta_{+}$ is as follows: for positive values of $C_{2}$, the larger the value, the higher the constant value approached by $\beta_{+}$, which is always greater than the one attained for $C_{2} = 0$; for negative values of $C_{2}$, the smaller the value, the lower the constant value reached by $\beta_{+}$, and these are always below the value obtained for $C_{2} = 0$. For the scale function  $\beta_{-}$, we have: for positive values of $C_{2}$, the larger the value, the smaller the constant value approached by $\beta_{-}$, which is always less than the one attained for $C_{2} = 0$; for negative values of $C_{2}$, the more negative the value, the smaller the constant value reached by $\beta_{-}$, and these are always above the value obtained for $C_{2} = 0$ (see Figure~\ref{fig:combined_C2} and Table~\ref{tab:C2_values_full}).

\subsection{Varying the RRG parameters}

\begin{figure}[H]
	\centering
	\begin{subfigure}{0.45\textwidth}
		\centering
		\includegraphics[width=\textwidth]{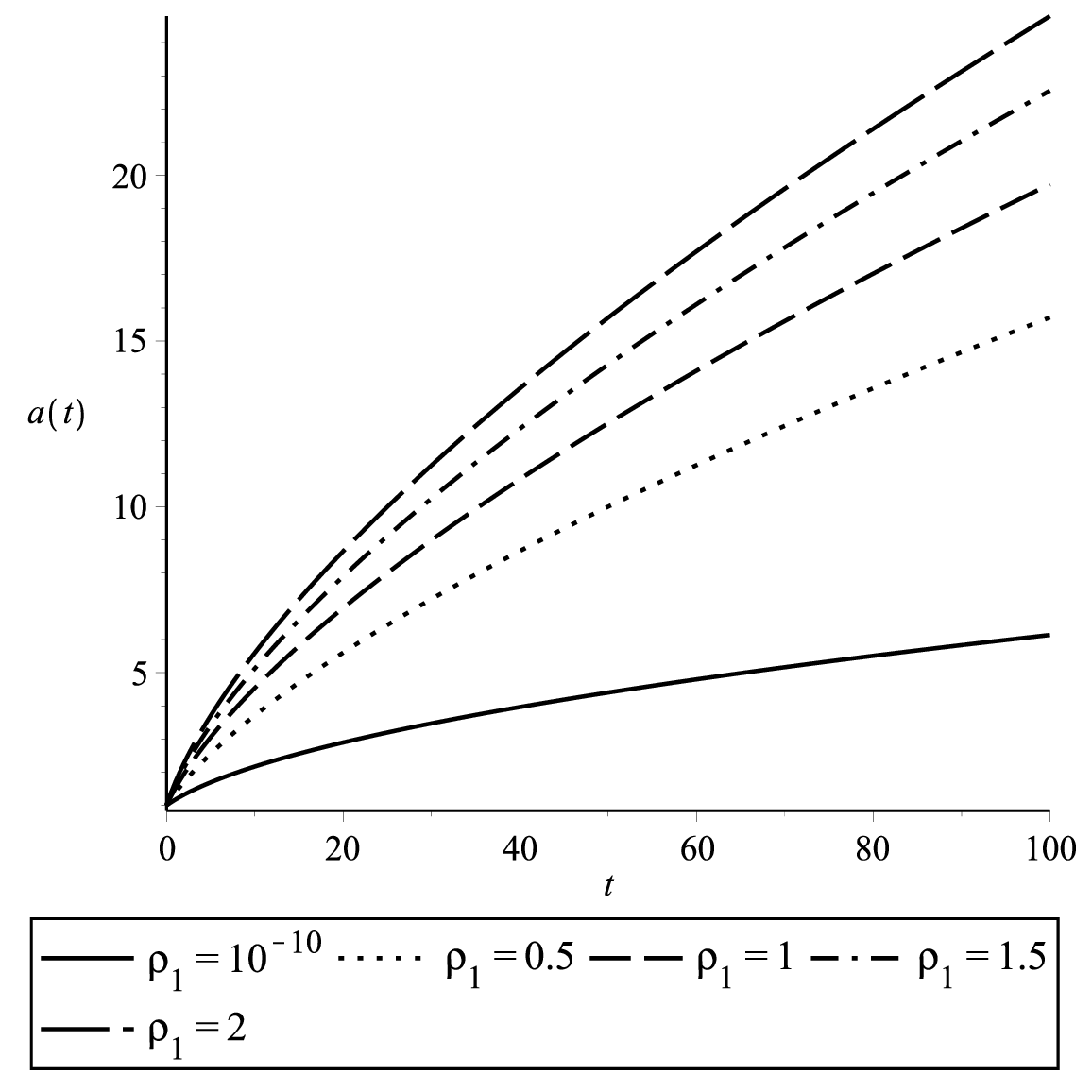}
		\label{fig:a_rho1_short}
	\end{subfigure}
	\hfill
	\begin{subfigure}{0.45\textwidth}
		\centering
		\includegraphics[width=\textwidth]{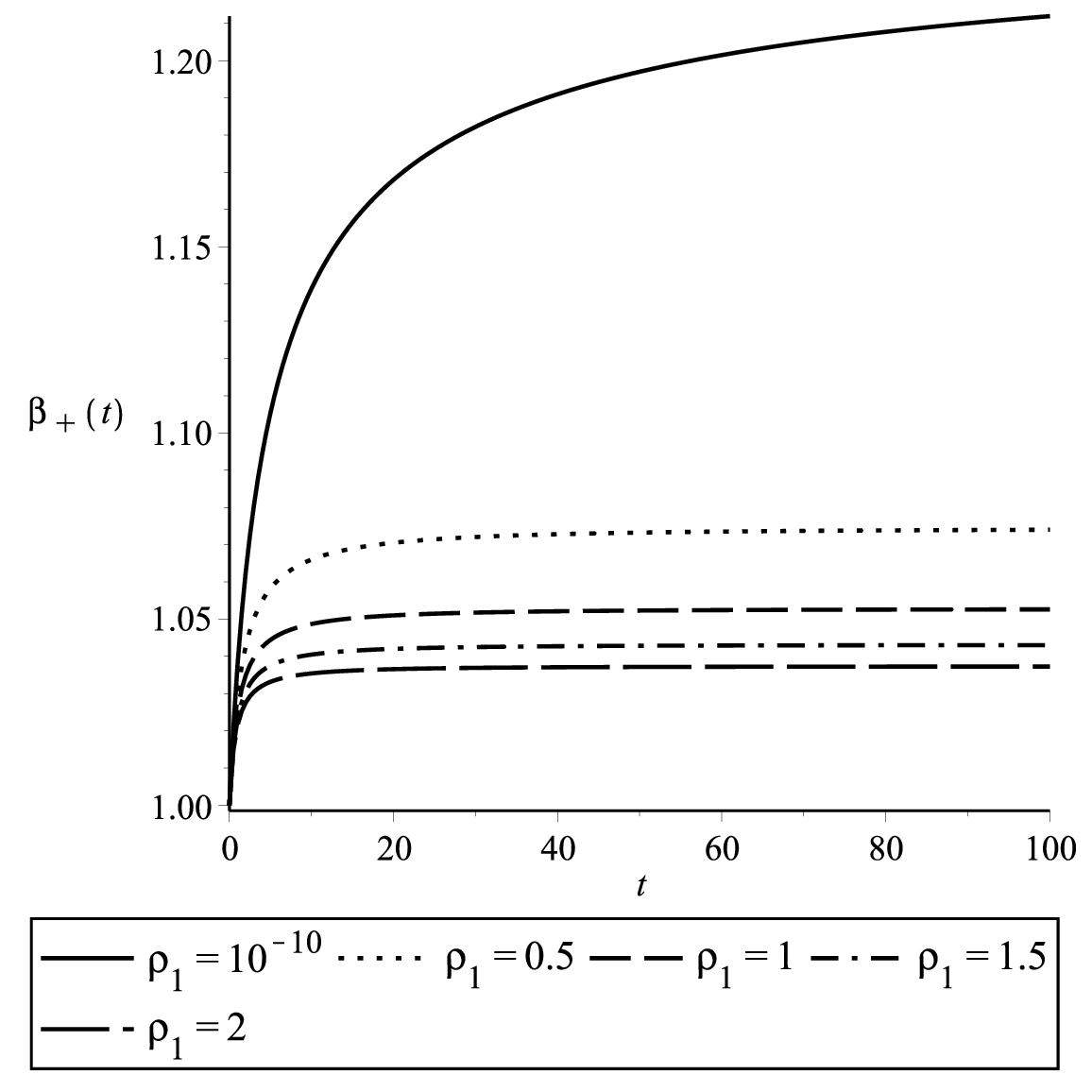}
		\label{fig:beta_plus_rho1_short}
	\end{subfigure}
	\hfill
	\begin{subfigure}{0.45\textwidth}
		\centering
		\includegraphics[width=\textwidth]{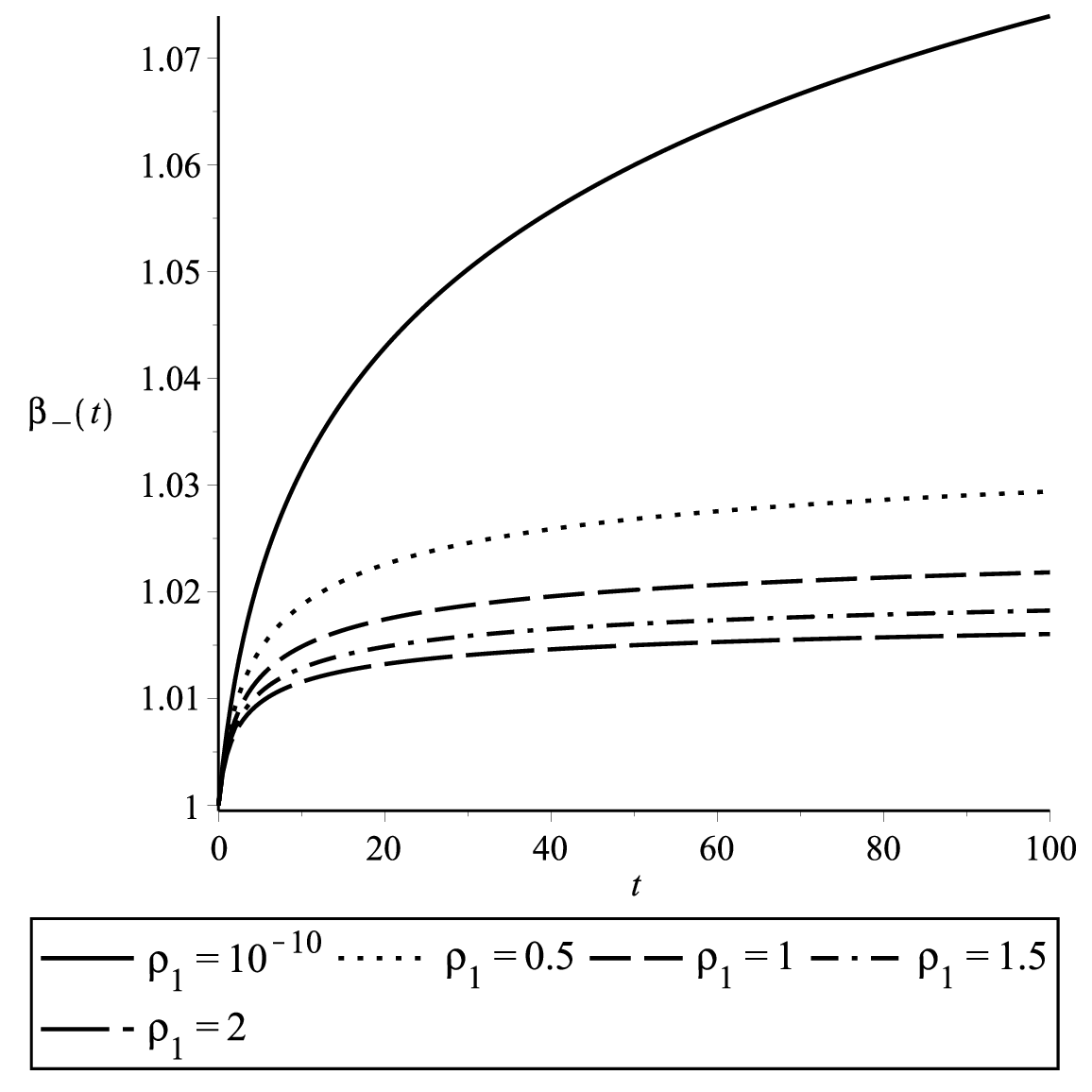}
		\label{fig:beta_minus_rho1_short}
	\end{subfigure}
	
	\caption{\small Behavior of $a$, $\beta_{+}$, and $\beta_{-}$ with the time $t$ for different values of $\rho_{1}$. We take $\rho_2$ = 0.1, $C_1$ = 0.1, $C_2$ = 0.2, $\chi$ = 0.1, $\sigma_{+}$ = 0.01, and $\sigma_{-}$ = -0.05.}
	\label{fig:combined_rho1}
\end{figure}

\begin{figure}[H]
	\centering
	\begin{subfigure}{0.45\textwidth}
		\centering
		\includegraphics[width=\textwidth]{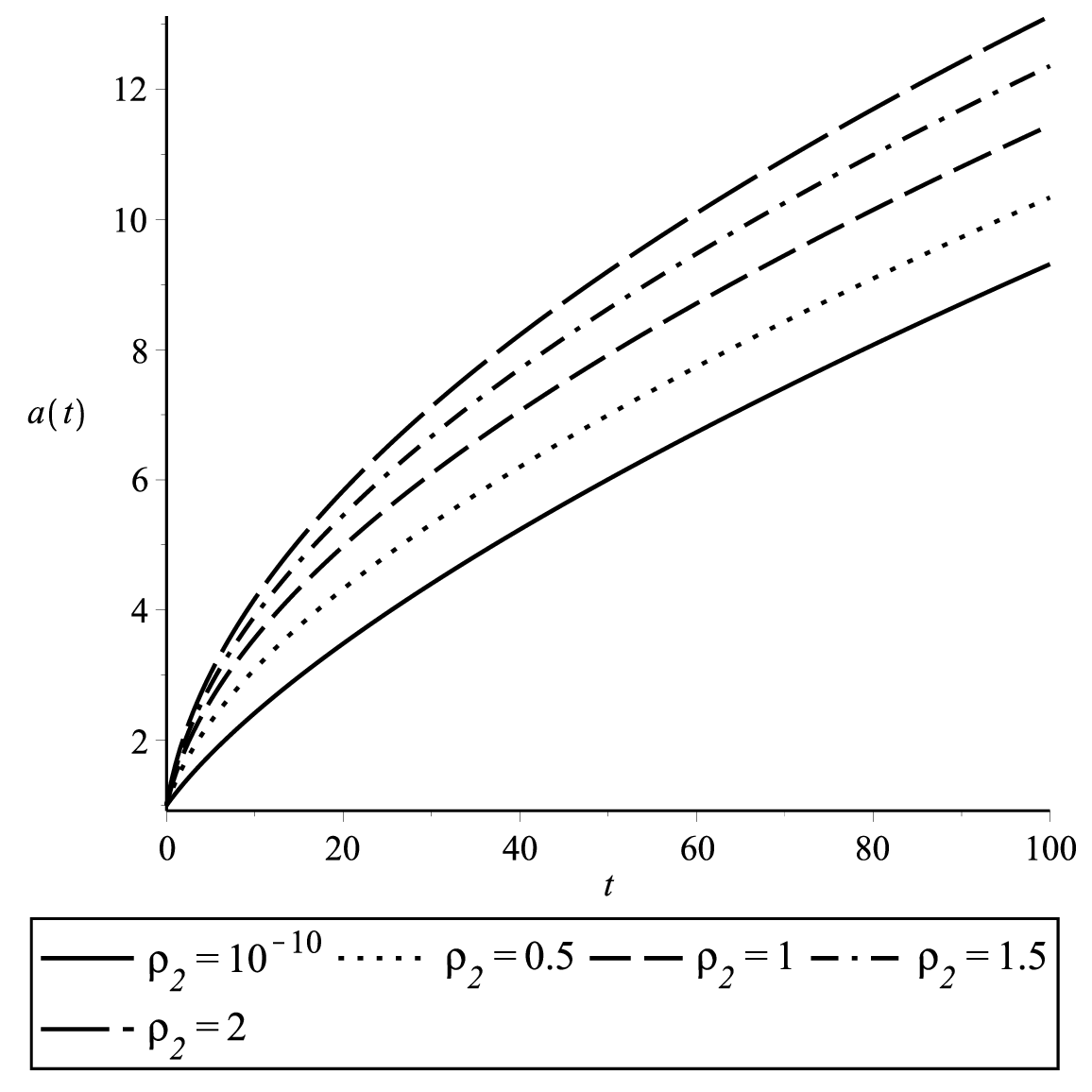}
		\label{fig:a_rho2_short}
	\end{subfigure}
	\hfill
	\begin{subfigure}{0.45\textwidth}
		\centering
		\includegraphics[width=\textwidth]{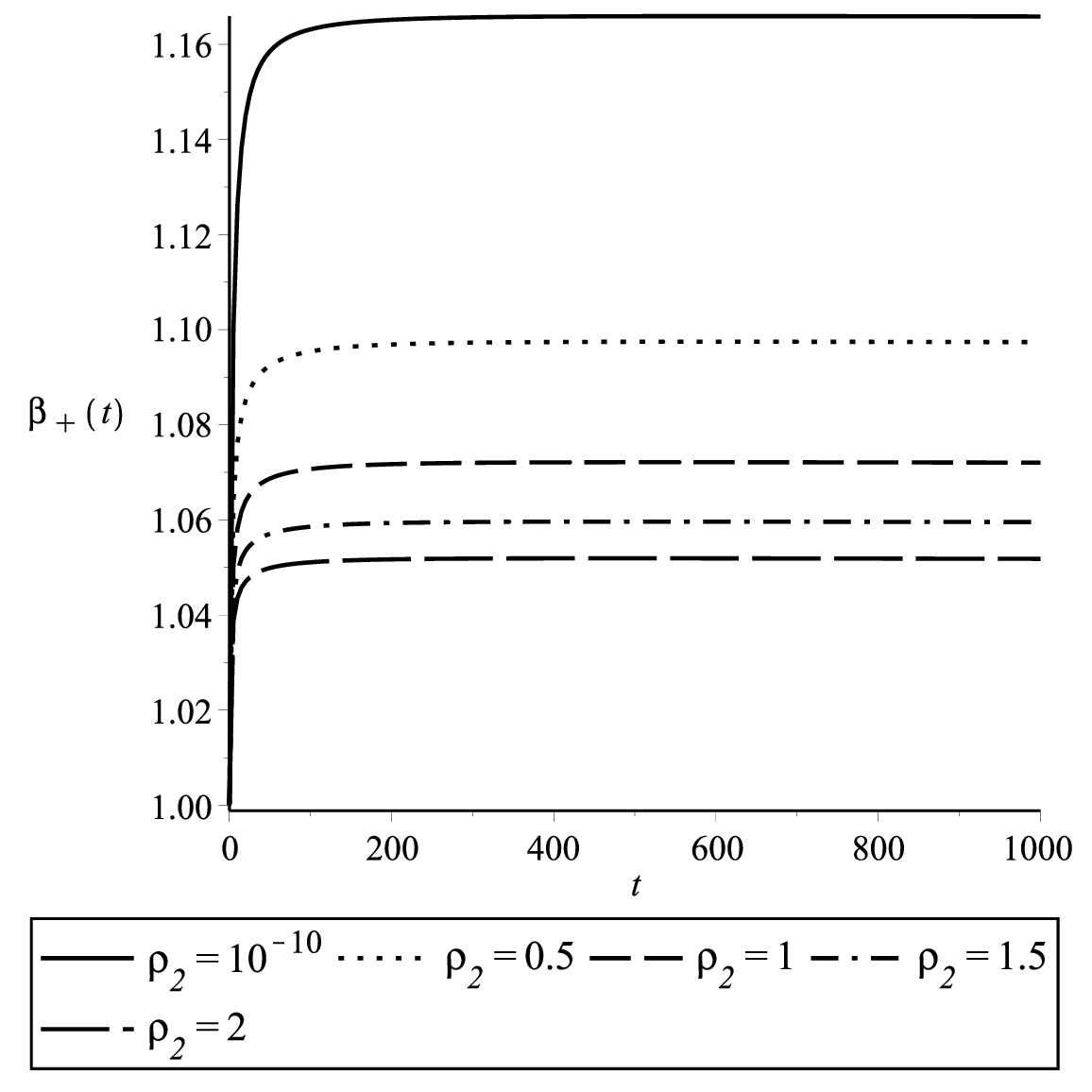}
		\label{fig:beta_plus_rho2_short}
	\end{subfigure}
	\hfill
	\begin{subfigure}{0.45\textwidth}
		\centering
		\includegraphics[width=\textwidth]{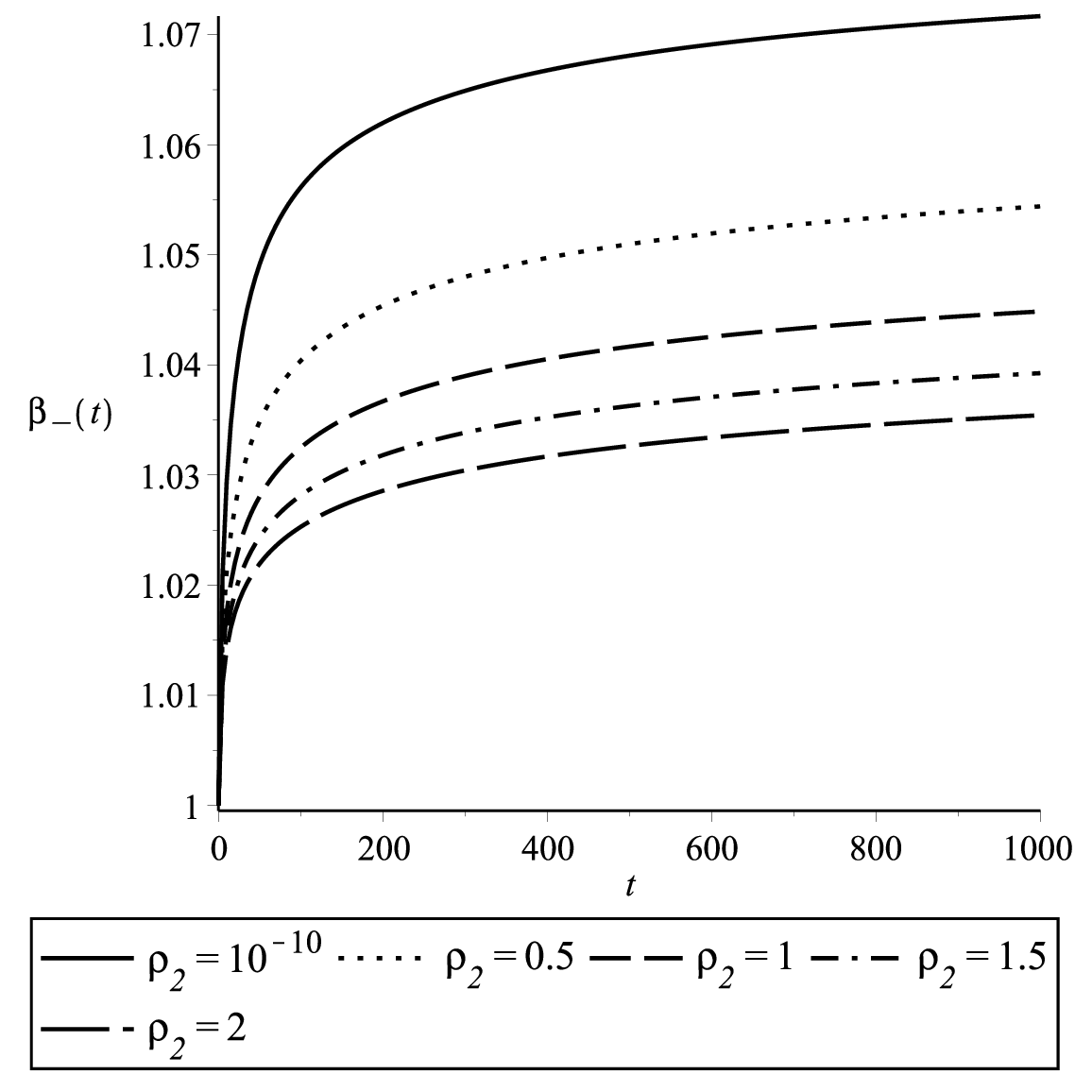}
		\label{fig:beta_minus_rho2_short}
	\end{subfigure}
	
	\caption{\small Behavior of $a$, $\beta_{+}$, and $\beta_{-}$ with the time $t$ for different values of $\rho_{2}$. We take $\rho_1$ = 0.1, $C_1$ = 0.1, $C_2$ = 0.2, $\chi$ = 0.1, $\sigma_{+}$ = 0.01, and $\sigma_{-}$ = -0.05.}
	\label{fig:combined_rho2}
\end{figure}

For the parameters $\rho_{1}$ and $\rho_{2}$, associated with the reduced relativistic gas, we obtained the following results: for $\rho_{1}$, the scale factor $a$ expands more rapidly as $\rho_{1}$ increases, while higher values of $\rho_{1}$ simultaneously lead to lower constant values of both $\beta_{+}$ and $\beta_{-}$ (see Figure \ref{fig:combined_rho1} and Table \ref{tab:rho1_values_full}). In the case of $\rho_{2}$, the scale factor $a$ expands more rapidly for higher values of $\rho_{2}$, while larger values of $\rho_{2}$ lead to lower constant values of both $\beta_{+}$ and $\beta_{-}$ (see Figure \ref{fig:combined_rho2} and Table \ref{tab:rho2_values_full}).

\subsection{Varying the initial values of scale functions}

\begin{figure}[H]
	\centering
	\begin{subfigure}{0.45\textwidth}
		\centering
		\includegraphics[width=\textwidth]{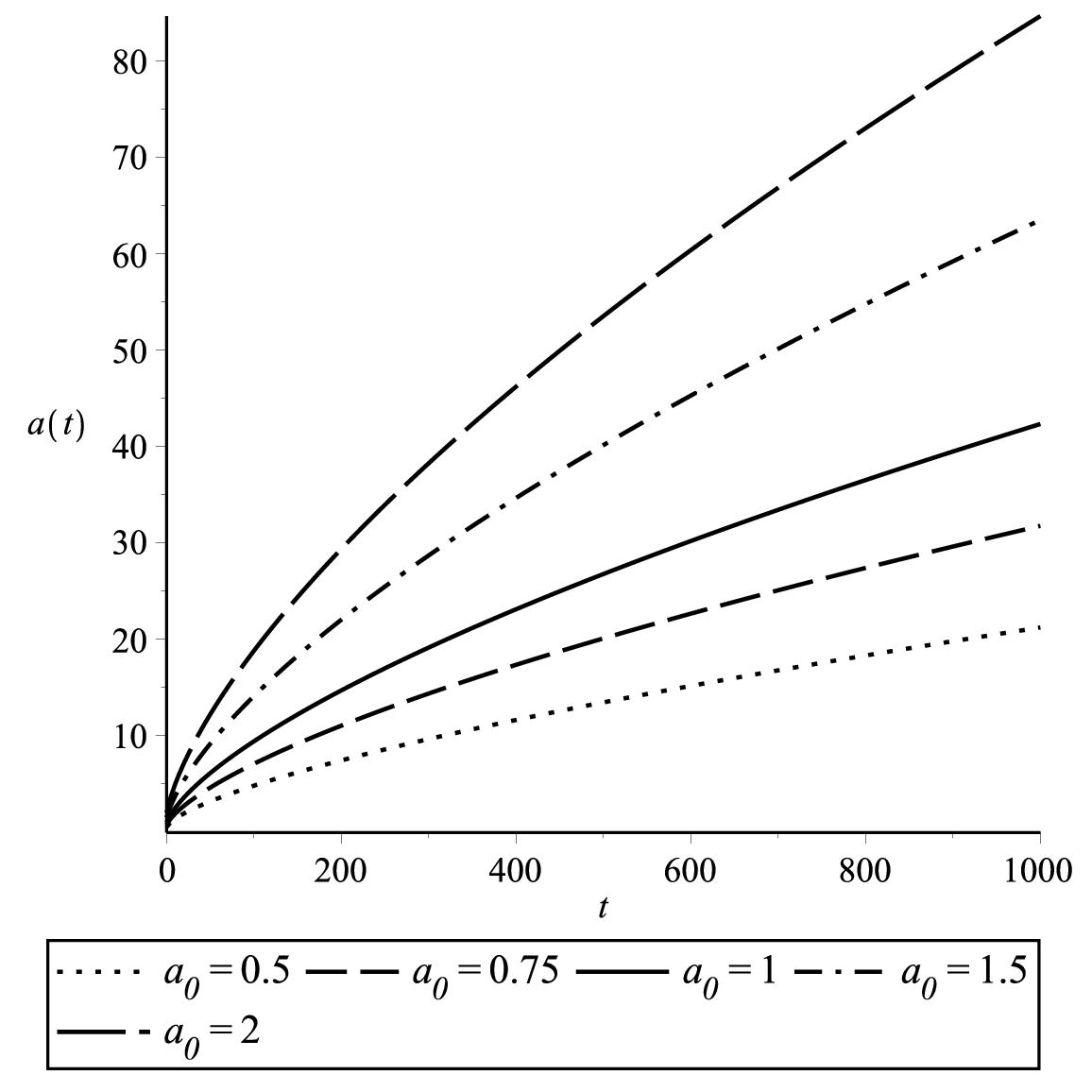}
		\label{fig:a_a0_short}
	\end{subfigure}
	\hfill
	\begin{subfigure}{0.45\textwidth}
		\centering
		\includegraphics[width=\textwidth]{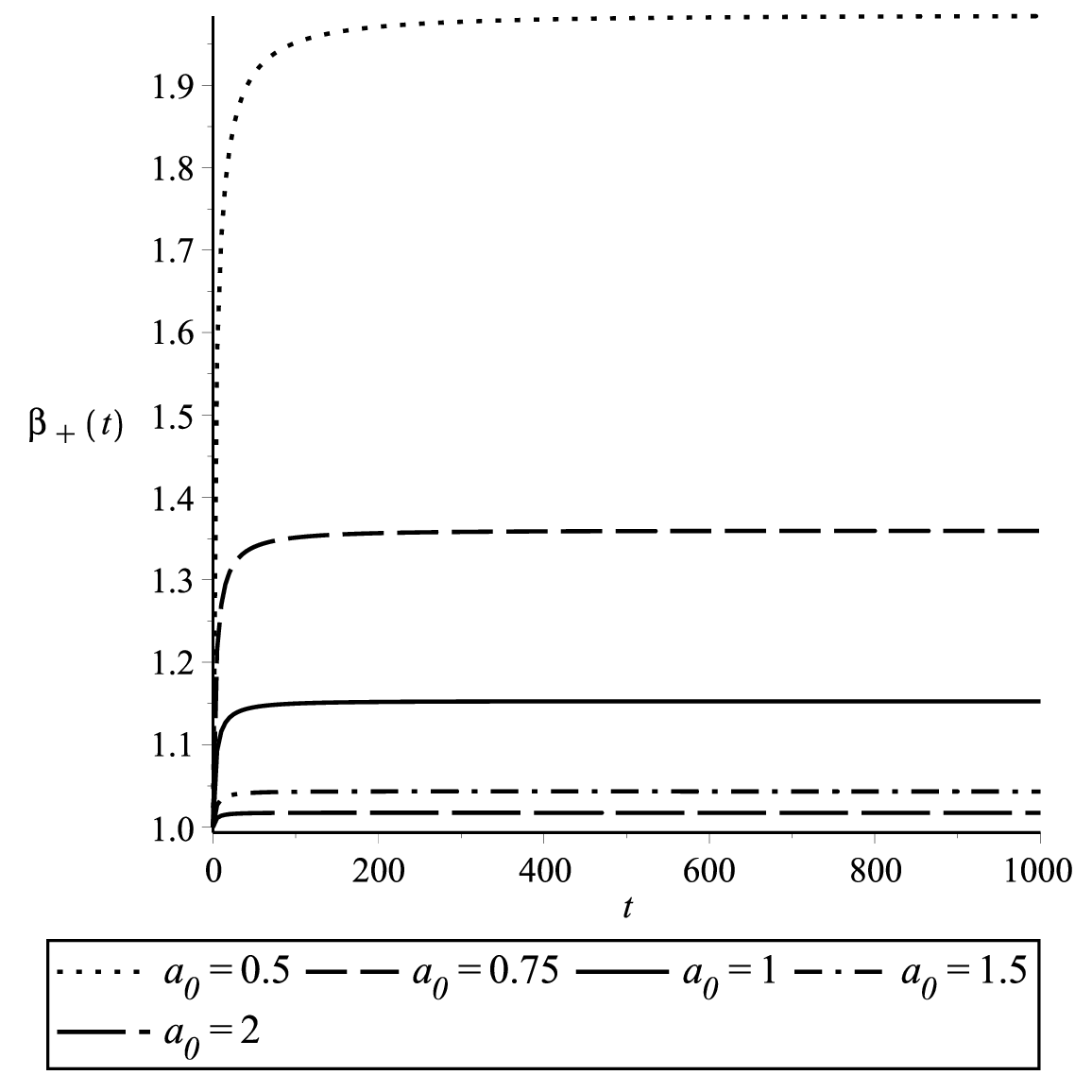}
		\label{fig:beta_plus_a0_short}
	\end{subfigure}
	\hfill
	\begin{subfigure}{0.45\textwidth}
		\centering
		\includegraphics[width=\textwidth]{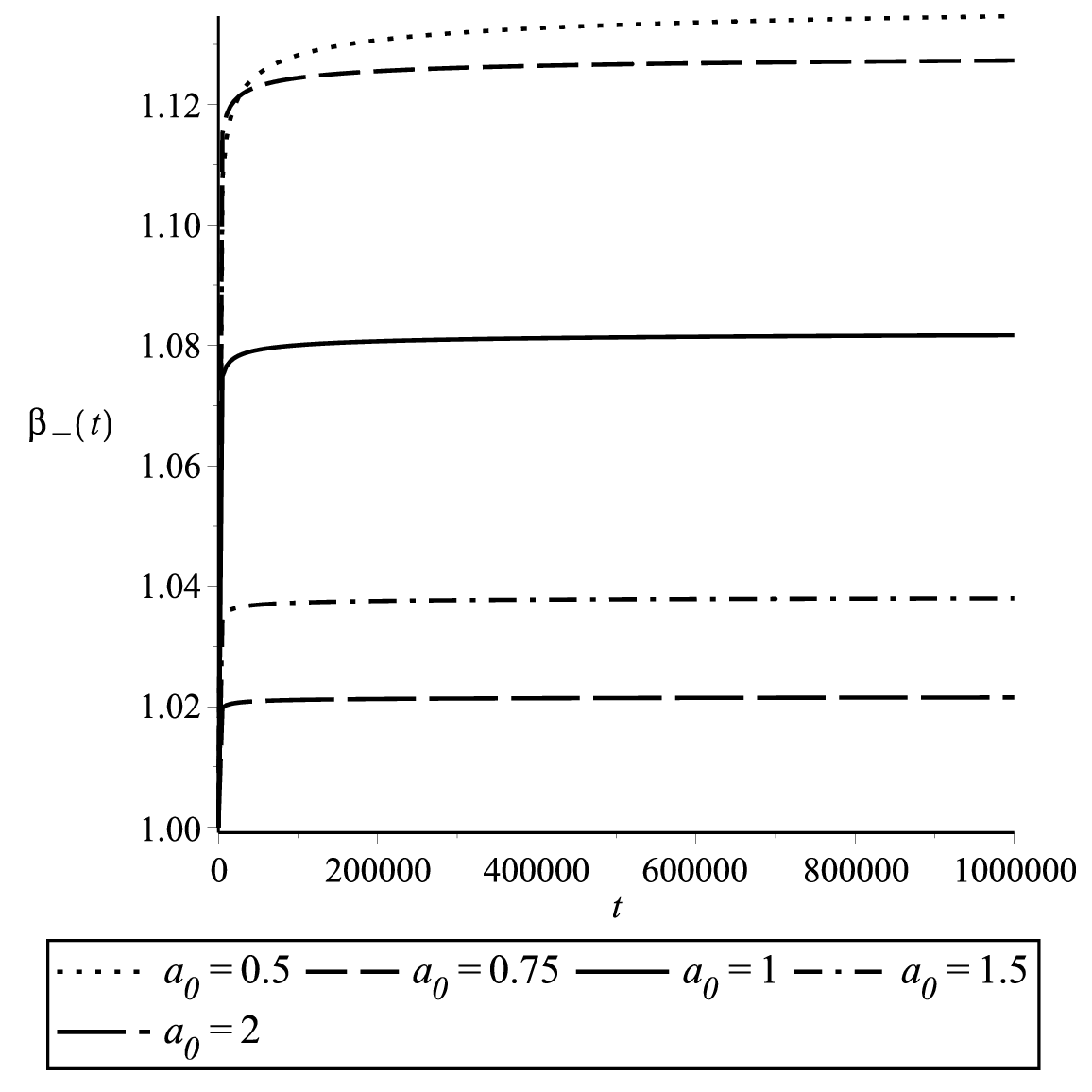}
		\label{fig:beta_minus_a0_short}
	\end{subfigure}
	
	\caption{\small Behavior of $a$, $\beta_{+}$, and $\beta_{-}$ with the time $t$ for different values of $a_{0}$. We take $\rho_1$ = $\rho_2$ = 0.1, $C_1$ = 0.1, $C_2$ = 0.2, $\chi$ = 0.1, $\sigma_{+}$ = 0.01, and $\sigma_{-}$ = -0.05.}
	\label{fig:combined_a0}
\end{figure}

\begin{figure}[H]
	\centering
	\begin{subfigure}{0.45\textwidth}
		\centering
		\includegraphics[width=\textwidth]{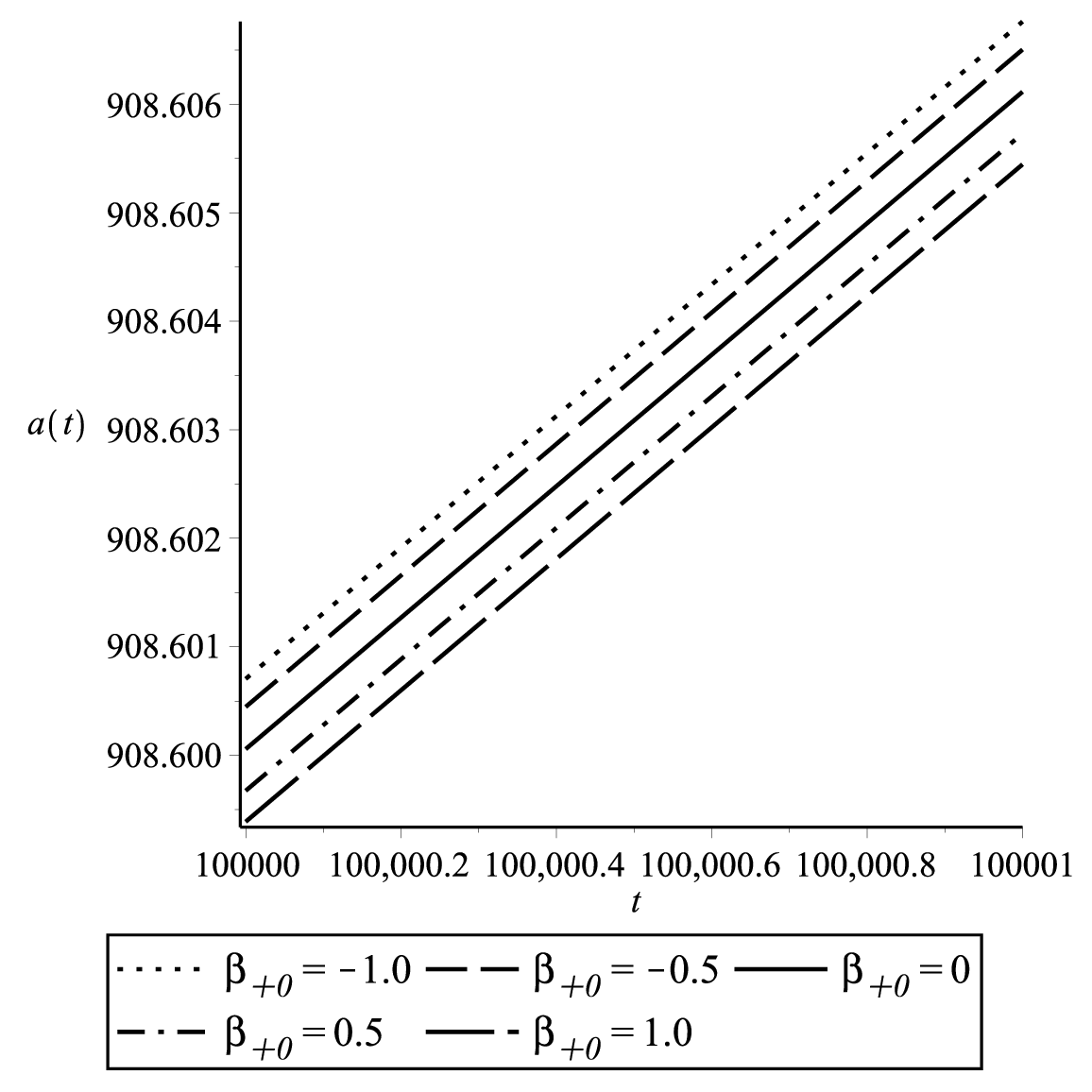}
		\label{fig:a_beta_plus0_short}
	\end{subfigure}
	\hfill
	\begin{subfigure}{0.45\textwidth}
		\centering
		\includegraphics[width=\textwidth]{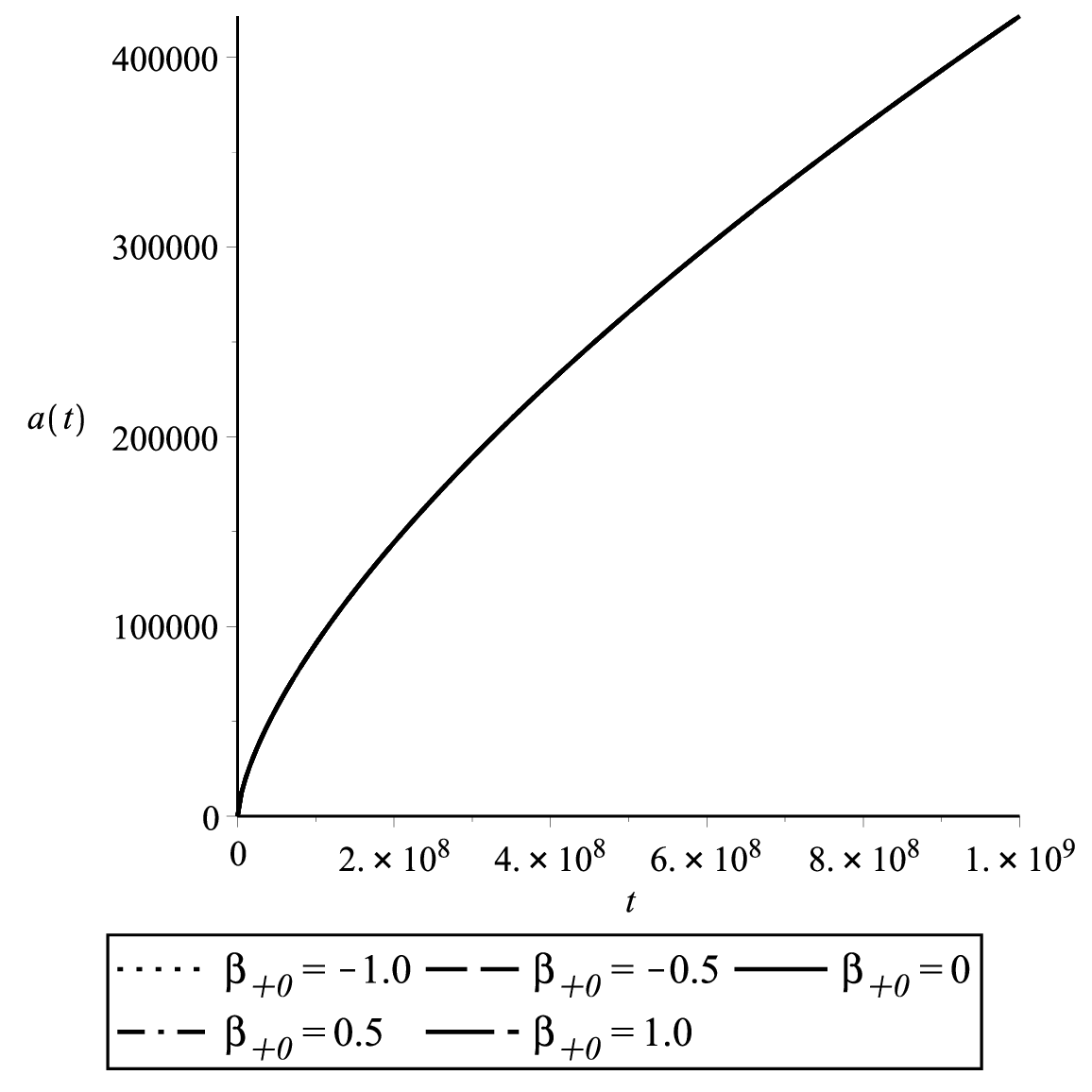}
		\label{fig:a_beta_plus0_long}
	\end{subfigure}
	\hfill
	\begin{subfigure}{0.45\textwidth}
		\centering
		\includegraphics[width=\textwidth]{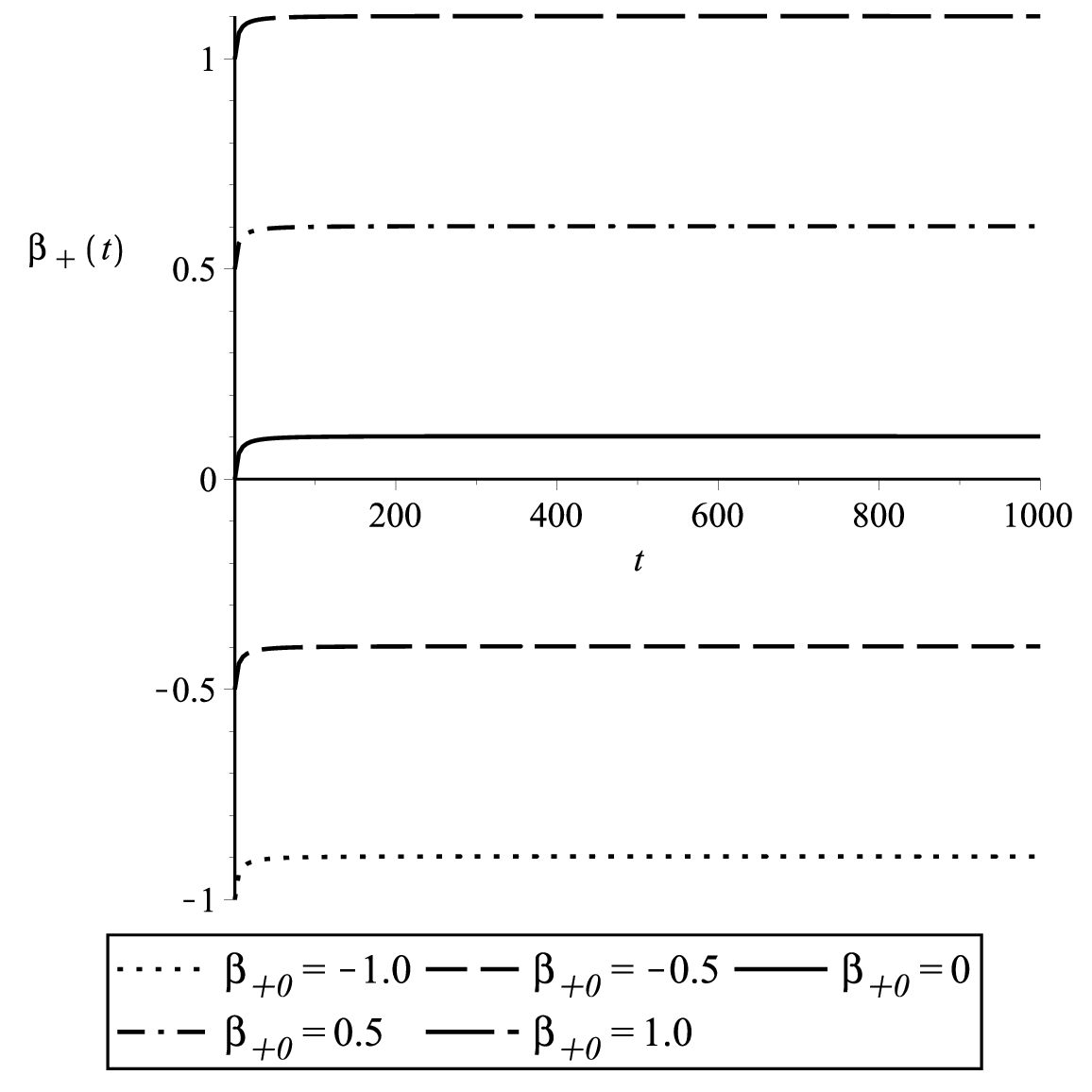}
		\label{fig:beta_plus_beta_plus0_short}
	\end{subfigure}
	\hfill
	\begin{subfigure}{0.45\textwidth}
		\centering
		\includegraphics[width=\textwidth]{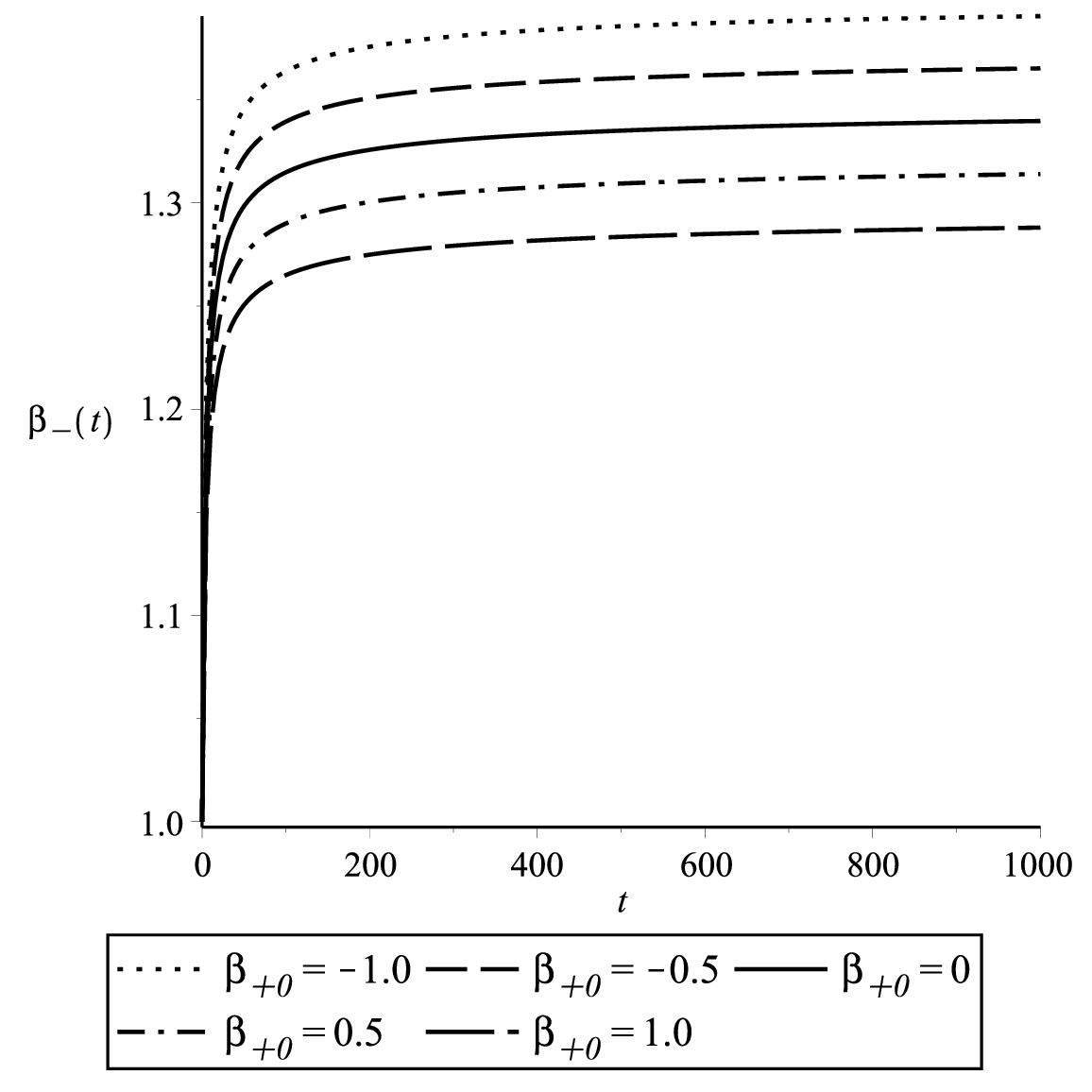}
		\label{fig:beta_minus_beta_plus0_short}
	\end{subfigure}
	
	\caption{\small Behavior of $a$, $\beta_{+}$, and $\beta_{-}$ with the time $t$ for different values of $\beta_{+0}$. We take $\rho_1$ = $\rho_2$ = 0.1, $C_1$ = 0.5, $C_2$ = 0.1, $\chi$ = 0.1, $\sigma_{+}$ = 0.01, and $\sigma_{-}$ = -0.05.}
	\label{fig:combined_beta_plus0}
\end{figure}

\begin{figure}[H]
	\centering
	\begin{subfigure}{0.45\textwidth}
		\centering
		\includegraphics[width=\textwidth]{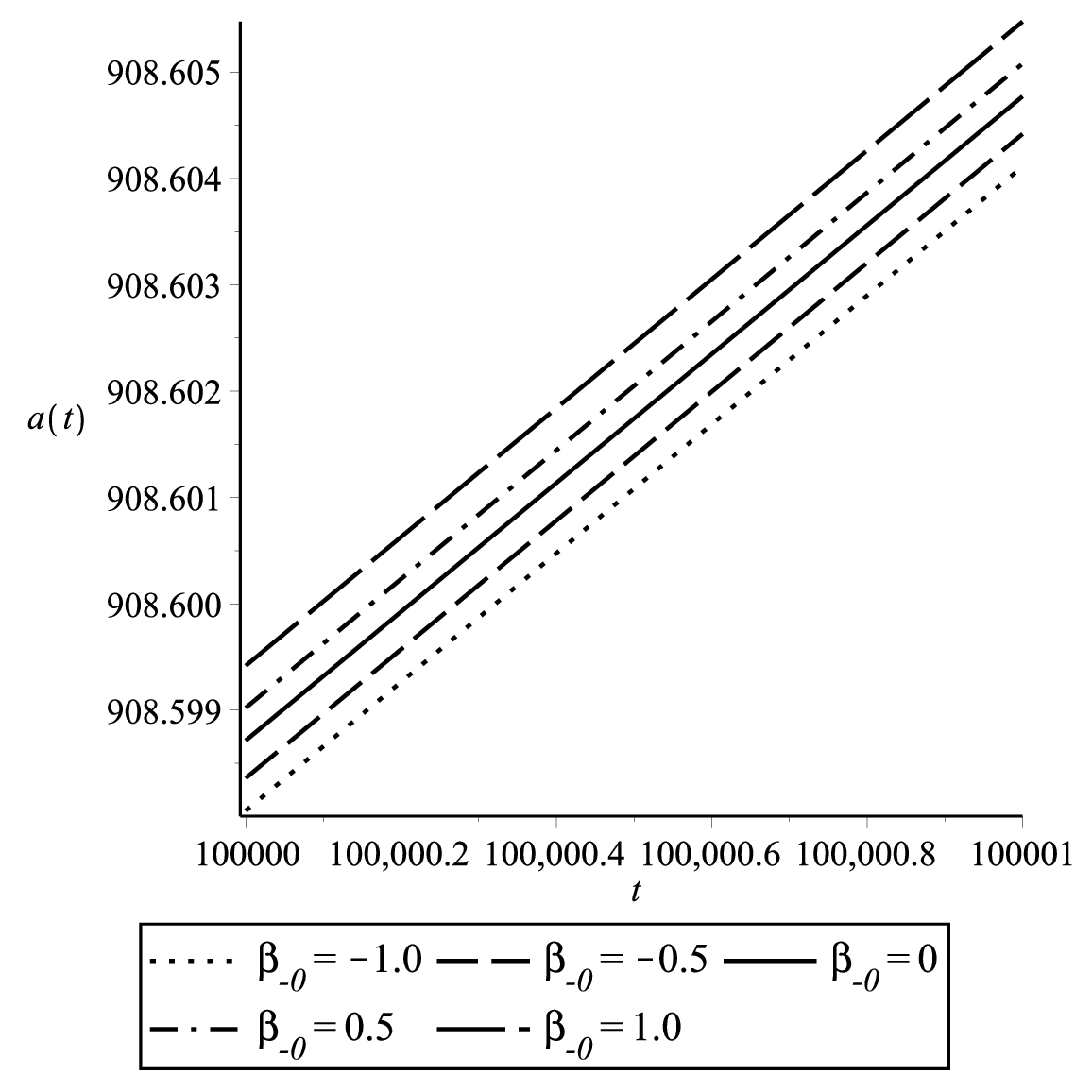}
		\label{fig:a_beta_minus0_short}
	\end{subfigure}
	\hfill
	\begin{subfigure}{0.45\textwidth}
		\centering
		\includegraphics[width=\textwidth]{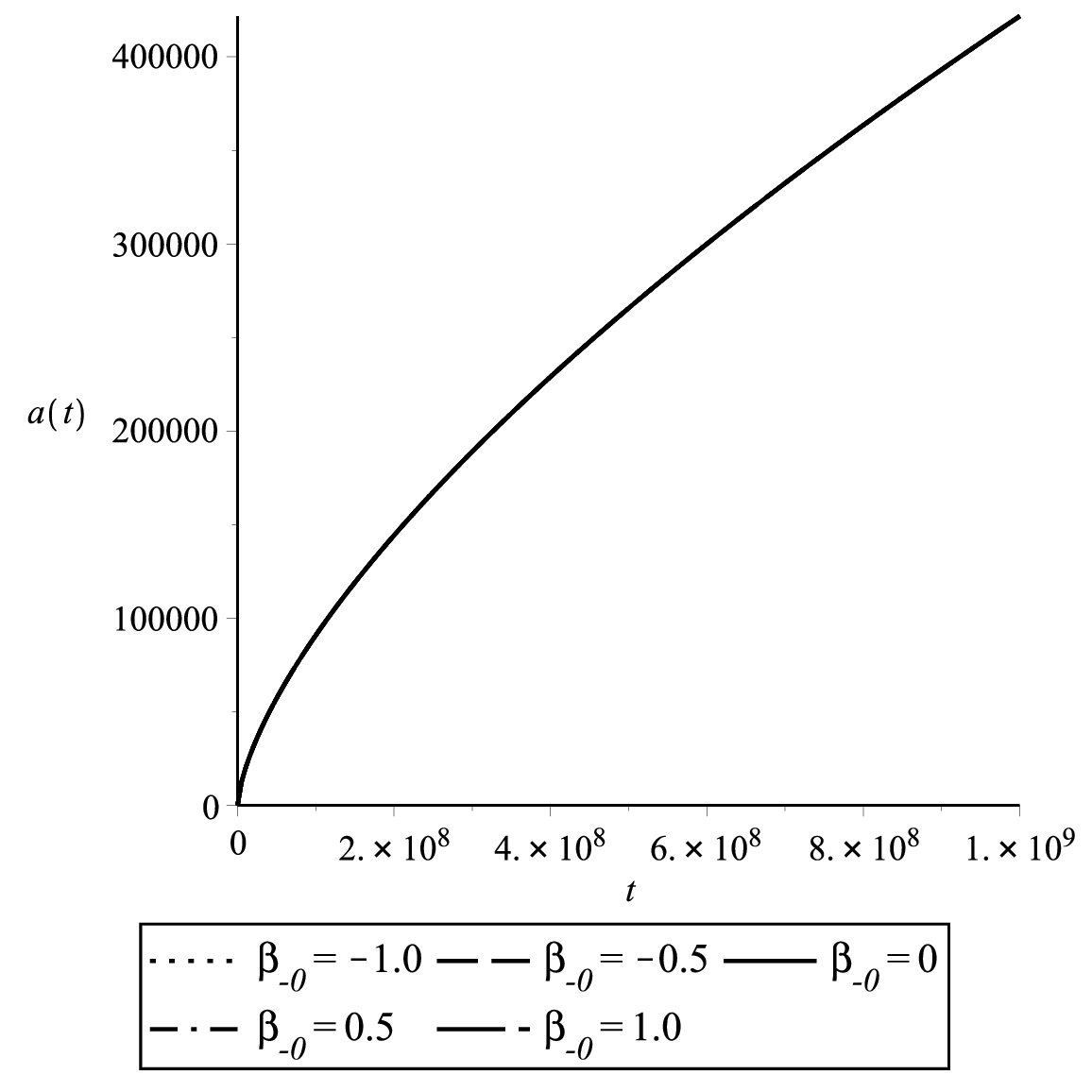}
		\label{fig:a_beta_minus0_long}
	\end{subfigure}
	\hfill
	\begin{subfigure}{0.45\textwidth}
		\centering
		\includegraphics[width=\textwidth]{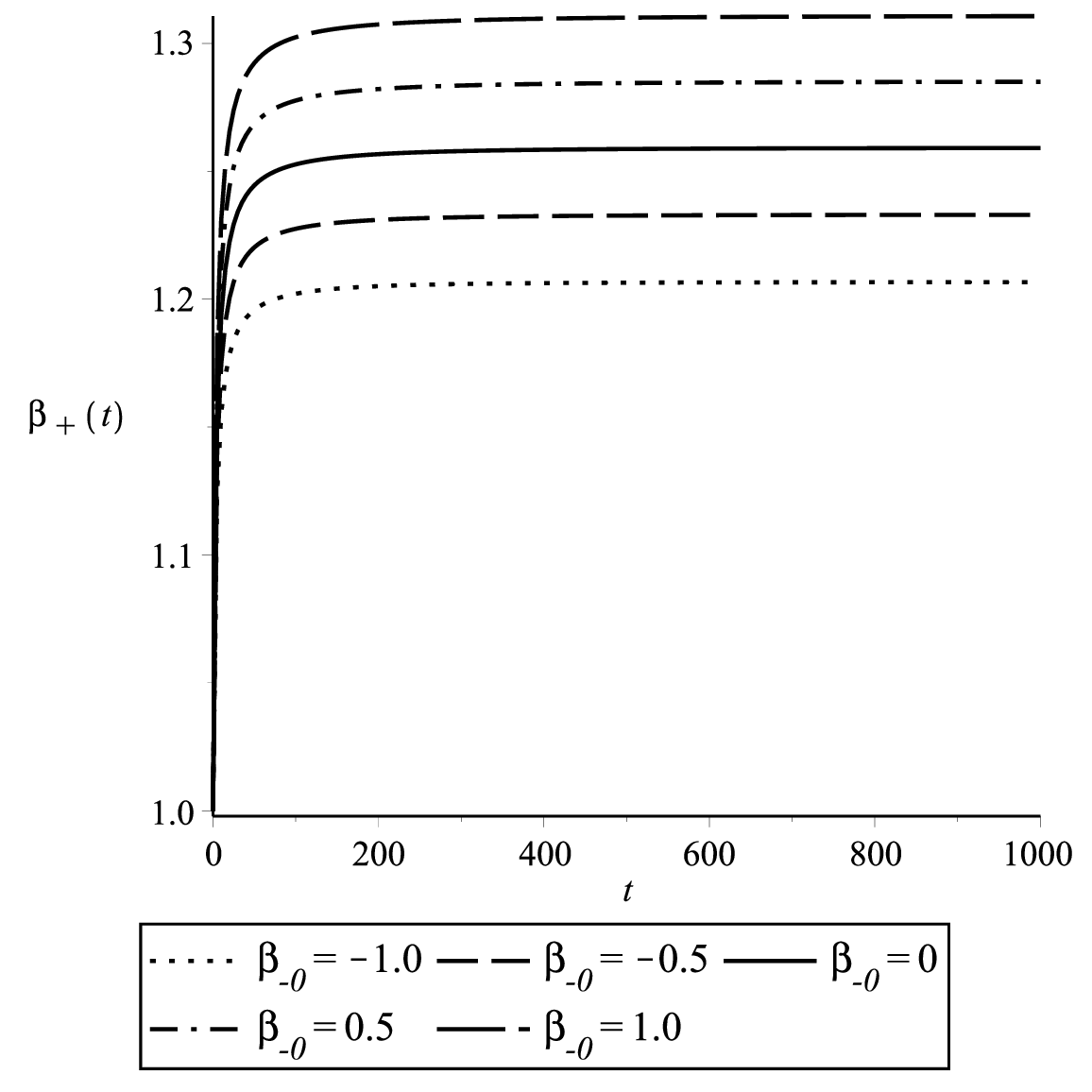}
		\label{fig:beta_plus_beta_minus0_short}
	\end{subfigure}
	\hfill
	\begin{subfigure}{0.45\textwidth}
		\centering
		\includegraphics[width=\textwidth]{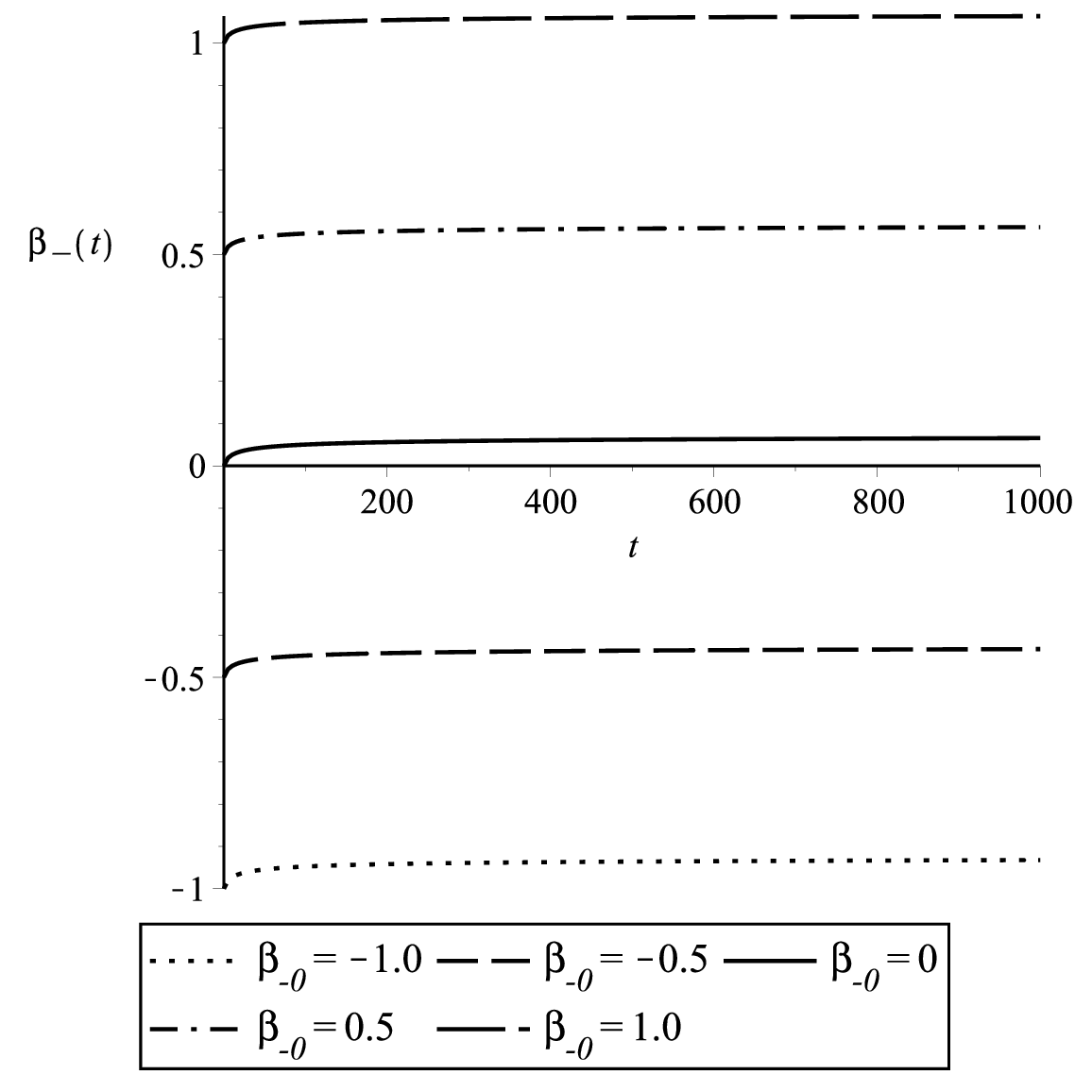}
		\label{fig:beta_minus_beta_minus0_short}
	\end{subfigure}
	
	\caption{\small Behavior of $a$, $\beta_{+}$, and $\beta_{-}$ with the time $t$ for different values of $\beta_{-0}$. We take $\rho_1$ = $\rho_2$ = 0.1, $C_1$ = 0.1, $C_2$ = 0.5, $\chi$ = 0.1, $\sigma_{+}$ = 0.01, and $\sigma_{-}$ = -0.05.}
	\label{fig:combined_beta_minus0}
\end{figure}

For the initial scale functions, we obtained the following results: the scale factor $a$ expands more rapidly for higher values of $a_{0}$. Moreover, larger values of $a_{0}$ also lead to lower constant values of both $\beta_{+}$ and $\beta_{-}$ (see Figure \ref{fig:combined_a0} and Table \ref{tab:a0_values_full}). Furthermore, a faster expansion of $a$ occurs for lower values of $\beta_{+0}$. Additionally, for higher values of $\beta_{+0}$, $\beta_{+}$ goes to higher constant values. On the other hand, for higher values of $\beta_{+0}$,  $\beta_{-}$ goes to smaller constant values (see Figure~\ref{fig:combined_beta_plus0} and Table~\ref{tab:beta_plus_0_values_full}). In the case of $\beta_{0}$, we obtain that: larger values of $\beta_{-0}$ result in a more rapid expansion of $a$, with both $\beta_{+}$ and $\beta_{-}$ going to greater constant values (see Figure \ref{fig:combined_beta_minus0} and Table \ref{tab:beta_minus_0_values_full}).

\section{Estimation of the values of noncommutative parameters}
\label{estimation}
In principle, it is possible to estimate values for the noncommutative parameters using the equation that describes the age of the universe,
\begin{equation}
t_{0} = \int_{0}^{a_{0}} \frac{1}{a H(a)} da \, ,
\label{age_universe}
\end{equation}
where $ t_0 $ denotes the age of the universe, $ a $ represents the scale factor, and $ H(a) = \left(\dot{a}/a\right) $ is the Hubble parameter. For our NC BI model with RRG, the Hubble parameter is obtained from Eq. \eqref{condition_a}, namely,
\begin{eqnarray}
	H^{2} (a) &=& \frac{1}{36 a^6} \left[ C_{1}\left( C_{1} - 2 \chi \beta_{+} \right) + C_{2} \left( C_{2} + 2 \chi \beta_{-} \right) \right] \nonumber \\
&-& \frac{1}{18 a^5} \left ( C_{1} \sigma_{-} + C_{2} \sigma_{+} \right )  
+ \frac{1}{3} \left[ \rho_{1}^2 \left( \frac{a_{0}}{a} \right)^{6} + \rho_{2}^2 \left( \frac{a_{0}}{a} \right)^{8} \right]^{1/2} .
\label{Hubble_parameter}
\end{eqnarray}

Note that in this case the Hubble parameter involves a fluid term due to the reduced relativistic gas, $ {\rho}/{3}  $, where $ \rho $ is the fluid density given by Eq. \eqref{eqRRG}, and contributions related to noncommutativity. We will not include a cosmological constant contribution because, in our model, we aim to describe any dark energy effects solely through the use of noncommutativity.

Thus, using the Hubble parameter provided by Eq. \eqref{Hubble_parameter} in Eq. \eqref{age_universe}, we derive the following equation for the age of the universe in our NC BI model: 
\begin{equation}
t_{0} = \int_{0}^{a_{0}} \frac{1}{a \sqrt{A + B}} \, da \, ,
\end{equation}
where
\begin{equation}
A = \frac{1}{36 a^6} \left[ C_1 (C_1 - 2 \chi \beta_+) + C_2 (C_2 + 2 \chi \beta_-) \right] - \frac{1}{18 a^5} \left ( C_{1} \sigma_{-} + C_{2} \sigma_{+} \right )  \, ,
\end{equation}
represents the contribution from noncommutativity, while
\begin{equation}
B = \frac{1}{3} \left[ \rho_1^2 \left( \frac{a}{a_0} \right)^6 + \rho_2^2 \left( \frac{a}{a_0} \right)^8 \right]^{1/2} \, 
\end{equation}
accounts for the contribution from the reduced relativistic gas fluid.

For our analysis, we will associate $\rho_1$ and $\rho_2$ with the densities of matter $\rho_m$ and radiation $\rho_r$, respectively, that is $\rho_1 = \rho_m$ and $\rho_2 = \rho_r$. Using the fact that the critical density in natural units is $\rho_c = 3 H_{0}^{2}$, we can express $\Omega_m = \rho_m / \rho_c$ and $\Omega_r = \rho_r / \rho_c$. Hence, we can rewrite the term $B$ as:
\begin{equation}
B = H_0^2 \left[ \Omega_m^2 \left( \frac{a}{a_0} \right)^6 + \Omega_r^2 \left( \frac{a}{a_0} \right)^8 \right]^{1/2} .
\end{equation}

To perform this estimation, we will use data from the Planck 2018 mission \cite{Planck2018results} for the values of cosmological parameters, which provide $H_0$ = $(67.4 \pm 0.5) \, \text{km/s/Mpc}$, $\Omega_m$ = $(0.315 \pm 0.007)$ and $\Omega_r  \approx 10^{-4}$. The approach we chose to estimate the noncommutative parameters is to consider them one by one separately, so that we calculate $\chi$, $\sigma_{+}$, and $\sigma_{-}$ individually, in turn, as non-zero constants using Eq. \eqref{age_universe} and assuming the other two to be zero. Additionally, we take $t_0$ = $13.8 \, \text{Gyr}$ = $(13.8 \times 10^{9}) \times (3.154 \times 10^{7}) \, \text{s} $ = $4.35132 \times 10^{17} \, \text{s}$ and consider $a_0 = 1$, $\beta_{0+} = 10^{-5}$, and $\beta_{0-} = - 10^{-5}$. This choice for the anisotropy parameters comes from assuming that their present values are of the order of CMB (cosmic microwave background) temperature fluctuations. Using the data from Planck Mission, we can write $H_0$ = $67.4 \times 10^3 / (3.0857 \times 10^{22})  \,\text{s}^{-1}$ = $2.18 \times 10^{-18} \,\text{s}^{-1}$, and therefore $\rho_1$ = $3 H_0^2 \Omega_m$ = $4.49 \times 10^{-36} \,\text{s}^{-2}$ while $\rho_2$ = $3 H_0^2 \Omega_r$ = $1.43 \times 10^{-39} \,\text{s}^{-2}$. Since $C_1$ and $C_2$ have the same dimension as $H_0$, we will assume them to be of the same order of magnitude as the latter and choose $C_1 = C_2 = 10^{-18} \, \text{s}^{-1}$. Using this approach and numerically solving Eq. \eqref{age_universe} with these values, we obtain the following estimates:
\begin{equation}
\begin{aligned}
\chi &= -4.073915452 \times 10^{-13} \,\text{s}^{-1} ,\\
\sigma_{+} &= \sigma_{-} = -1.214786244 \times 10^{-17} \,\text{s}^{-1}.
\end{aligned}
\end{equation} 
These results confirm the consistency of our approach, as they align with previous studies on noncommutative cosmological models\cite{oliveira2017b,abreu2019,oliveira2021,oliveira2024}.

\section{Conclusions}
\label{conclusions}

This work investigated the effects of noncommutativity in the cosmological Bianchi I (BI) model with a reduced relativistic gas (RRG). To introduce noncommutative parameters into the equations of motion of the model, we made use of a generalized symplectic formalism of Faddeev-Jackiw. Through this approach, we analyzed the time evolution of the isotropic scale factor $a$ and the anisotropic scale functions $\beta_{+}$ and $\beta_{-}$ arising from Misner's parameterization.

Our results reveal that the presence of noncommutativity alters the dynamics of the universe's expansion and isotropization of the commutative model, modifying both the expansion rate of the scale factor $a$ and the time required for isotropization, given by the stabilization of the anisotropic functions $\beta_{+}$ and $\beta_{-}$. Numerical analysis indicated that, regardless of the specific values chosen for the noncommutative parameters, the scale factor $a$ always remains increasing, corroborating the expectation of an expanding universe, while the anisotropy functions tend to constant values after an initial period of growth or decay, evidencing the tendency of isotropization even in the presence of noncommutativity. In particular, the analysis of the noncommutative parameters $\chi$, $\sigma_{+}$, and $\sigma_{-}$ revealed that smaller values result in a faster expansion of $a$, as well as modify the isotropization time of the functions $\beta_{+}$ and $\beta_{-}$. These results suggest that noncommutative parameters could, in principle, be used to describe the current accelerated expansion of the universe without invoking dark energy, as they can be chosen to increase the expansion rate of the isotropic scale factor $a$, mimicking the effect of dark energy.

Additionally, we obtained estimates for the values of the noncommutative parameters using the age of the universe and the Planck 2018 cosmological data. To achieve this, we assumed a simplified approach in which each noncommutative parameter was analyzed individually while setting the others to zero. By numerically solving the equation for the age of the universe, we obtained the estimates $\chi = -4.073915452 \times 10^{-13} \,\text{s}^{-1}$ and $\sigma_{+} = \sigma_{-} = -1.214786244 \times 10^{-17} \,\text{s}^{-1}$. These values align with results from previous works on noncommutative cosmological models, reinforcing the consistency of our approach.

In summary, the study demonstrated that the incorporation of noncommutativity into the BI model with RRG not only preserves the fundamental features of cosmic expansion and isotropization but also reveals additional aspects about the influence of noncommutative parameters in these processes. Future research may explore extensions of this research by investigating other anisotropic models and testing different configurations for the noncommutative parameters based on recent cosmological observations.

\section*{Acknowledgments}

T. M. Abreu thanks Programa de Pós-Graduação em Física (PPG-Física), Universidade Federal de Juiz de Fora (UFJF) and CAPES (Coordenação de Aperfeiçoamento de Pessoal de Nível Superior) for financial support. G. Oliveira-Neto thanks FAPEMIG (APQ-06640-24) for partial financial support.

\clearpage

\begin{appendices}

	\section{List of Tables}
	\label{appendix}

	\begin{table}[H]
		\centering
		
		\caption{\footnotesize Values of $a(t)$, $\beta_{+}(t)$, and $\beta_{-}(t)$ at different times $t$ for various values of $\chi$. We consider $\rho_1$ = $\rho_2$ = 0.1, $C_1$ = 0.2, $C_2$ = 0.1, $\sigma_{+}$ = 0.01 and $\sigma_{-}$ = -0.05.}
		
		\renewcommand{\arraystretch}{1.0} 
		\setlength{\tabcolsep}{12pt} 
		\tiny 
		\begin{tabular}{c c c c c}
			\hline
			$\chi$ & $t$ & $a(t)$ & $\beta_{+}(t)$ & $\beta_{-}(t)$ \\
			\hline
			$-0.5$ & $10^{2}$ & $9.41042802963272$ & $7.24368174303990  \times 10^{-1}$ & $1.39542863027622$ \\
			$-0.5$ & $10^{5}$ & $9.08598193127700 \times 10^{2}$ & $7.06314117648214 \times 10^{-1}$& $1.43388319085617$ \\
			$-0.5$ & $10^{10}$ & $1.95743387717530 \times 10^{6}$ & $7.05707755360914 \times 10^{-1}$ & $1.43685870558427$ \\
			$-0.5$ & $10^{50}$ & $9.08560314387074 \times 10^{32}$ & $7.05694705648835 \times 10^{-1}$ & $1.43692395358037$ \\
			$-0.5$ & $10^{80}$ & $9.08560959521214 \times 10^{52}$ & $7.05694705648835 \times 10^{-1}$ & $1.43692395358037$ \\
			\hline
			
			$-0.25$ & $10^{2}$ & $9.40947006136555$ & $8.90242434959212 \times 10^{-1}$ & $1.29073247708939$ \\
			$-0.25$ & $10^{5}$ & $9.08598009734253 \times 10^{2}$ & $8.80152939639179 \times 10^{-1}$ & $1.32631510601686$ \\
			$-0.25$ & $10^{10}$ & $1.95743403608163 \times 10^{6}$ & $8.79555144141166 \times 10^{-1}$ & $1.32928766582382$ \\
			$-0.25$ & $10^{50}$ & $9.08560421886239 \times 10^{32}$ & $8.79542094522764 \times 10^{-1}$ & $1.32935291375100$ \\
			$-0.25$ & $10^{80}$ & $9.08560349505895 \times 10^{52}$ & $8.79542094522764 \times 10^{-1}$ & $1.32935291375100$ \\
			\hline
			
			$0$ & $10^{2}$ & $9.40698824622729$ & $1.04197167569020$ & $1.16480508211536$ \\
			$0$ & $10^{5}$ & $9.08597838042156 \times 10^{2}$ & $1.03873259751173$ & $1.19573169737328$ \\
			$0$ & $10^{10}$ & $1.95743387659648 \times 10^{6}$ & $1.03814217442786$ & $1.19869936721878$ \\
			$0$ & $10^{50}$ & $9.08560313979026 \times 10^{32}$ & $1.03812912487542$ & $1.19876461513652$ \\
			$0$ & $10^{80}$ & $9.08558975316375 \times 10^{52}$ & $1.03812912487542$ & $1.19876461513652$ \\
			\hline
			
			$0.25$ & $10^{2}$ & $9.40297472671239$ & $1.17712712586782$ & $1.01889638508121$ \\
			$0.25$ & $10^{5}$ & $9.08597320339495 \times 10^{2}$ & $1.17928599839870$ & $1.04362323794785$ \\
			$0.25$ & $10^{10}$ & $1.95743394637394 \times 10^{6}$ & $1.17870137651463$ & $1.04658436014418$ \\
			$0.25$ & $10^{50}$ & $9.08560400360462 \times 10^{32}$ & $1.17868832702436$ & $1.04664960797635$ \\
			$0.25$ & $10^{80}$ & $9.08560207949522 \times 10^{52}$ & $1.17868832702436$ & $1.04664960797635$ \\
			\hline
			
			$0.5$ & $10^{2}$ & $9.39743066565372$ & $1.29320919548767$ & $8.54404009447011 \times 10^{-1}$ \\
			$0.5$ & $10^{5}$ & $9.08596785075582 \times 10^{2}$ & $1.29900299362236$ & $8.71680943593564 \times 10^{-1}$ \\
			$0.5$ & $10^{10}$ & $1.95743386471424 \times 10^{6}$ & $1.29842226310440$ & $8.74634185059349 \times 10^{-1}$ \\
			$0.5$ & $10^{50}$ & $9.08560319191661 \times 10^{32}$ & $1.29840921364877$ & $8.74699432834718 \times 10^{-1}$ \\
			$0.5$ & $10^{80}$ & $9.08559541099842 \times 10^{52}$ & $1.29840921364877$ & $8.74699432834719 \times 10^{-1}$ \\
			\hline
		\end{tabular}

		\label{tab:chi_values_full}
	\end{table}

	\begin{table}[H]
		\centering    	\caption{\footnotesize Values of $a(t)$, $\beta_{+}(t)$, and $\beta_{-}(t)$ at different times $t$ for various values of $\sigma_{+}$. We consider $\rho_1$ = $\rho_2$ = 0.1, $C_1$ = 0.1, $C_2$ = 0.1, $\chi$ = 0.1 and $\sigma_{-}$ = -0.05.}
		
		\label{tab:sigma_plus_values_full}
		\renewcommand{\arraystretch}{1.0} 
		\setlength{\tabcolsep}{12pt} 
		\tiny 
		\begin{tabular}{c c c c c}
			\hline
			$\sigma_{+}$ & $t$ & $a(t)$ & $\beta_{+}(t)$ & $\beta_{-}(t)$ \\
			\hline
			$-0.5$ & $10^{2}$ & $9.43665123789371$ & $1.72466219142303$ & $1.04257852269666$ \\
			$-0.5$ & $10^{5}$ & $9.08602316710674 \times 10^{2}$ & $1.99770351691716$ & $1.06756535655440$ \\
			$-0.5$ & $10^{10}$ & $1.95743389194868 \times 10^{6}$ & $2.02734253387754$ & $1.07052634267056$ \\
			$-0.5$ & $10^{50}$ & $9.08560371972301 \times 10^{32}$ & $2.02799501260640$ & $1.07059159051379$ \\
			$-0.5$ & $10^{80}$ & $9.08561138918551 \times 10^{52}$ & $2.02799501260640$ & $1.07059159051379$ \\
			\hline
			
			$-0.25$ & $10^{2}$ & $9.42201887499948$ & $1.44534891988293$ & $1.04788176941480$ \\
			$-0.25$ & $10^{5}$ & $9.08600018680703 \times 10^{2}$ & $1.58521832144715$ & $1.07355395199512$ \\
			$-0.25$ & $10^{10}$ & $1.95743398577917 \times 10^{6}$ & $1.60004127211432$ & $1.07651586705457$ \\
			$-0.25$ & $10^{50}$ & $9.08560413332137 \times 10^{32}$ & $1.60036751140506$ & $1.07658111488570$ \\
			$-0.25$ & $10^{80}$ & $9.08560666594664 \times 10^{52}$ & $1.60036751140505$ & $1.07658111488570$ \\
			\hline
			
			$0$ & $10^{2}$ & $9.40703749057558$ & $1.16140235514515$ & $1.05342558512222$ \\
			$0$ & $10^{5}$ & $9.08597715513859 \times 10^{2}$ & $1.16784345541523$ & $1.07979695334432$ \\
			$0$ & $10^{10}$ & $1.95743408229731 \times 10^{6}$ & $1.16785030061627$ & $1.08275980755320$ \\
			$0$ & $10^{50}$ & $9.08560414594794 \times 10^{32}$ & $1.16785030068478$ & $1.08282505537457$ \\
			$0$ & $10^{80}$ & $9.08561452843611 \times 10^{52}$ & $1.16785030068478$ & $1.08282505537457$ \\
			\hline
			
			$0.25$ & $10^{2}$ & $9.39167890513298$ & $8.72481062037855 \times 10^{-1}$ & $1.05923377764994$ \\
			$0.25$ & $10^{5}$ & $9.08595328752850 \times 10^{2}$ & $7.45227965360418 \times 10^{-1}$ & $1.08631917313842$ \\
			$0.25$ & $10^{10}$ & $1.95743400083624 \times 10^{6}$ & $7.30418666443052 \times 10^{-1}$ & $1.08928297939653$ \\
			$0.25$ & $10^{50}$ & $9.08560416941104 \times 10^{32}$ & $7.30092427305644 \times 10^{-1}$ & $1.08934822724373$ \\
			$0.25$ & $10^{80}$ & $9.08562555932953 \times 10^{52}$ & $7.30092427305644 \times 10^{-1}$ & $1.08934822724373$ \\
			\hline
			
			$0.5$ & $10^{2}$ & $9.37591329284476$ & $5.78195871730801 \times 10^{-1}$ & $1.06533405336862$ \\
			$0.5$ & $10^{5}$ & $9.08592950888611 \times 10^{2}$ & $3.16972244354758 \times 10^{-1}$ & $1.09314940026232$ \\
			$0.5$ & $10^{10}$ & $1.95743408800537 \times 10^{6}$ & $2.87346768892812 \times 10^{-1}$ & $1.09611416986244$ \\
			$0.5$ & $10^{50}$ & $9.08560409695774 \times 10^{32}$ & $2.86694290719273 \times 10^{-1}$ & $1.09617941770251$ \\
			$0.5$ & $10^{80}$ & $9.08561039564722 \times 10^{52}$ & $2.86694290719273 \times 10^{-1}$ & $1.09617941770251$ \\
			\hline
		\end{tabular}
		
	\end{table}
	
	\begin{table}[H]
		\centering
		\caption{\footnotesize Values of $a(t)$, $\beta_{+}(t)$, and $\beta_{-}(t)$ at different times $t$ for various values of $\sigma_{-}$. We consider $\rho_1$ = $\rho_2$ = 0.1, $C_1$ = 0.1, $C_2$ = 0.2, $\chi$ = 0.1 and $\sigma_{+}$ = 0.01.}
		
		\label{tab:sigma_minus_values_full}
		\renewcommand{\arraystretch}{1.0} 
		\setlength{\tabcolsep}{12pt} 
		\tiny 
		\begin{tabular}{c c c c c}
			\hline
			$\sigma_{-}$ & $t$ & $a(t)$ & $\beta_{+}(t)$ & $\beta_{-}(t)$ \\
			\hline
			$-0.5$ & $10^{2}$ & $9.42056586845038$ & $1.15735081165928$ & $1.56870069668505$ \\
			$-0.5$ & $10^{5}$ & $9.08599867939072 \times 10^{2}$ & $1.15965991516414$ & $1.83530594978088$ \\
			$-0.5$ & $10^{10}$ & $1.95743399670609 \times 10^{6}$ & $1.15907580983154$ & $1.86493781538651$ \\
			$-0.5$ & $10^{50}$ & $9.08560326878763 \times 10^{32}$ & $1.15906276034789$ & $1.86559029385974$ \\
			$-0.5$ & $10^{80}$ & $9.08560707107241 \times 10^{52}$ & $1.15906276034789$ & $1.86559029385975$ \\
			\hline
			
			$-0.25$ & $10^{2}$ & $9.41275445316399$ & $1.15327261145669$ & $1.28358617057231$ \\
			$-0.25$ & $10^{5}$ & $9.08598598955447 \times 10^{2}$ & $1.15491204291855$ & $1.41679858533293$ \\
			$-0.25$ & $10^{10}$ & $1.95743408985037 \times 10^{6}$ & $1.15432699857548$ & $1.43161436890215$ \\
			$-0.25$ & $10^{50}$ & $9.08560407312769 \times 10^{32}$ & $1.15431394908694$ & $1.43194060801887$ \\
			$-0.25$ & $10^{80}$ & $9.08560117601180 \times 10^{52}$ & $1.15431394908694$ & $1.43194060801887$ \\
			\hline
			
			$0$ & $10^{2}$ & $9.40483907444557$ & $1.14910159719786$ & $9.95906883815341 \times 10^{-1}$ \\
			$0$ & $10^{5}$ & $9.08597391778347 \times 10^{2}$ & $1.15006412905849$ & $9.95589467391686 \times 10^{-1}$ \\
			$0$ & $10^{10}$ & $1.95743409147710 \times 10^{6}$ & $1.14947813997389$ & $9.95589134268101 \times 10^{-1}$ \\
			$0$ & $10^{50}$ & $9.08560403713589 \times 10^{32}$ & $1.14946509047568$ & $9.95589134264780 \times 10^{-1}$ \\
			$0$ & $10^{80}$ & $9.08561202735366 \times 10^{52}$ & $1.14946509047568$ & $9.95589134264780 \times 10^{-1}$ \\
			\hline
			
			$0.25$ & $10^{2}$ & $9.39681450284962$ & $1.14483306989309$ & $7.05561936402713 \times 10^{-1}$ \\
			$0.25$ & $10^{5}$ & $9.08596221357758 \times 10^{2}$ & $1.14511122719217$ & $5.71574935426187 \times 10^{-1}$ \\
			$0.25$ & $10^{10}$ & $1.95743387539776 \times 10^{6}$ & $1.14452428728798$ & $5.56758470686469 \times 10^{-1}$ \\
			$0.25$ & $10^{50}$ & $9.08560314050930 \times 10^{32}$ & $1.14451123777010$ & $5.56432231315845 \times 10^{-1}$ \\
			$0.25$ & $10^{80}$ & $9.08561966098247 \times 10^{52}$ & $1.14451123777010$ & $5.56432231315845 \times 10^{-1}$ \\
			\hline
			
			$0.5$ & $10^{2}$ & $9.38867625838012$ & $1.14046195618689$ & $4.12443195411216 \times 10^{-1}$ \\
			$0.5$ & $10^{5}$ & $9.08595002633641 \times 10^{2}$ & $1.14004794692086$ & $1.44644052393304 \times 10^{-1}$ \\
			$0.5$ & $10^{10}$ & $1.95743389768829 \times 10^{6}$ & $1.13946004961744$ & $1.15011440181691 \times 10^{-1}$ \\
			$0.5$ & $10^{50}$ & $9.08560317898204 \times 10^{32}$ & $1.13944700009062$ & $1.14358961488025 \times 10^{-1}$ \\
			$0.5$ & $10^{80}$ & $9.08560481804203 \times 10^{52}$ & $1.13944700009062$ & $1.14358961488025 \times 10^{-1}$ \\
			\hline
		\end{tabular}
		
	\end{table}

	\begin{table}[H]
		\centering
		\caption{\footnotesize Values of $a(t)$, $\beta_{+}(t)$, and $\beta_{-}(t)$ at different times $t$ for various values of $C_{1}$. We consider $\rho_1$ = $\rho_2$ = 0.1, $C_2$ = 0.2, $\chi$ = 0.1, $\sigma_{+}$ = 0.01 and $\sigma_{-}$ = -0.05.}
		
		\label{tab:C1_values_full}
		\renewcommand{\arraystretch}{1.0} 
		\setlength{\tabcolsep}{12pt} 
		\tiny 
		\begin{tabular}{c c c c c}
			\hline
			$C_{1}$ & $t$ & $a(t)$ & $\beta_{+}(t)$ & $\beta_{-}(t)$ \\
			\hline
			$-5$ & $10^{2}$ & $10.0459496101976$ & $1.05569851755328$ & $-3.57749796131748 \times 10^{-1}$ \\
			$-5$ & $10^{5}$ & $9.08667094446427 \times 10^{2}$ & $1.05359794306293$ & $-4.29131439171937 \times 10^{-1}$ \\
			$-5$ & $10^{10}$ & $1.95743387288468 \times 10^{6}$ & $1.05300881238436$ & $-4.26281758498628 \times 10^{-1}$ \\
			$-5$ & $10^{50}$ & $9.08560354799414 \times 10^{32}$ & $1.05299576284506$ & $-4.26216511761807 \times 10^{-1}$ \\
			$-5$ & $10^{80}$ & $9.08559090814433 \times 10^{52}$ & $1.05299576284506$ & $-4.26216511761807 \times 10^{-1}$ \\
			\hline
			
			$-2.5$ & $10^{2}$ & $9.66109338972946$ & $1.08699230940393$ & $4.94271516245870 \times 10^{-2}$ \\
			$-2.5$ & $10^{5}$ & $9.08623681635766 \times 10^{2}$ & $1.08583188038260$ & $2.30236905572444 \times 10^{-2}$ \\
			$-2.5$ & $10^{10}$ & $1.95743406772157 \times 10^{6}$ & $1.08524374050937$ & $2.59289119518341 \times 10^{-2}$ \\
			$-2.5$ & $10^{50}$ & $9.08560360691514 \times 10^{32}$ & $1.08523069098814$ & $2.59941592034492 \times 10^{-2}$ \\
			$-2.5$ & $10^{80}$ & $9.08562155595090 \times 10^{52}$ & $1.08523069098814$ & $2.59941592034492 \times 10^{-2}$ \\
			\hline
			
			$0$ & $10^{2}$ & $9.40559942423711$ & $1.14855886624150$ & $1.00017040752812$ \\
			$0$ & $10^{5}$ & $9.08597467816733 \times 10^{2}$ & $1.14954323483822$ & $1.02446958347823$ \\
			$0$ & $10^{10}$ & $1.95743408906086 \times 10^{6}$ & $1.14895731161066$ & $1.02743025686642$ \\
			$0$ & $10^{50}$ & $9.08560408470774 \times 10^{32}$ & $1.14894426211311$ & $1.02749550466501$ \\
			$0$ & $10^{80}$ & $9.08560406869727 \times 10^{52}$ & $1.14894426211311$ & $1.02749550466501$ \\
			\hline
			
			$0.25$ & $10^{2}$ & $9.67821597244958$ & $1.12470686001941$ & $1.97061252307846$ \\
			$0.25$ & $10^{5}$ & $9.08629075032140 \times 10^{2}$ & $1.12751051683548$ & $2.04497494023390$ \\
			$0.25$ & $10^{10}$ & $1.95743386120812 \times 10^{6}$ & $1.12692687141964$ & $2.04799115855271$ \\
			$0.25$ & $10^{50}$ & $9.08560324838872 \times 10^{32}$ & $1.12691382193447$ & $2.04805640695885$ \\
			$0.25$ & $10^{80}$ & $9.08561778835941 \times 10^{52}$ & $1.12691382193447$ & $2.04805640695885$ \\
			\hline
			
			$0.5$ & $10^{2}$ & $1.00783585402990 \times 10^{1}$ & $1.09330795698491$ & $2.37921585602254$ \\
			$0.5$ & $10^{5}$ & $9.08677639192374 \times 10^{2}$ & $1.09655387086791$ & $2.49731283151759$ \\
			$0.5$ & $10^{10}$ & $1.95743408262577 \times 10^{6}$ & $1.09597124591564$ & $2.50038457816292$ \\
			$0.5$ & $10^{50}$ & $9.08560415414005 \times 10^{32}$ & $1.09595819645059$ & $2.50044982707496$ \\
			$0.5$ & $10^{80}$ & $9.08560896128015 \times 10^{52}$ & $1.09595819645059$ & $2.50044982707496$ \\
			\hline
		\end{tabular}

	\end{table} 
	
	\begin{table}[H]
		\centering
		
		\caption{\footnotesize Values of $a(t)$, $\beta_{+}(t)$, and $\beta_{-}(t)$ at different times $t$ for various values of $C_{2}$. We consider $\rho_1$ = $\rho_2$ = 0.1, $C_1$ = 0.2, $\chi$ = 0.1, $\sigma_{+}$ = 0.01 and $\sigma_{-}$ = -0.05.}
		
		\label{tab:C2_values_full}
		
		\renewcommand{\arraystretch}{1.0} 
		\setlength{\tabcolsep}{12pt} 
		\tiny 
		\begin{tabular}{c c c c c}
			\hline
			$C_{2}$ & $t$ & $a(t)$ & $\beta_{+}(t)$ & $\beta_{-}(t)$ \\
			\hline
			$-2$ & $10^{2}$ & $9.58409485486387$ & $1.86558991031420 \times 10^{-1}$ & $1.10967267121014$ \\
			$-2$ & $10^{5}$ & $9.08617239968484 \times 10^{2}$ & $1.42657848057070 \times 10^{-1}$ & $1.13985799672665$ \\
			$-2$ & $10^{10}$ & $1.95743406569289 \times 10^{6}$ & $1.42023303573352 \times 10^{-1}$ & $1.14282531897339$ \\
			$-2$ & $10^{50}$ & $9.08560420589568 \times 10^{32}$ & $1.42010253588681 \times 10^{-1}$ & $1.14289056684279$ \\
			$-2$ & $10^{80}$ & $9.08560683617993 \times 10^{52}$ & $1.42010253588680 \times 10^{-1}$ & $1.14289056684279$ \\
			\hline
			
			$-1$ & $10^{2}$ & $9.45569793094089$ & $5.42189678068712 \times 10^{-1}$ & $1.11662143958044$ \\
			$-1$ & $10^{5}$ & $9.08603186643463 \times 10^{2}$ & $5.18332976612503 \times 10^{-1}$ & $1.14632665159738$ \\
			$-1$ & $10^{10}$ & $1.95743387226677 \times 10^{6}$ & $5.17720661646618 \times 10^{-1}$ & $1.14929316133707$ \\
			$-1$ & $10^{50}$ & $9.08560314899101 \times 10^{32}$ & $5.17707611875129 \times 10^{-1}$ & $1.14935840924404$ \\
			$-1$ & $10^{80}$ & $9.08560706079767 \times 10^{52}$ & $5.17707611875129 \times 10^{-1}$ & $1.14935840924404$ \\
			\hline
			
			$0$ & $10^{2}$ & $9.40385725325067$ & $1.04469518483916$ & $1.11044727336720$ \\
			$0$ & $10^{5}$ & $9.08597454975933 \times 10^{2}$ & $1.04170645008748$ & $1.13918203503830$ \\
			$0$ & $10^{10}$ & $1.95743408613669 \times 10^{6}$ & $1.04111633760827$ & $1.14214739228934$ \\
			$0$ & $10^{50}$ & $9.08560411727468 \times 10^{32}$ & $1.04110328806875$ & $1.14221264013515$ \\
			$0$ & $10^{80}$ & $9.08558331113058 \times 10^{52}$ & $1.04110328806875$ & $1.14221264013515$ \\
			\hline
			
			$1$ & $10^{2}$ & $9.47169107721698$ & $1.52598044422872$ & $1.09030059982321$ \\
			$1$ & $10^{5}$ & $9.08604620658715 \times 10^{2}$ & $1.54377098978625$ & $1.11788111177764$ \\
			$1$ & $10^{10}$ & $1.95743405285729 \times 10^{6}$ & $1.54320305290712$ & $1.12084534029065$ \\
			$1$ & $10^{50}$ & $9.08560350468533 \times 10^{32}$ & $1.54319000358729$ & $1.12091058813560$ \\
			$1$ & $10^{80}$ & $9.08560329503041 \times 10^{52}$ & $1.54319000358730$ & $1.12091058813560$ \\
			\hline
			
			$2$ & $10^{2}$ & $9.60463535545249$ & $1.85665620809048$ & $1.07275479870357$ \\
			$2$ & $10^{5}$ & $9.08619047765233 \times 10^{2}$ & $1.89435050423508$ & $1.09940945318153$ \\
			$2$ & $10^{10}$ & $1.95743392513310 \times 10^{6}$ & $1.89380475039630$ & $1.10237287818814$ \\
			$2$ & $10^{50}$ & $9.08560320760793 \times 10^{32}$ & $1.89379170129257$ & $1.10243812605387$ \\
			$2$ & $10^{80}$ & $9.08562072933600 \times 10^{52}$ & $1.89379170129257$ & $1.10243812605387$ \\
			\hline
		\end{tabular}
		
	\end{table}   
	
	\begin{table}[H]
		\centering
		
		\caption{\footnotesize Values of $a(t)$, $\beta_{+}(t)$, and $\beta_{-}(t)$ at different times $t$ for various values of $\rho_{1}$. We consider $\rho_2$ = 0.1, $C_1$ = 0.1, $C_{2}$ = 0.2, $\chi$ = 0.1, $\sigma_{+}$ = 0.01 and $\sigma_{-}$ = -0.05.}
		
		\label{tab:rho1_values_full}
		\renewcommand{\arraystretch}{1.0} 
		\setlength{\tabcolsep}{12pt} 
		\tiny 
		\begin{tabular}{c c c c c}
			\hline
			$\rho_{1}$ & $t$ & $a(t)$ & $\beta_{+}(t)$ & $\beta_{-}(t)$ \\
			\hline
			$10^{-10}$ & $10^{2}$ & $6.13537541337349$ & $1.21193551556382$ & $1.07394095192046$ \\
			$10^{-10}$ & $10^{5}$ & $1.91094467035759 \times 10^{2}$ & $1.22535785575048$ & $1.22760498425854$ \\
			$10^{-10}$ & $10^{10}$ & $6.04275163522618 \times 10^{4}$ & $1.17436744264121$ & $1.49024501901334$ \\
			$10^{-10}$ & $10^{50}$ & $9.08560388140065 \times 10^{29}$ & $1.06903487808059$ & $2.01693429922999$ \\
			$10^{-10}$ & $10^{80}$ & $9.08560282570388 \times 10^{49}$ & $1.06903487807998$ & $2.01693429923302$ \\
			\hline
			
			$0.5$ & $10^{2}$ & $1.57097229921255 \times 10^{1}$ & $1.07399284234006$ & $1.02939545733188$ \\
			$0.5$ & $10^{5}$ & $1.55363381879641 \times 10^{3}$ & $1.07346388474391$ & $1.03862733235199$ \\
			$0.5$ & $10^{10}$ & $3.34716512859322 \times 10^{6}$ & $1.07326255030632$ & $1.03964072483693$ \\
			$0.5$ & $10^{50}$ & $1.55361646796602 \times 10^{33}$ & $1.07325808743086$ & $1.03966303928147$ \\
			$0.5$ & $10^{80}$ & $1.55361773264867 \times 10^{53}$ & $1.07325808743086$ & $1.03966303928147$ \\
			\hline
			
			$1$ & $10^{2}$ & $1.97261204044842 \times 10^{1}$ & $1.05261471761680$ & $1.02182550119397$ \\
			$1$ & $10^{5}$ & $1.95744914396313 \times 10^{3}$ & $1.05210875223729$ & $1.02766110479724$ \\
			$1$ & $10^{10}$ & $4.21716341327917 \times 10^{6}$ & $1.05198174129581$ & $1.02829951211929$ \\
			$1$ & $10^{50}$ & $1.95743391614805 \times 10^{33}$ & $1.05197892985686$ & $1.02831356934756$ \\
			$1$ & $10^{80}$ & $1.95743470002773 \times 10^{53}$ & $1.05197892985686$ & $1.02831356934756$ \\
			\hline
			
			$1.5$ & $10^{2}$ & $2.25483345334953 \times 10^{1}$ & $1.04301309193384$ & $1.01824891136023$ \\
			$1.5$ & $10^{5}$ & $2.24071658695753 \times 10^{3}$ & $1.04256262154419$ & $1.02270839043128$ \\
			$1.5$ & $10^{10}$ & $4.82744721871633 \times 10^{6}$ & $1.04246562831717$ & $1.02319558936594$ \\
			$1.5$ & $10^{50}$ & $2.24070262440667 \times 10^{33}$ & $1.04246348278769$ & $1.02320631703563$ \\
			$1.5$ & $10^{80}$ & $2.24071006042442 \times 10^{53}$ & $1.04246348278769$ & $1.02320631703563$ \\
			\hline
			
			$2$ & $10^{2}$ & $2.47966191439827 \times 10^{1}$ & $1.03726655232483$ & $1.01604632976990$ \\
			$2$ & $10^{5}$ & $2.46622581261623 \times 10^{3}$ & $1.03686109071109$ & $1.01973038867486$ \\
			$2$ & $10^{10}$ & $5.31329348656977 \times 10^{6}$ & $1.03678099044448$ & $1.02013256356785$ \\
			$2$ & $10^{50}$ & $2.46621222916853 \times 10^{33}$ & $1.03677921934914$ & $1.02014141906127$ \\
			$2$ & $10^{80}$ & $2.46621529513349 \times 10^{53}$ & $1.03677921934914$ & $1.02014141906127$ \\
			\hline
		\end{tabular}
		
	\end{table}   
	
	\begin{table}[H]
		\centering
		
		\caption{\footnotesize Values of $a(t)$, $\beta_{+}(t)$, and $\beta_{-}(t)$ at different times $t$ for various values of $\rho_{2}$. We consider $\rho_1$ = 0.1, $C_1$ = 0.1, $C_2$ = 0.2, $\chi$ = 0.1, $\sigma_{+}$ = 0.01 and $\sigma_{-}$ = -0.05.}
		
		\label{tab:rho2_values_full}
		
		\renewcommand{\arraystretch}{1.0} 
		\setlength{\tabcolsep}{12pt} 
		\tiny 
		\begin{tabular}{c c c c c}
			\hline
			$\rho_{2}$ & $t$ & $a(t)$ & $\beta_{+}(t)$ & $\beta_{-}(t)$ \\
			\hline
			$10^{-10}$ & $10^{2}$ & $9.31251450041558$ & $1.16317929402583$ & $1.05612525143250$ \\
			$10^{-10}$ & $10^{5}$ & $9.08583244792929 \times 10^{2}$ & $1.16435557014773$ & $1.08265766956361$ \\
			$10^{-10}$ & $10^{10}$ & $1.95743386019092 \times 10^{6}$ & $1.16376977156232$ & $1.08562055622949$ \\
			$10^{-10}$ & $10^{50}$ & $9.08560338252777 \times 10^{32}$ & $1.16375672205595$ & $1.08568580410055$ \\
			$10^{-10}$ & $10^{80}$ & $9.08560658601393 \times 10^{52}$ & $1.16375672205595$ & $1.08568580410055$ \\
			\hline
			
			$0.5$ & $10^{2}$ & $1.03365221764273 \times 10^{1}$ & $1.09544380032210$ & $1.04037098658098$ \\
			$0.5$ & $10^{5}$ & $9.08801509926965 \times 10^{2}$ & $1.09587753596146$ & $1.06525538870264$ \\
			$0.5$ & $10^{10}$ & $1.95743386859638 \times 10^{6}$ & $1.09529177001095$ & $1.06821805864287$ \\
			$0.5$ & $10^{50}$ & $9.08560319950136 \times 10^{32}$ & $1.09527872050409$ & $1.06828330651598$ \\
			$0.5$ & $10^{80}$ & $9.08561549663051 \times 10^{52}$ & $1.09527872050409$ & $1.06828330651598$ \\
			\hline
			
			$1$ & $10^{2}$ & $1.14448040949290 \times 10^{1}$ & $1.07070564685937$ & $1.03248179752514$ \\
			$1$ & $10^{5}$ & $9.09210075290272 \times 10^{2}$ & $1.07054451573514$ & $1.05550164419676$ \\
			$1$ & $10^{10}$ & $1.95743389385094 \times 10^{6}$ & $1.06995886246497$ & $1.05846367549512$ \\
			$1$ & $10^{50}$ & $9.08560315055366 \times 10^{32}$ & $1.06994581295842$ & $1.05852892336613$ \\
			$1$ & $10^{80}$ & $9.08561168317314 \times 10^{52}$ & $1.06994581295842$ & $1.05852892336613$ \\
			\hline
			
			$1.5$ & $10^{2}$ & $1.23573221723894 \times 10^{1}$ & $1.05858233172539$ & $1.02813652949940$ \\
			$1.5$ & $10^{5}$ & $9.09725545614107 \times 10^{2}$ & $1.05805933413832$ & $1.04966808115506$ \\
			$1.5$ & $10^{10}$ & $1.95743410173794 \times 10^{6}$ & $1.05747383808440$ & $1.05262926017210$ \\
			$1.5$ & $10^{50}$ & $9.08560367387947 \times 10^{32}$ & $1.05746078858645$ & $1.05269450799974$ \\
			$1.5$ & $10^{80}$ & $9.08561048684997 \times 10^{52}$ & $1.05746078858645$ & $1.05269450799974$ \\
			\hline
			
			$2$ & $10^{2}$ & $1.31237110429439 \times 10^{1}$ & $1.05110067783972$ & $1.02528587180071$ \\
			$2$ & $10^{5}$ & $9.10321669422148 \times 10^{2}$ & $1.05034602059369$ & $1.04561929149445$ \\
			$2$ & $10^{10}$ & $1.95743405041348 \times 10^{6}$ & $1.04976071347423$ & $1.04857946416862$ \\
			$2$ & $10^{50}$ & $9.08560330232812 \times 10^{32}$ & $1.04974766397312$ & $1.04864471201181$ \\
			$2$ & $10^{80}$ & $9.08559990322096 \times 10^{52}$ & $1.04974766397312$ & $1.04864471201181$ \\
			\hline
		\end{tabular}
		
	\end{table}

	\begin{table}[H]
		\centering
		
		\caption{\footnotesize Values of $a(t)$, $\beta_{+}(t)$, and $\beta_{-}(t)$ at different times $t$ for various values of $a_{0}$. We consider $\rho_1$ = $\rho_2$ = 0.1, $C_1$ = 0.1, $C_2$ = 0.2, $\chi$ = 0.1, $\sigma_{+}$ = 0.01 and $\sigma_{-}$ = -0.05.}
		
		\label{tab:a0_values_full}
		
		\renewcommand{\arraystretch}{1.0} 
		\setlength{\tabcolsep}{12pt} 
		\tiny 
		\begin{tabular}{c c c c c}
			\hline
			$a_{0}$ & $t$ & $a(t)$ & $\beta_{+}(t)$ & $\beta_{-}(t)$ \\
			\hline
			$0.5$ & $10^{2}$ & $4.80204761297754$ & $1.95181190277162$ & $1.03844566171505$ \\
			$0.5$ & $10^{5}$ & $4.54310136790279 \times 10^{2}$ & $1.98058793327090$ & $1.12821705130857$ \\
			$0.5$ & $10^{10}$ & $9.78717024745127 \times 10^{5}$ & $1.97827304191631$ & $1.14005239030363$ \\
			$0.5$ & $10^{50}$ & $4.54280211071282 \times 10^{32}$ & $1.97822084420954$ & $1.14031338145478$ \\
			$0.5$ & $10^{80}$ & $4.54279450873148 \times 10^{52}$ & $1.97822084420954$ & $1.14031338145478$ \\
			\hline
			
			$0.75$ & $10^{2}$ & $7.07180295765039$ & $1.35131472552600$ & $1.07878073558528$ \\
			$0.75$ & $10^{5}$ & $6.81450112382086 \times 10^{2}$ & $1.35719199222652$ & $1.12444425690487$ \\
			$0.75$ & $10^{10}$ & $1.46807550290792 \times 10^{6}$ & $1.35615486547585$ & $1.12971033007366$ \\
			$0.75$ & $10^{50}$ & $6.81420313150436 \times 10^{32}$ & $1.35613166640850$ & $1.12982632621592$ \\
			$0.75$ & $10^{80}$ & $6.81420228037576 \times 10^{52}$ & $1.35613166640850$ & $1.12982632621592$ \\
			\hline
			
			$1$ & $10^{2}$ & $9.40643072064142$ & $1.14994344357657$ & $1.05365266875245$ \\
			$1$ & $10^{5}$ & $9.08597634039731 \times 10^{2}$ & $1.15104194490590$ & $1.08005230377670$ \\
			$1$ & $10^{10}$ & $1.95743409120472 \times 10^{6}$ & $1.15045614526105$ & $1.08301519552582$ \\
			$1$ & $10^{50}$ & $9.08560404500688 \times 10^{32}$ & $1.15044309576478$ & $1.08308044334640$ \\
			$1$ & $10^{80}$ & $9.08561578521900 \times 10^{52}$ & $1.15044309576478$ & $1.08308044334640$ \\
			\hline
			
			$1.5$ & $10^{2}$ & $1.41020528834880 \times 10^{1}$ & $1.04289234554498$ & $1.02542629678675$ \\
			$1.5$ & $10^{5}$ & $1.36289557831965 \times 10^{3}$ & $1.04240690761140$ & $1.03727807720020$ \\
			$1.5$ & $10^{10}$ & $2.93615113203478 \times 10^{6}$ & $1.04214550949215$ & $1.03859503961123$ \\
			$1.5$ & $10^{50}$ & $1.36284061467459 \times 10^{33}$ & $1.04213970970556$ & $1.03862403864395$ \\
			$1.5$ & $10^{80}$ & $1.36284197721022 \times 10^{53}$ & $1.04213970970556$ & $1.03862403864395$ \\
			\hline
			
			$2$ & $10^{2}$ & $1.88018609413325 \times 10^{1}$ & $1.01735626756389$ & $1.01444591590451$ \\
			$2$ & $10^{5}$ & $1.81719399779706 \times 10^{3}$ & $1.01681413748484$ & $1.02112330364217$ \\
			$2$ & $10^{10}$ & $3.91486817229359 \times 10^{6}$ & $1.01666681526657$ & $1.02186410605438$ \\
			$2$ & $10^{50}$ & $1.81712082341809 \times 10^{33}$ & $1.01666355288374$ & $1.02188041801046$ \\
			$2$ & $10^{80}$ & $1.81712107574496 \times 10^{53}$ & $1.01666355288374$ & $1.02188041801046$ \\
			\hline
		\end{tabular}
		
	\end{table}

	\begin{table}[H]
		\centering
		
		\caption{\footnotesize Values of $a(t)$, $\beta_{+}(t)$, and $\beta_{-}(t)$ at different times $t$ for various values of $\beta_{{+}_0}$. We consider $\rho_1$ = $\rho_2$ = 0.1, $C_1$ = 0.5, $C_2$ = 0.1, $\chi$ = 0.1, $\sigma_{+}$ = 0.01 and $\sigma_{-}$ = -0.05.}
		
		\label{tab:beta_plus_0_values_full}
		
		\renewcommand{\arraystretch}{1.0} 
		\setlength{\tabcolsep}{12pt} 
		\tiny 
		\begin{tabular}{c c c c c}
			\hline
			$\beta_{{+}_0}$ & $t$ & $a(t)$ & $\beta_{+}(t)$ & $\beta_{-}(t)$ \\
			\hline
			$-1.0$ & $10^{2}$ & $9.43237044372862$ & $-8.99043186232376 \times 10^{-1}$ & $1.36382178687463$ \\
			$-1.0$ & $10^{5}$ & $9.08600704718301 \times 10^{2}$ & $-8.99396005099798 \times 10^{-1}$ & $1.40285878107158$ \\
			$-1.0$ & $10^{10}$ & $1.95743409151642 \times 10^{6}$ & $-8.99983308497155 \times 10^{-1}$ & $1.40583511208698$ \\
			$-1.0$ & $10^{50}$ & $9.08560403814084 \times 10^{32}$ & $-8.99996358008473 \times 10^{-1}$ & $1.40590036004201$ \\
			$-1.0$ & $10^{80}$ & $9.08559013717371 \times 10^{52}$ & $-8.99996358008473 \times 10^{-1}$ & $1.40590036004201$ \\
			\hline
			
			$-0.5$ & $10^{2}$ & $9.42929721877525$ & $-3.99195839385121 \times 10^{-1}$ & $1.33943316905596$ \\
			$-0.5$ & $10^{5}$ & $9.08600446621962 \times 10^{2}$ & $-3.99599410049934 \times 10^{-1}$ & $1.37743317290128$ \\
			$-0.5$ & $10^{10}$ & $1.95743396410633 \times 10^{6}$ & $-4.00186769776125 \times 10^{-1}$ & $1.38040839237232$ \\
			$-0.5$ & $10^{50}$ & $9.08560322294976 \times 10^{32}$ & $-4.00199819294202 \times 10^{-1}$ & $1.38047364034721$ \\
			$-0.5$ & $10^{80}$ & $9.08559945542423 \times 10^{52}$ & $-4.00199819294202 \times 10^{-1}$ & $1.38047364034721$ \\
			\hline
			
			$0$ & $10^{2}$ & $9.42618440907028$ & $1.00651226628816 \times 10^{-1}$ & $1.31483842364292$ \\
			$0$ & $10^{5}$ & $9.08600057552586 \times 10^{2}$ & $1.00196469729132 \times 10^{-1}$ & $1.35180063670056$ \\
			$0$ & $10^{10}$ & $1.95743386040079 \times 10^{6}$ & $9.96090526781551 \times 10^{-2}$ & $1.35477474759664$ \\
			$0$ & $10^{50}$ & $9.08560325259010 \times 10^{32}$ & $9.95960031553549 \times 10^{-2}$ & $1.35483999558118$ \\
			$0$ & $10^{80}$ & $9.08559763611078 \times 10^{52}$ & $9.95960031553549 \times 10^{-2}$ & $1.35483999558118$ \\
			\hline
			
			$0.5$ & $10^{2}$ & $9.42303283969463$ & $6.00498273941076 \times 10^{-1}$ & $1.29003130223167$ \\
			$0.5$ & $10^{5}$ & $9.08599673782568 \times 10^{2}$ & $5.99991838762176 \times 10^{-1}$ & $1.32595475249997$ \\
			$0.5$ & $10^{10}$ & $1.95743396314812 \times 10^{6}$ & $5.99404363984586 \times 10^{-1}$ & $1.32892775439970$ \\
			$0.5$ & $10^{50}$ & $9.08560406509050 \times 10^{32}$ & $5.99391314466451 \times 10^{-1}$ & $1.32899300234694$ \\
			$0.5$ & $10^{80}$ & $9.08563442780767 \times 10^{52}$ & $5.99391314466452 \times 10^{-1}$ & $1.32899300234694$ \\
			\hline
			
			$1.0$ & $10^{2}$ & $9.41984124621913$ & $1.10034518359955$ & $1.26500409726278$ \\
			$1.0$ & $10^{5}$ & $9.08599388493651 \times 10^{2}$ & $1.09978655088542$ & $1.29988777691884$ \\
			$1.0$ & $10^{10}$ & $1.95743408783435 \times 10^{6}$ & $1.09919901832531$ & $1.30285966742779$ \\
			$1.0$ & $10^{50}$ & $9.08560383314576 \times 10^{32}$ & $1.09918596881132$ & $1.30292491534033$ \\
			$1.0$ & $10^{80}$ & $9.08560985311799 \times 10^{52}$ & $1.09918596881132$ & $1.30292491534033$ \\
			\hline
		\end{tabular}

	\end{table}

	\begin{table}[H]
		\centering
		
		\caption{\footnotesize Values of $a(t)$, $\beta_{+}(t)$, and $\beta_{-}(t)$ at different times $t$ for various values of $\beta_{{-}_0}$. We consider $\rho_1$ = $\rho_2$ = 0.1, $C_1$ = 0.1, $C_2$ = 0.5, $\chi$ = 0.1, $\sigma_{+}$ = 0.01 and $\sigma_{-}$ = -0.05.}
		
		\label{tab:beta_minus_0_values_full}
		
		\renewcommand{\arraystretch}{1.0} 
		\setlength{\tabcolsep}{12pt} 
		\tiny 
		\begin{tabular}{c c c c c}
			\hline
			$\beta_{{-}_0}$ & $t$ & $a(t)$ & $\beta_{+}(t)$ & $\beta_{-}(t)$ \\
			\hline
			$-1.0$ & $10^{2}$ & $9.41019591628443$ & $1.20210372472720$ & $-9.47925260912452 \times 10^{-1}$ \\
			$-1.0$ & $10^{5}$ & $9.08598053865428 \times 10^{2}$ & $1.20529952553825$ & $-9.21643283937094 \times 10^{-1}$ \\
			$-1.0$ & $10^{10}$ & $1.95743387771909 \times 10^{6}$ & $1.20471594452625$ & $-9.18680514580093 \times 10^{-1}$ \\
			$-1.0$ & $10^{50}$ & $9.08560314605763 \times 10^{32}$ & $1.20470289504231$ & $-9.18615266711498 \times 10^{-1}$ \\
			$-1.0$ & $10^{80}$ & $9.08560096805416 \times 10^{52}$ & $1.20470289504231$ & $-9.18615266711499 \times 10^{-1}$ \\
			\hline
			
			$-0.5$ & $10^{2}$ & $9.41344003207248$ & $1.22751716287134$ & $-4.48741660052993 \times 10^{-1}$ \\
			$-0.5$ & $10^{5}$ & $9.08598360443362 \times 10^{2}$ & $1.23175891931113$ & $-4.22519198153319 \times 10^{-1}$ \\
			$-0.5$ & $10^{10}$ & $1.95743408460860 \times 10^{6}$ & $1.23117644737100$ & $-4.19556487491320 \times 10^{-1}$ \\
			$-0.5$ & $10^{50}$ & $9.08560378393213 \times 10^{32}$ & $1.23116339790743$ & $-4.19491239669683 \times 10^{-1}$ \\
			$-0.5$ & $10^{80}$ & $9.08559986577534 \times 10^{52}$ & $1.23116339790743$ & $-4.19491239669683 \times 10^{-1}$ \\
			\hline
			
			$0$ & $10^{2}$ & $9.41664124748264$ & $1.25269906778350$ & $5.04576203721868 \times 10^{-2}$ \\
			$0$ & $10^{5}$ & $9.08598714176510 \times 10^{2}$ & $1.25798578557081$ & $7.66211806993265 \times 10^{-2}$ \\
			$0$ & $10^{10}$ & $1.95743386886611 \times 10^{6}$ & $1.25740442287600$ & $7.95838324338266 \times 10^{-2}$ \\
			$0$ & $10^{50}$ & $9.08560316355025 \times 10^{32}$ & $1.25739137341396$ & $7.96490803026848 \times 10^{-2}$ \\
			$0$ & $10^{80}$ & $9.08560853399209 \times 10^{52}$ & $1.25739137341396$ & $7.96490803026848 \times 10^{-2}$ \\
			\hline
			
			$0.5$ & $10^{2}$ & $9.41980128642747$ & $1.27765744681799$ & $5.49671931224922 \times 10^{-1}$ \\
			$0.5$ & $10^{5}$ & $9.08599021964904 \times 10^{2}$ & $1.28398813141633$ & $5.75777176378408 \times 10^{-1}$ \\
			$0.5$ & $10^{10}$ & $1.95743385967244 \times 10^{6}$ & $1.28340787787475$ & $5.78739770406624 \times 10^{-1}$ \\
			$0.5$ & $10^{50}$ & $9.08560327471523 \times 10^{32}$ & $1.28339482842359$ & $5.78805018275960 \times 10^{-1}$ \\
			$0.5$ & $10^{80}$ & $9.08560580783918 \times 10^{52}$ & $1.28339482842359$ & $5.78805018275960 \times 10^{-1}$ \\
			\hline
			
			$1.0$ & $10^{2}$ & $9.42292228767818$ & $1.30239975881252$ & $1.04890066009200$ \\
			$1.0$ & $10^{5}$ & $9.08599417736373 \times 10^{2}$ & $1.30977338701646$ & $1.07494814811381$ \\
			$1.0$ & $10^{10}$ & $1.95743394379859 \times 10^{6}$ & $1.30919424301502$ & $1.07791068277867$ \\
			$1.0$ & $10^{50}$ & $9.08560321949883 \times 10^{32}$ & $1.30918119357823$ & $1.07797593063107$ \\
			$1.0$ & $10^{80}$ & $9.08560689144900 \times 10^{52}$ & $1.30918119357823$ & $1.07797593063107$ \\
			\hline
		\end{tabular}
		
	\end{table}

\end{appendices}

\end{document}